\documentclass[twocolumn]{aastex6}
\usepackage{graphicx}
\usepackage{xcolor}
\usepackage{amsmath}
\usepackage{soul}

\shorttitle{Properties of trans-fast Magnetosonic Jets in  BH magnetospheres}
\shortauthors{Pu \& Takahashi}

\begin{document}

\title{Properties of Trans-fast Magnetosonic Jets in  Black Hole Magnetospheres}

\author{Hung-Yi Pu\altaffilmark{1} \& Masaaki Takahashi\altaffilmark{2}}
\altaffiltext{1}{Perimeter Institute for Theoretical Physics, 31 Caroline Street North, Waterloo, ON, N2L 2Y5, Canada}
\altaffiltext{2}{Department of Physics and Astronomy, Aichi University of Education, Kariya, Aichi 448-8542, Japan}

\begin{abstract}
Traveling across several order of magnitude in distance, relativistic jets from strong gravity region 
to asymptotic flat spacetime region are believed to consist of several  general relativistic
magnetohydrodynamic (GRMHD) processes. We present a semi-analytical approach for 
modeling the global  structures of a trans-fast magnetosonic 
relativistic jet, which should be ejected from a plasma source nearby a black hole    in a  funnel region  enclosed by  dense accreting  flow and  also disk corona 
around  the black hole. 
Our model consistently includes the inflow and outflow part of the GRMHD solution along  the magnetic
field lines penetrating the black hole horizon. After the rotational 
energy of the black hole is extracted electromagnetically by the  negative energy
GRMHD inflow, the  huge electromagnetic energy flux then propagates from the inflow to the outflow 
region across the plasma source, and in the outflow region the electromagnetic energy
converts to the fluid kinetic energy.  Eventually, the accelerated
outflow  must exceed the fast-magnetosonic  wave speed. We apply the 
semi-analytical  trans-fast magnetosonic flow model to the black hole magnetosphere 
for both parabolic and split-monopole magnetic field configurations, and discuss the general flow 
properties;  that is,  jet acceleration, jet magnetization, and the locations of some 
characteristic surfaces of the black hole magnetosphere. We have confirmed that, at large distance, the GRMHD jet solutions are in good agreement with the previously known
trans-fast special relativistic magnetohydrodynamic (SRMHD) jet properties, as expected. 
The flexibility of the model provides a prompt and heuristic way to approximate the global GRMHD 
trans-fast magnetosonic  jet properties.

\end{abstract}

\keywords{accretion, accretion disks --- black hole physics --- galaxies: jets --- magnetic fields --- magnetohydrodynamics (MHD)}

\section{Introduction}
Black holes with accreting matters are believed to be the central engines of the observed relativistic jets from micorquasars \citep[][]{fen04,mil12,rus17}, active galactic nuclei (AGN) \citep[][]{hom15,had16,bru17,pus17,mck18}, and  presumably gamma-ray bursts \citep[][]{cha12,nav17,ryd17}.
Travelling across several order of magnitude in distance from the black hole  horizon ($r\sim r_{\rm g}$, where $r_{\rm g}=GM_{\bullet}/c^{2}$ is the gravitational radius and $M_{\bullet}$ is the black hole mass) to large distance, a relativistic jet formulated in the magnetosphere of a  black hole is  among the most efficient way to accelerate particles and redistribute energy and angular momentum from small to large scale. For example, jets from AGN can extends to a scale larger than the Bondi radius ($\sim 10^{5-6} r_{\rm g}$) \citep[e.g.][]{alg17}, and even larger than the size of the host galaxy ( $>10^{8-9} r_{\rm g}$), providing mechanical feedback to the galaxy clusters \citep[e.g.][]{fab12} . 

Supported by several observational evidences \citep[e.g.,][]{hov12,kin17}, it is believed that large-scale magnetic field plays an important role in extracting the energy from the central region and in accelerating and collimating the jet. Current understanding for  the relativistic jets in the magnetohydrodynamical framework,  including both the special relativistic magnetohydrodynamics (SRMHD; excluding the effect of gravity) and the general relativistic magnetohydrodynamics (GRMHD; including the effect of gravity), provides the following pictures.
At the footpoint region of the jet,  where the strong gravity of the central black hole must be considered, large-scale magnetic filed penetrates the black hole  horizon at least near the funnel region of accreting gas, and the rotational energy of the black hole would be electromagnetically extracted outward by the GRMHD flow \citep[e.g.,][]{bla77,tak90,koi03,mck06,haw06}. At far region from the central black hole, where the spacetime becomes almost flat and SRMHD become a good approximation, the magnetic energy gradually converts to particle kinetic energy, and the flow accelerates to its terminal velocity \citep[][]{cam86a,cam86b,cam87,bes98,fen95,fen96,tak98,fen01,vla04,bes06,tch08,tch09,lyu09,kom09,lyu10,tch10}. 

More specifically, a full and consistent model of  black hole jet formation across several order of magnitude in distance, from near zone ($r\sim r_{g}$) to far zone ($r> 10^{4-5} r_{g}$), is challenging due to at least the following four reasons.  The {\it First} one is about the configuration of the magnetic fields. The equation of motion for the MHD flow, 
\begin{equation}\label{eq:eom}
T^{\mu\nu}_{;\nu}=0\;,
\end{equation}
where the stress energy tensor $T^{\mu\nu}=T^{\mu\nu}_{\rm EM}+T^{\mu\nu}_{\rm FL}$ consists of the eletromangetic part $T^{\mu\nu}_{\rm EM}$ and the fluid part $T^{\mu\nu}_{\rm FL}$,
 can be decomposed into the {\it force-balance equation} between magnetic field lines (the Grad-Shafranov equation) and 
the {\it wind equation} along a magnetic field line (the relativistic Bernoulli equation).
The former describes the magnetic field configuration, while the latter describes the jet acceleration \citep[][]{nit91,bes09}. However, in general, solving the force-balance equation analytically  is complicated and usually computational demanding \citep[e.g.,][]{fen95,fen96,nat14,pan17,hua19}. 
The {\it Second} one is the condition at the fast-magnetosonic surface (FMS) of the jet, where the  jet velocity becomes the fast-magnetosonic wave speed. 
 The FMS should be located at a finite radius on the way to a distant region  in the black hole magnetosphere \citep[][]{fen97,bes98,tom01,bes06}.  Generally, a complicated critical condition analysis at the FMS is required.
 The {\it third} one is related to the plasma source region of the jet. 
 In the black hole magentosphere, due to the dominant gravity near the black hole and the dominant centrifugal force by the Lorentz force away from the black hole,
the inflow  and outflow regions and therefore a stagnation surface must coexist\citep{tak90,mck04,mck06}, where the stagnation surface 
separates the inflow/outflow regions.
 At the stagnation surface  the matching condition for the two zones is necessary \citep[][]{pu15}, where additional discussion of the state of the plasma source  should be required  (e.g. the electron-position pair creation etc.). 
With these difficulties, to date, the insight of the global feature includes both near and far regions, was only possible by performing a large-scale GRMHD simulation \citep[e.g.][]{mck06, lis18,chat19}.
Furthermore, in the funnel region close to the black hole axis which we are interested in here, the  radially self-similar approach \citep[][]{bla82,vla00,pol13,pol14} is not applicable, while the meridional self-similarity can provide a more suitable alternative \citep[][]{sau94,mel06,tsi10,glo14,chan18}.

In this paper, we present a semi-analytical approach, an attractive alternative, to include the key features of the above mentioned physical processes.
Especially, we focus on magnetic-energy-dominated flows which are capable to extract the black hole rotational energy  near the plasma source and kinetic-energy-dominated jet structure at a far distant region. 
The presented model  
provides a prompt, flexible, and heuristic way to investigate the  trans-fast magnetosonic jet structures and properties semi-analytically.
Our model is an application of 
solving the relativistic Bernoulli equation along magnetic field lines in an algebraic way via prescriptions of the poloidal and toroidal (azimuthal) magnetic fields \citep[][hereafter TT03 and TT08, respectively]{tom03,tak08}, and  of consistently matching  the inflow/outflow flow solutions \citep[][]{pu15}.
In contract to the standard approaches for trans-magnetosonic flow, which employ the  regularity condition on the FMS to solve the relativistic Bernoulli equation \citep[e.g.][]{tak90, tak02}, the novel approach provided by the former work,  TT03 and TT08, is  to solve the jet's Bernoulli equation by introducing a {\it regular function} of the poloidal electic-to-toroidal megnatic field amplitutdes ration $\xi$ in all regions of the jet. This allows us to obtain easily trans-fast magnetosonic flow solution for relativistic jet without the regularity condition analysis. The funciton $\xi$ has
sophisticated constraints of the magnetic field components at several characteristic surfaces such as the particle injection surface, Alfv\'en surface, event horizon, etc (see also 
\S2.2).
The idea that the outflow along the magnetic field lines in a magnetosphere can be determined by the distribution of poloidal and toroidal field has also been discussed in \citet[][]{con95,con99}. 

 The purpose of this paper is to demonstrate that the basic GRMHD jet structure can actually be well approximated from the above mentioned approaches. The underlying physical motivation is to mimic the GRMHD flow solution of equation (\ref{eq:eom}) with the known knowledge for  GRMHD theory, including some  solutions of the force-balance equation, and how the magnetosonic points are related to the solution of wind equations (see \S\ref{sec:model}).
 We ignore the gas pressure  in the flow, which is minor for the global flow structure \citep{cam86b}, and adopt the cold limit in our computations. Under such a limit, the stagnation surface can be solely determined by the locus across field lines where the force balance between gravity force and megneto-centrifugal force when the flow has vanishing initial velocity (or very slow sub-Alfv\'enic velocity). As the plasma loading is conserved along each field line (see \S\ref{sec:model}), the stagnation surface is treated as the plasma source region. The plasma source in the jet funnel region, which is beyond the scope of the current paper, is currently poorly understood \citep[e.g.][]{tak90, lev11,mos11,bro15,hir16,ori18}. Nevertheless, comparison of flow velocity and magnetic field configuration between numerical simulation\footnote{In numerical simulation, to avoid numerical issues, a density floor is usually set to ensure a minimum density in the simulation; materials are therefore arbitrary injected in the funnel region, which usually taking place close to the central black hole.} and semi-analytical solution along a large-scale magnetic field line in the funnel region shows similar properties \citep{pu15}, indicating the GRMHD flow velocity does not significantly affected by the plasma source.

 We do not solve the force-balance equation for the configuration of magnetic field, but instead assume a likely magnetic field shape. We are especially interested in the parabolic parabolic and split-monopole poloidal magnetic fields, because there are extensive SRMHD studies for the trans-fast megnetosonic flow in these two magnetic configurations  \citep[e.g. TT03;][]{bes98, bes06}, and we can compare our GRMHD flow solutions with the SRMHD flow features. Note also that a parabolic magnetic field geometry is commonly indicated by VLBI observations of AGN jets\citep[e.g.][]{had13,alg17,nak18},  and it is a common scene
of GRMHD numerical simulations of an accreting black hole system \citep[e.g.,][]{mck04,mck06,haw06}.
Therefore, we  choose to apply our model to these magnetic fields, and explore the general flow properties near the black hole, the jet acceleration of the outflow, as well as the characteristic surfaces, together with their dependence of black hole spin, field angular velocity, and outflow energy.
In the region far away from the black hole, the resulting semi-analytical  GRMHD outflow acceleration are in good qualitative and quantitate agreement with   previous analysis of semi-analytical MHD flow acceleration properties \citep[TT03;][]{bes98, bes06}.

The remainder of the paper is organized as follows. In sec \S\ref{sec:model} we describe the details of the model. The model parameters considered in this paper is given in \S\ref{sec:model_par}.
We then present result and the GRMHD flow properties in a black hole magnetosphere with parabolic  and split-monopole magnetic field lines are presented in \S\ref{sec:para} and \S\ref{sec:mono}, respectively. Comments on the limitation of the model in \S\ref{sec:limit}. Finally, the summary and future application of the model is given in \S\ref{sec:summary}.

\section{Trans-fast Magnetosonic Flows in a Black Hole Magnetosphere}\label{sec:model}
Our goal is develop a semi-analytical approach to model a  trans-fast magnetosonic flow  with a very large total specific energy along magnetic field lines attached on to the horizon, and include all the related key physics.
While it is well known that a mild plasma loading  can result in a slight deformation of the magnetic field lines and  the existence of the FMS by carefully solving both the trans-field and wind equations,  the following working compromise is adopted. 
First, a  poloidal force-free magnetic field configuration is applied   (i.e., the deformation of the magnetic fields due to the perturbation via plasma loading is ignored). Second, to preserve the existence of the FMS for a MHD flow, a sophisticated relation between the poloidal and toroidal components of the magnetic field is prescribed {\it in prior} (TT08).

As the model is an extension of the method presented in TT08, in the following we adopt the same signature [+,\;--,\;--,\;--]  for the Boyer-Lindquist metric, with $c=G=1$.  The dimensionless black hole spin parameter is denoted by $a$.

\subsection{Basic GRMHD Flow Properties}\label{sec:basic}
We assume a cold ideal GRMHD flow  that the gas pressure is  negligible, Then, there are four conserved quantities along the magnetic field line given by magnetic stream function $\Psi(r,\theta)=$ constant \citep[][]{cam86a,cam86b,cam87,tak90} : 
the angular velocity of the field line $\Omega_{F} (\Psi)$, the particle number flux per unit electromagnetic flux (mass loading) $\eta(\Psi)$, the total energy of the flow $E(\Psi)$, and the total angular momentum $L (\Psi)$
\begin{equation}\label{eq:omega_def}
    \Omega_{F} (\Psi)  =  -\frac{F_{t\theta}}{F_{\theta\phi}}\;,
\end{equation} 

\begin{equation} 
\label{eq:eta_def}
    \mu\eta(\Psi)  =  \frac{n \mu u_{p}}{\bar{B}_{p}}\;,
\end{equation}
\begin{equation}
\label{eq:E_def}
  \hat{E}(\Psi)\equiv \frac{E(\Psi)}{\mu}=u_{t}-\frac{\Omega_{\mathrm{F}}B_{\phi}}{4\pi\mu\eta}\;, 
\end{equation}

\begin{equation}
\label{eq:L_def}
\hat{L} (\Psi)\equiv \frac{L(\Psi)}{\mu}= -u_{\phi}-\frac{B_{\phi}}{4\pi\mu\eta}\;,
\end{equation}
where $F_{\mu\nu}$ is the electromagnetic tensor, and the magnetic field $B_{\alpha}\equiv(1/2)\epsilon_{\alpha\beta\gamma\delta}k^{\beta}F^{\gamma \delta}$ is defined by the time-like Killing vector $k^{\alpha}=(1,0,0,0)$, and  $\epsilon_{\alpha\beta\gamma\delta}$ is the Levi-Civita tensor. 
The hat symbols for $\hat{E}$ and $\hat{L}$ represents the physical quantities per specific enthalpy, $\mu$, which is given by $\mu=m_{\rm p}c^{2}$ (with the speed of light $c$ is momentary recovered here) and $m_{\rm p}$ is the particle's rest mass.
In addition, the poloidal velocity  and poloidal magnetic field are respectively defined by $u_{p}^{2}\equiv-(u^r u_{r}+u^{\theta} u_{\theta})$ and $B_{p}^{2}\equiv-(B^r B_{r}+B^{\theta} B_{\theta})$, and the rescaled poloidal magnetic field is defined by 
\begin{equation}
\bar{B}^2_{p}\equiv B^{2}_{p}/\rho^{2}_{w}\;, 
\end{equation}
where $\rho^{2}_{w}=g^{2}_{t\phi}-g_{tt}g_{\phi\phi}$. Similarly, we define the rescaled toroidal magnetic field by 
\begin{equation}
\bar{B}_{\phi}\equiv B_{\phi}/\rho^{2}_{w}\;.
\end{equation}

Along the large-scale magnetic field immersed in the black hole, there must exist inflow and outflow regions, divided by the location of the stagnation surface $r_{s}(\Psi)$.
In cold limit, 
the conservative quantities ($\hat{E},\hat{L}$) can be alternatively determined by ($r_{\rm s}, r_{\rm A}$), where  $r_{\rm A}=r_{\rm A}(\Psi)$ is the location of Alfv\'en surface, where the flow velocity equals to the poloidal Alfv\'en speed \citep[][]{tak90}. 
Hereafter we denote the inflow (or outflow) properties by the superscript ``--" (or ``'+'), and use the unsigned parameters for the base for both inflow and outflow.
We focus on cases when black hole rotational energy is extracted outward as the energy budget of the GRMHD flow, and the model applies for all rotating black holes ($a>0$), with $0<\Omega_{F}<\Omega_{\rm H}$ (i.e., the ``type II" flow defined in \citet[][]{tak90}), where $\Omega_{\rm H}$ is the angular velocity of the black hole.

\subsection{ Overview  of  TT08}\label{sec:TT08}
For a given streamline function and therefore the poloidal magnetic field configuration, a typical procedure for solving the wind equation along a stream line function then requires a fine-tune of the set of the conserved quantities: $\Omega_{F}, \eta, \hat{E}, \hat{L}$, such that a physical cold flow solution  pass both the Alfv\'en surface and the FMS, where the so-called critical condition should be satisfied. When the physical flow solution is obtained, the toroidal magnetic field structure is uniquely determined, that should be regular in all region of the flow. Note that, without the critical condition at the FMS, the toroidal magnetic field diverges there; i.e. such a solution is unphysical.

To always get a physical trans-fast magnetosonic solution, we focus on regularity of toroidal magnetic field. Now we introduce a regular function by relating the the ratio between the poloidal and toroidal magnetic field by the parameter $\beta$,\begin{equation}\label{eq:beta_def}
\beta(r; \Psi)\equiv\frac{B_{\phi}}{B_{p}}=\frac{\bar{B}_{\phi}}{\bar{B}_{p}}\;.
\end{equation}
A regular trans-magnetosonic flow solution can therefore be obtained; such
 a new analytical method without the critical conditions is proposed in TT08.  
The parameter $\beta$ can be interpreted as the inverse of the pitch angle,  or the bending angle of the magnetic field line. Alternatively, related to $\beta$, the poloidal electric-to-toroidal magnetic field amplitude ratio seen by a zero angular momentum observer (ZAMO) can be defined by
\begin{equation}\label{eq:beta2xi_def}
\xi^{2}(r; \Psi)=g_{\phi\phi}\frac{(\Omega_{F}-\omega)^{2}}{\beta^{2}}\;,
\end{equation}
where $\omega\equiv-g_{t\phi}/g_{\phi\phi}$. 

As a result, by defining the Alfv\'en Mach number
\begin{equation}
M^{2}=4\pi\mu n \dfrac{u_{p}^{2}}{\bar{B}_{p}^{2}}=4\pi\mu\eta \dfrac{u_{p}}{\bar{B}_{p}^{2}}\;,
\end{equation}
the wind equation can be rewritten with the following quadratic equation
\begin{equation}\label{eq:mach_eqn}
\mathcal{A} M^{4}-2\mathcal{B} M^{2}+\mathcal{C}=0\;,
\end{equation}
where the coefficients $\mathcal{A}$, $\mathcal{B}$, and $\mathcal{C}$ are just functions of the conserved quantities $\Omega_{F}$, $\hat{E}$, $\hat{L}$, magnetic field pitch angle, $\beta$, and the background metric,  $g_{\mu\nu}$. Readers can refer to TT08 for the details. The location of the Alfv\'en and the FMSs\footnote{It is expected that, close to the axis ($\theta\to0$), $\beta\to0$ and hence $M^{2}_{\rm FM}=M^{2}_{\rm AW}$, resulting a closer distance between the Alfv\'en surface and FMS \citep[see also][]{bes98,bes09,tch09}. Such effect is caused by the modification of the force-free magnetosphere due to plasma effect, and our approach can at most only provide an artificial mimic to such effect since the poloidal magnetic field configuration is prescribed and fixed in the computation, as explained and described in \S\ref{sec:model_psi}.} of the flow can be found at where the
Mach number equals
\begin{equation}
M^{2}=M^{2}_{\rm AW}\equiv\alpha\;,
\end{equation}
\begin{equation}
M^{2}=M^{2}_{\rm FM}\equiv\alpha+\beta^{2}\;,
\end{equation}
where $\alpha=g_{tt}+2g_{t\phi}\Omega_{F}+g_{\phi\phi}\Omega_{F}^{2}$. The poloidal velocity is
\begin{equation}
u_{p}^2=\dfrac{\hat{e}^{2}-\alpha}{\alpha+\beta^{2}}\;,
\end{equation}
with the Jacobian constant $\hat{e}\equiv\hat{E}-\Omega_{F}\hat{L}$.

\begin{table}
\caption{Restrictions on the regular function $\xi$ at several characteristic locations, for a physical solution of wind equation which passes a FMS (Fast-Magnetosonic Surface).}
\label{tab:list}
\begin{center}
\begin{tabular}{c c}
\hline
characteristic locations & $\xi^{2}$  \\
\hline
\hline
Event Horizon                                                       &        1                   \\
corotation point$^{\dag}$                                                      &      0                     \\
Alfv\'en point                                                        &         finite                 \\
Separation point & finite \\
\hline
\end{tabular}
\end{center}
$^{\dag}$ where $\Omega_{F}=-g_{t\phi}/g_{\phi\phi}$
\end{table}

By solving the wind equation in terms of $\beta$ (or, alternatively, $\xi$), the restriction for $\beta$ and $\xi$ when FMS exist in the solution are found. For our interest, $0<\Omega_{F}<\Omega_{\rm H}$, the conditions are summarized in Table \ref{tab:list} (see also Appendix A of TT08).
Along a magnetic field line $\Psi=$ constant, the function $\xi^{2}(r; \Psi)$ would have different form in the inner and outer region of the separation surface (the plasma source). 
For the outflow,  by considering the reasonable shape of the magnetic field at a distant region, we apply the following function form
\begin{equation}\label{eq:xi_def}
(\xi^{+})^{2}=1-\frac{1}{(\hat{E}^{+})^{2}}+\zeta_{0}\;,
\end{equation}
 where $\zeta_{0}$ is a constant associated with the flow acceleration in the super fast-magnetosonic regime. 
 Previous extensive studies for the SRMHD flow acceleration indicates a dependence of  different magnetic field geometry \citep[e.g.,][]{bes98, bes06,tch08, tch09,kom09}. 
We explored different choices of $\zeta_{0}$  and identify that $\zeta_{0}\cong0$ corresponds to a {\it linear acceleration} regime of the flow (in contrast to a slower, {\it logarithmic acceleration}; see more details in Appendix \ref{app:geo_effect}).
We therefore adopt 
\begin{equation}
\zeta_{0}=0 
\end{equation}
as the default value throughout the paper.
Although we consider a constant $(\xi^{+})^{2}$  along a flow, generally it is not necessary. Note that a constant $(\xi^{+})^{2}$ along a magnetic field line recovers that the ratio of polodial to toroidal filed is well-fitted by $1/(\Omega_{F}\sqrt{g_{\phi\phi}})$ (see also  \S \ref{sec:result_inout} for examples), as expected in the SRMHD jet  studies \citep[e.g.][]{lyu09}, and also found in previous GRMHD  simulations \citep[][]{mck06}. 

For the inflow,  from both the requirement at the horizon and the corotation point listed in Table \ref{tab:list},  a sophisticated form of $\xi$  has been suggested in TT08:
\begin{equation}\label{eq:xi_def2}
(\xi^{-})^{2}=\left[1+C\frac{\Delta }{\Sigma}\right]\left( \frac{\omega-\Omega_{F}}{\Omega_{\rm H}-\Omega_{F}}\right)^{2}\,,
\end{equation}
where $C$ is a constant. For our interest, $C$ is to be determined by a smooth connect for $\xi^{+}$ and $\xi^{-}$ (see also \S\ref{sec:model_mc} for the matching condition for outflow and inflow). Unlike $(\xi^{+})^{2}$, $(\xi^{-})^{2}$ cannot be a constant along the  magnetic field line,  as also can be seen in Table \ref{tab:list}.

\subsection{Magnetic Field Configurations} \label{sec:model_psi}
We focus on magnetically dominated flow at least in the jet formation region, and therefore assume the force-free magnetosphere is a good approximation for the  magnetic field configuration\footnote{Nevertheless, the semi-analytical method described in this section can be applied to $any$ given physical magnetic configuration $\Psi$.}. A simple approximation of force-free  magnetic field is found in \citet[][]{tch08}:\begin{equation}\label{eq:ff_pol}
\Psi(r, \theta; p)=r^{p}(1-\cos\theta)
\end{equation}
where $0\le p\le 1.25$. When $p=1$ ($p=0$), the magnetic field has a parabolic (split-monopole) configuration. In general, the magnetosphere depends on parameters like black hole spin. Recent GRMHD numerical simulations implies that a single $p$ value for the outmost streamline may apply to simulation results of different black hole spin \citep[][]{nak18}, and that the resulting opening angle of the magnetosphere is closely related the total magnetic flux finally accumulated on the the event horizon \citep[e.g.][]{nar12}, suggesting to treat the black hole spin and magnetosphere as independent parameters for possible combinations.

One of the features of the above mentioned force-free magnetic field is the absence of the FMS  of the outflow. The resulting magnetic flux $\Phi\equiv \bar{B}_{\rm p}R^{2}$ of equation (\ref{eq:ff_pol}) is roughly constant at large distance, which against the condition $d\Phi/dR<0$ for an efficient MHD acceleration and the existence of FMS \citep[][]{tak98,tch09}. As a result, the outflow along the force-free magnetic field will remain sub-fast-magnetosonic if we simply solve the Bernoulli equation and obtain the toroidal field from the solution. 
We therefore overcome the non-existence of a FMS for poloidal magnetic field described in equation (\ref{eq:ff_pol}) by prescribing relation between the poloidal and toroidal magnetic field of the resulting flow, a method introduced in TT03 and TT08, to mimic the effect of efficient acceleration for the outflow. By using this method, the  the solution for the trans-fast magnetosonic flow equation can be easily obtained without the critical condition analysis. However, as a cost for this approach, it is expected that $\bar{B}_{p}$ computed directly from Equation (\ref{eq:ff_pol}) would not be consistent with the solution of the force-balance equation due to the reason mentioned before. We will therefore obtain $\bar{B}_{p}$ from the trans-fast magnetosonic flow solution, as will be described in \S \ref{sec:model_B}.

\subsection{Boundary Condition}\label{sec:model_bc}
The energy of the outflow $\hat{E}^{+}$ is assigned for each magnetic field line as the outer boundary. 
Across the black hole magnetosphere, $\hat{E}^{+}(\Psi)$  is  a free function, which is not necessary a constant  across the magnetosphere.
 For each stream line, with {known $\Omega_{F}$}, a specified flow energy, $\hat{E}^{+}$, and the location of flow launching with zero velocity, $r_{\rm s}$, the location of Alfv\'en surface, $r_{\rm A}^{+}$, can be  solved by equation (\ref{eq:mach_eqn}). In terms, the angular momentum for the outer flow $\hat{L}^{+}$ are also determined\footnote{Recall that ($\hat{E}(\Psi),\hat{L}(\Psi)$) can be alternatively determined by ($r_{\rm s}(\Psi), r_{\rm A}(\Psi)$) in the cold limit. If any two out of  these four parameters in these pairs are known, the rest two parameters are also known.} \citep[e.g., see Equations (43) and (44) of][]{tak90}  (see also Appendix \ref{app:flow_chart} for a flow chart).

\subsection{Matching Condition}
\label{sec:model_mc}
Two criteria are required to be satisfied in order to match the inflow solution and outflow solution along each magnetic field line.

First, to ensure $B^{2}=B_{p}^{2}+B_{\phi}^{2}$ is continuous across the stagnation surface $r_{\rm s}$, the constant $C$ is determined by 
\begin{equation}\label{eq:matching}
\xi^{-}(r_{\rm s};\Psi)=\xi^{+}(r_{\rm s};\Psi)\;.
\end{equation}

Second, in addition to the continuity of magnetic field strength, it is also expected that the outward electromagnetic energy flux is continuous across $r_{\rm s}$. However, such condition is degenerate and leaves an undetermined ratio  \citep{pu15},
\begin{equation}
\delta=\left|\frac{\hat{E}^{+}_{\rm EM}}{\hat{E}^{-}_{\rm EM}}\right|=\left|\frac{\eta^{-}}{\eta_{+}}\right|\;,
\end{equation}
note that $\hat{E}^{+}_{\rm EM}>0$ and $\hat{E}^{-}_{\rm EM}<0$ (i.e., negative energy GRMHD inflows) have different signs.
The nature of the degeneracy of the choice of $\delta$ lies in that the inflow velocity does not sensitive to the $\delta$, provided that the flow is magnetically dominated. 
In the above notation, $\hat{E}_{\rm EM}$ the electromagnetic part of the total energy $\hat{E}$ (and the fluid part is $\hat{E}_{\rm FL}=\hat{E}-\hat{E}_{\rm EM}=u_{t}$).

At the stagnation surface, it is a good approximation to adopt $\hat{E}_{\rm FL}=u_{t}(r=r_{\rm s})\approx1$ (the specific fluid energy is roughly equal to its rest-mass energy), and therefore we have
\begin{equation}
\hat{E}^{-}(\Psi)=1+\hat{E}_{\rm EM}^{-}<0
\end{equation}
\begin{equation}
\hat{E}^{+}(\Psi)=1+\hat{E}_{\rm EM}^{+}>0
\end{equation}
for the inflow and outflow solutions near $r=r_{\rm s}$, respectively.

For the simple and straightforward case $\delta=1$ and 
and $\hat{E}_{\rm EM}^{+}\approx-\hat{E}_{\rm EM}^{-}$ \citep[][]{pu15}, the following
useful matching condition for the outflow and inflow solution is obtained
\begin{equation}\label{eq:matching_1}
\hat{E}^{-}(\Psi)=2-\hat{E}^{+}(\Psi)\;.
\end{equation}
To satisfy equation (\ref{eq:matching}), the value of $C$ in the equation (\ref{eq:xi_def2}) can then be specified by equations (\ref{eq:xi_def}), (\ref{eq:xi_def2}), and (\ref{eq:matching_1}).
The extraction of BH rotational energy (by the $\hat{E}^{-}(\Psi)<0$ flows) implies a minimal outflow energy
\begin{equation}\label{eq:E_limit}
\hat{E}^{+}(\Psi)>2\;.
\end{equation}
With known $E^{-}$ and $\xi^{-}$, similar to the outflow case, it is sufficient to solve the Mach number  and the poloidal velocity of the inflow, as described in equation (\ref{eq:mach_eqn}).
A flow chart for the above procedure is presented and discussed in the Appendix \ref{app:flow_chart}.

\subsection{Flow Velocity}\label{sec:flow_acc}
The approximate flow velocity components
 $u^{r}$ and $u^{\theta}$ can be obtained by the relation
\begin{equation}
\frac{u^{r}}{F_{\theta\phi}}=\frac{u^{\theta}}{F_{r\phi}}\;,
\end{equation}
together with the definition of the poloidal velocity $u_{p}$.

The rest components of the four-velocity, $u^{t}$ and $u^{\phi}$
are obtained by the relation of
\begin{equation}
u_{t}+\Omega_{F}u_{\phi}=\hat{E}-\Omega_{F}\hat{L}\;,
\end{equation}
together with  
\begin{equation}
u^{\alpha}u_{\alpha}=1\;.
\end{equation}

Once that four-velocity of the flow is obtained, the magnetization parameter
\begin{equation}\label{eq:sigma_def}
\sigma(r; \Psi)\equiv \frac{\hat{E}_{\rm EM}}{\hat{E}_{\rm FL}}=\frac{\hat{E}(\Psi)-\hat{E}_{\rm FL}}{\hat{E}_{\rm FL}}=\frac{\hat{E}(\Psi)-u_{t}}{u_{t}}
\end{equation}
is also determined with $\hat{E}_{\rm FL}=u_{t}$. 
The profile of $\sigma(\Psi; r)$ is associated with the energy conversion from $\hat{E}_{\rm EM}$ to $\hat{E}_{\rm FL}$, and therefore the flow acceleration efficiency. The initial magnetization at the stagnation surface has a good approximation with the flow energy of the outflow by 
\begin{equation}\label{eq:Eatrs}
\sigma_{s}\equiv\sigma(r_{s};\Psi)\approx \hat{E}_{\rm EM}\approx\hat{E}^{+}(\Psi), 
\end{equation}
because $\hat{E}_{\rm FL}(r=r_{s})\approx1$.

For an efficient acceleration, at large scale $\sigma(r\to\infty)=\sigma_{\infty}\approx0$, and the terminal Lorentz factor of the jet $\gamma(r\to\infty)=\gamma_{\infty}\approx \hat{E}^{+}(\Psi)$.  Here we define the jet Lorentz factor including the gravitational redshift factor of the outflow by \citep[e.g.][]{mck06}
\begin{equation}\label{eq:def_lorentz}
\gamma\equiv\sqrt{g_{tt}}u^{t}\;.
\end{equation}
It is intriguing to note that, from equations (\ref{eq:beta_def})-(\ref{eq:xi_def}), for a specific field line, a faster outflow (a large value of $\gamma$, and therefore $\hat{E}^{+}$) corresponds to a increase of pitch angle (a smaller value of $\beta$) at the stagnation surface.

\subsection{Field Strength}\label{sec:model_B}
As the deformation of the magnetic field due to the mass loading is ignored,  the flow solution is constructed by only three of the four stream line conserved quantities, ($\Omega_{F}({\Psi})$, $\hat{E}({\Psi})$, $\hat{L}({\Psi})$) (see also Appendix \ref{app:flow_chart}), and the fourth conserved quantities, $\eta({\Psi})$ does not affect the resulting flow solution.
For a given free function $\eta(\Psi)$, the magnteic field components have the form
\begin{equation}\label{eq:bp}
\bar{B}_{p}(r; \Psi)=b_{0}\frac{u_{p}}{M^{2}}\eta(\Psi)\;,
\end{equation}
\begin{equation}\label{eq:bphi}
\bar{B}_{\phi}(r; \Psi)=\beta(r; \Psi) \bar{B}_{p}(r; \Psi)\;,
\end{equation}
where $b_{0}$ is a free nomalization parameter.
Equations (\ref{eq:bp}) and (\ref{eq:bphi}) also provide another way to understand our matching condition for the inflow and outflow solution along the same  magnetic field line. Provided that both poloidal and toroidal magnetic field are smooth and continuous at the stagnation surface $r_{s}$, we simply require a smooth and continuous  behavior of $\beta$ and $\eta$ at there. Accordingly,  we have $\xi^{-}(r_{\rm s};\Psi)=\xi^{+}(r_{\rm s};\Psi)$, equation (\ref{eq:matching}), and $\eta^{-}(\Psi)\cong\eta^{+}(\Psi)$ (i.e. $\delta\cong1$) at $r_{s}$, as described in \S \ref{sec:model_mc}. To avoid the singular behaviour $u^{r}\to0$ as $r\to r_{\rm s}$, in practice an interpolation of $\bar{B}_{p}(r\to r_{\rm s}; \Psi)$ near $r=r_{\rm s}$ can be applied to obtain the ratio between $\eta^{-}(\Psi)$ and $\eta^{+}(\Psi)$. One can also numerically verify $\delta\cong1$ can be consistently obtained when the energy matching condition, equation (\ref{eq:matching_1}), is adopted.

The number density, in both inflow and outflow region, can be obtained by the continuity relation
\begin{equation}\label{eq:n}
n=n_{0} \frac{\bar{B}_{p}}{u_{p}}\eta(\Psi)=\bar{n}_{0}\frac{\eta(\Psi)^{2}}{M^2}\;,
\end{equation}
where $\bar{n}_{0}=4\pi \mu^{2} n_{0}$ is a constant for the normalization.

\section{Model Parameters}\label{sec:model_par}

\begin{figure}
 \includegraphics[width=0.5\textwidth]{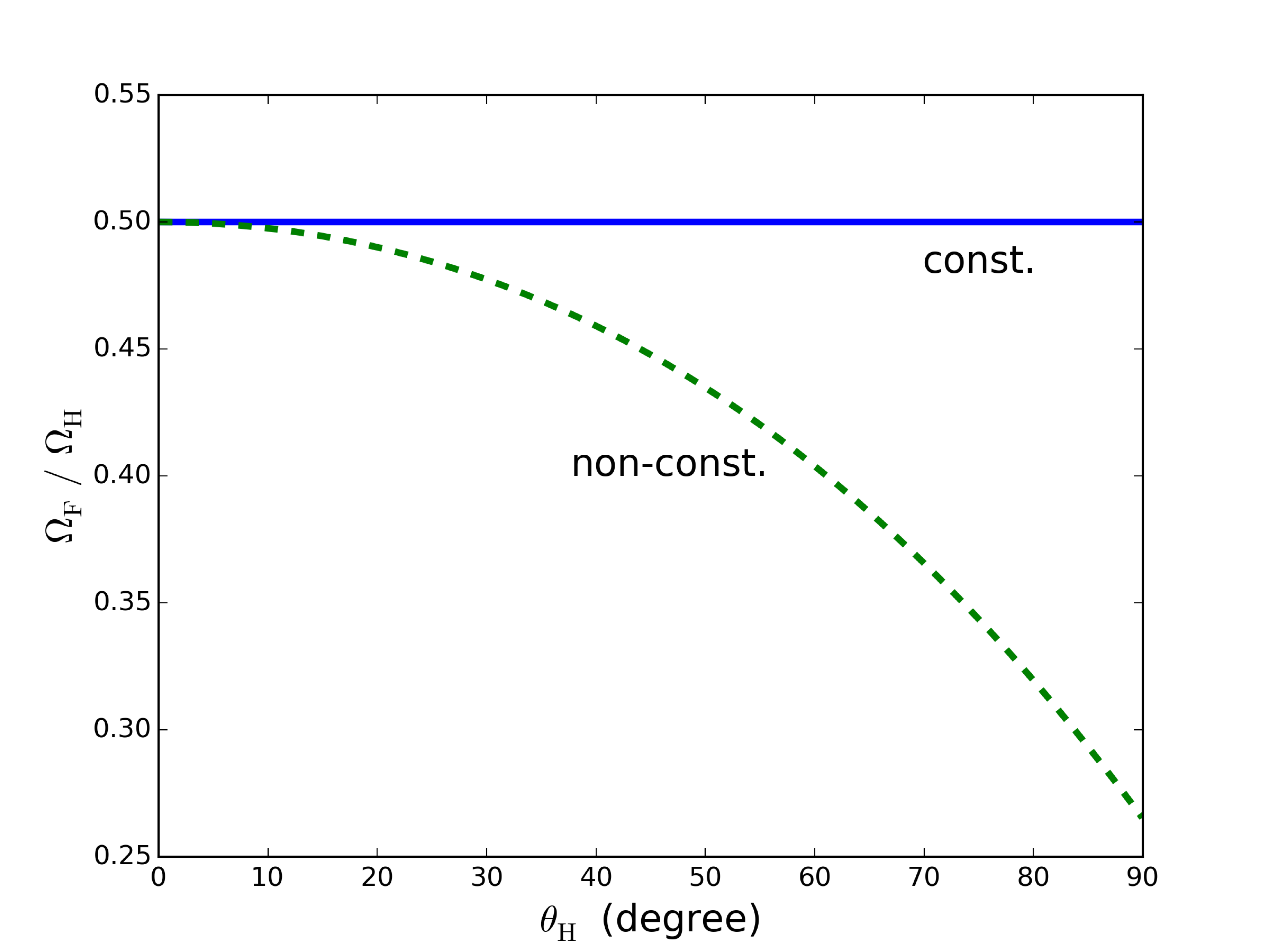}
\caption{Possible profiles of the angular velocity of magnetic fields $\Omega_{F}$ versus polar angle on black hole event horizon $\theta_{\rm H}$, in terms of the angular velocity of the hole $\Omega_{\rm H}$: constant field angular velocity (solid line; Equation (\ref{eq:field_vel_1})) and non-constant field angular velocity (dashed line; Equation (\ref{eq:field_vel_2})). }  \label{fig:para_omega_choice}
\end{figure}
Our semi-analytical model provides a  flexible way to explore the parameter space of the black hole spin $a$, magnetic field configuration $\Psi(p)$, field angular velocity $\Omega_{F}(\Psi)$, and the outflow energy $\hat{E}^{+}(\Psi)$.

We consider two qualitatively different functions of $\Omega_{F}(\Psi)$. The first one is a constant field angular velocity,  which is applied to both the parabolic and monopole fields:
\begin{equation}\label{eq:field_vel_1}
\frac{\Omega_{F}}{\Omega_{\rm H}}=0.5\;,
\end{equation}
where $\Omega_{\rm H}=a/(2 r_{\rm H})$ is the magnetic field angular velocity and  $r_{\rm H}=1+\sqrt{1-a^2}$ is the horizon radius.
The second one is a non-constant distribution of the magnetic field angular velocity $\Omega_{F}$ \citep[][]{bla77, bes09,mck07}, applied to the parabolic fields:
\begin{equation}\label{eq:field_vel_2}
\frac{\Omega_{F}}{\Omega_{\rm H}}=\frac{\sin^{2}\theta_{\rm H}[1+\ln\mathcal{G}]}{4\ln2+\sin^2\theta_{\rm H}+[\sin^{2}\theta_{\rm H}-2\mathcal{G}]\ln\mathcal{G}}\;,
\end{equation}
where $\mathcal{G}=(1+\cos\theta_{\rm H})$.
The profile of the two field angular velocities as function of penetrated horizon latitudes are plotted in Figure \ref{fig:para_omega_choice}.
For the non-constant field angular velocity, the value varies from $\Omega_{F}=0.5\Omega_{\rm H}$ for $\theta_{\rm H}=0$, to $\Omega_{F}\approx0.265\Omega_{\rm H}$ for $\theta_{\rm H}=\pi/2$.

In the following, we presented flow solutions along parabolic magnetic field lines ($p=1$) in \S\ref{sec:para},  and solutions along split-monopole magnetic field lines ($p=0$) in \S\ref{sec:mono}.

\section{Trans-fast magnetosonic Flow along Parabolic Magnetic Field Lines}\label{sec:para}
\subsection{Example of solutions along a magnetic field line}\label{sec:result_inout}
\begin{figure}
\begin{center}
\includegraphics[width=0.5\textwidth]{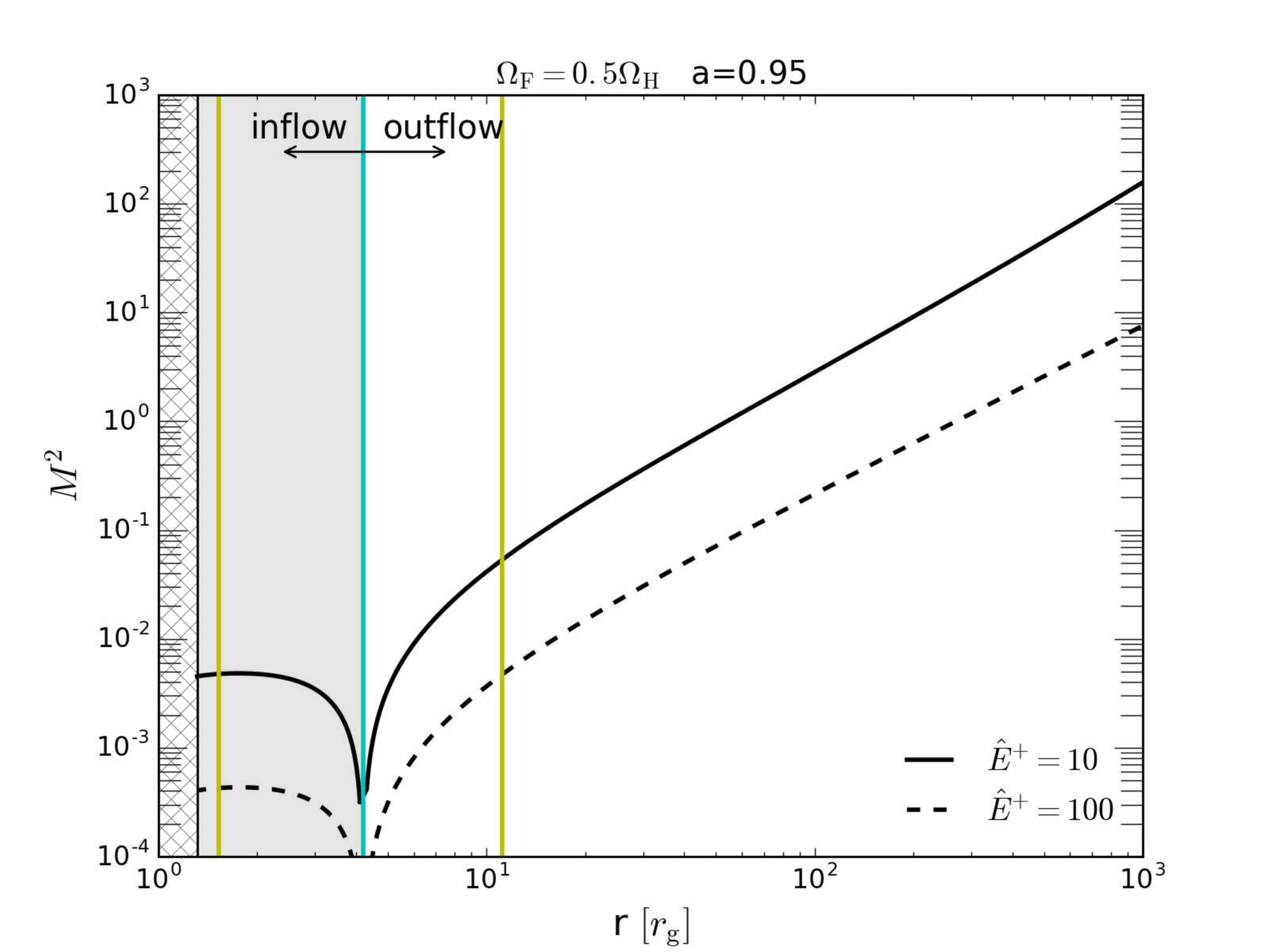}\\
\includegraphics[width=0.5\textwidth]{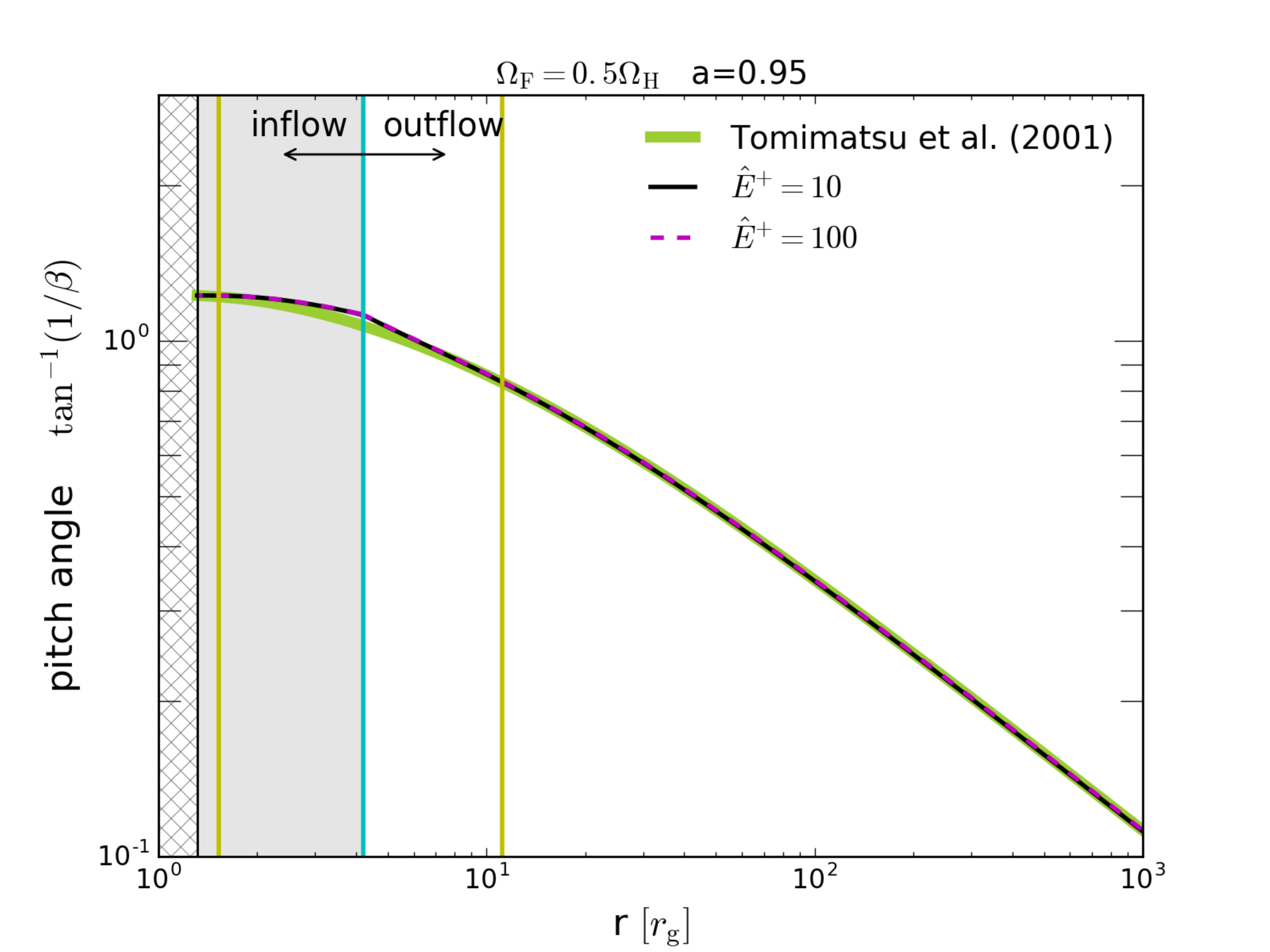}\\
\includegraphics[width=0.5\textwidth]{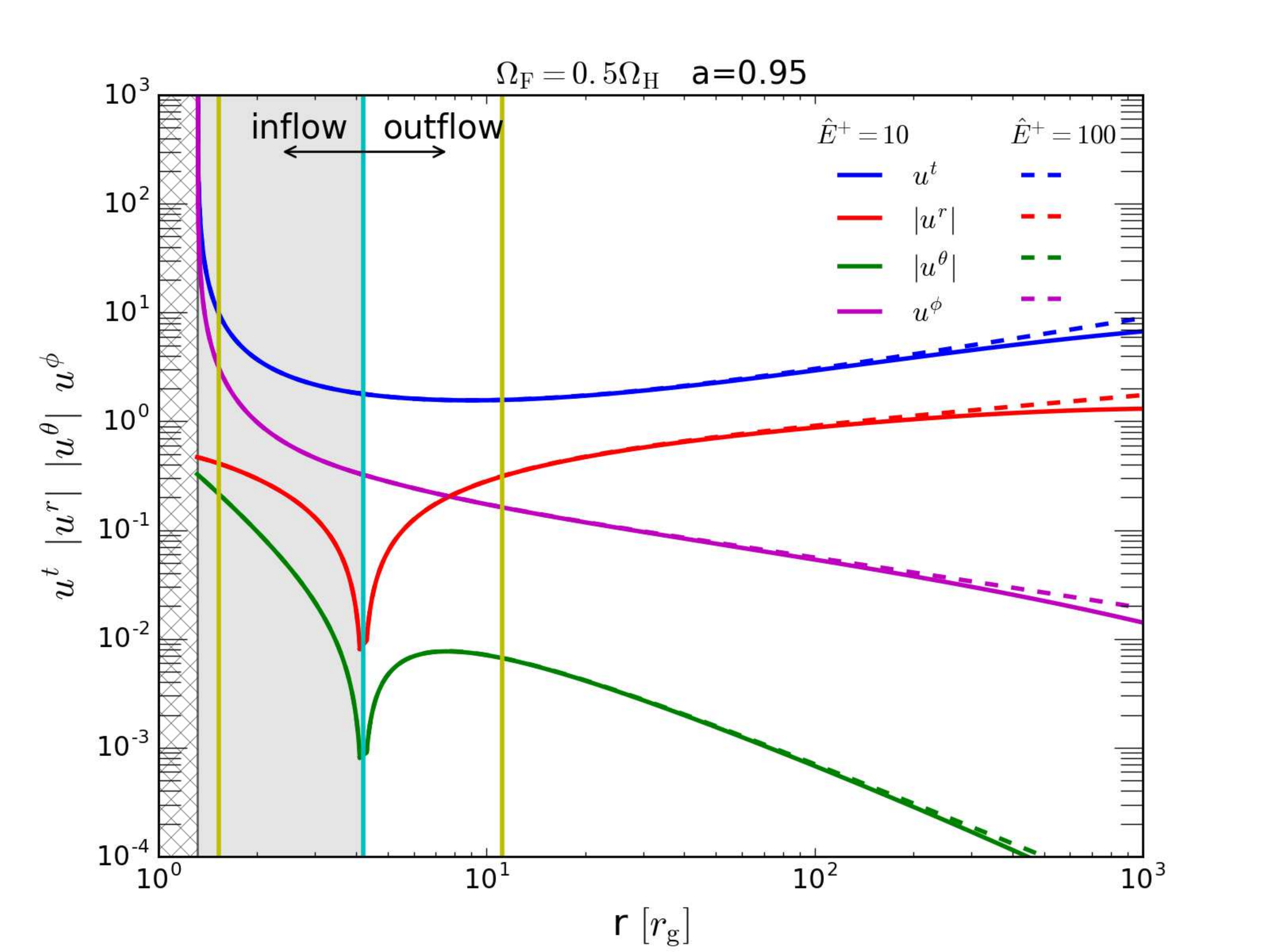}\\
\end{center}
\caption{Example flow solutions of different outflow energy $\hat{E}^{+}$ along a parabolic magnetic field line  ($\theta_{\rm H}=85^{\circ}$) in the magnetosphere of a rotating black hole. Top panel: Mach number square $M^{2}$. Middle Panel: Pitch angle, which is well approximated by the profile $(1/g_{\phi\phi}\Omega_{F}$), below which the kink instability take place \citep{tom01}. Bottom panel: flow four velocity $u^{\alpha}$.
The locations of the stagnation surface (cyan vertical line) and the inner/outer light surfaces (yellow vertical lines) are indicated. The shaded grey area indicates the inflow region, and the hatched area indicates the  black hole.}  \label{fig:sol_example}
\end{figure}

 To demonstrate our semi-analytical approach, let us start with flow solutions along one single magnetic field line. In Figure \ref{fig:sol_example} the solutions for the flow along a magnetic field line with  $\Omega_{F}=0.5\Omega_{\rm H}$ and $\theta_{\rm H}=85^{\circ}$ of a spinning black hole $a=0.95$ is shown. The solid and dashed profiles corresponds to two different boundary condition $\hat{E}^{+}=10$ and $\hat{E}^{+}=100$, respectively. Note that the corresponding $\hat{L}^{+}$ are uniquely determined with the given $\hat{E}^{+}$  and $\Omega_{F}$ (see \S\ref{sec:model_bc}). For this specific setup, $\hat{L}^{+}\simeq52.2$ for the case $\hat{E}^{+}=10$, and $\hat{L}^{+}\simeq549.4$ for the case $\hat{E}^{+}=100$.

The square of the mach number $M^{2}$,  which is a solution of equation (\ref{eq:mach_eqn}) for $\hat{E}=(10, 100)$, is shown in the top panel of Figure \ref{fig:sol_example}. The resulting pitch angles  by the matching condition introduced in \S2, as plotted in the second panel, roughly follow the the  criteria of the kink instability: $(B^{r}/B^{\phi})\approx(1/g_{\phi\phi}\Omega_{F})$ \citep[][]{tom01}, consistent with results of GRMHD simulation \citep[][]{mck06} and semi-analytical computations \citep[][]{pu15}. For the outflow,  the poloidal and toroidal component of the magnetic field become comparable ($\beta\approx1$) near the outer light surface (the vertical yellow solid line in the outflow region). This is a well-known property of a magnetically dominated MHD flow.
The components of the flow four-velocity are shown in the bottom panel of Figure \ref{fig:sol_example}. Note that the flow four-velocity of the two solutions has noticeable differences only at the larger scale ($r>100 r_{g}$ in the plot). For the inflow, as discussed in \citet[][]{pu15}, the four-velocity of a magnetically dominated inflow does not sensitive to the flow energy because that the Alfv\'en surfaces is always located close to the inner light surface and the FMS is always located close to the horizon, as will be shown later  in \S\ref{sec:near}. These characteristic surfaces will be shown later in this section. 

The physical reason for the existence  of the outer and inner light surfaces can also be recognized from the flow solution. To satisfy the requirement of causality, the outer surface marks the boundary beyond which the flow motion must be mostly poloidal, as can be seen by that $u^{r}>u^{\phi}$ beyond the outer light surface. The inner light surface, on the other hand, marks the boundary beyond which the flow is close enough to the central rotating black hole and must rotate faster enough due to the 
 the gravitational redshift effects. As we will also see in \S\ref{sec:near}, close to the horizon, the flow corotates with the black hole.

\subsection{Jet acceleration and Energy Conversion}\label{sec:result_out}
\begin{figure}
\includegraphics[width=0.45\textwidth]{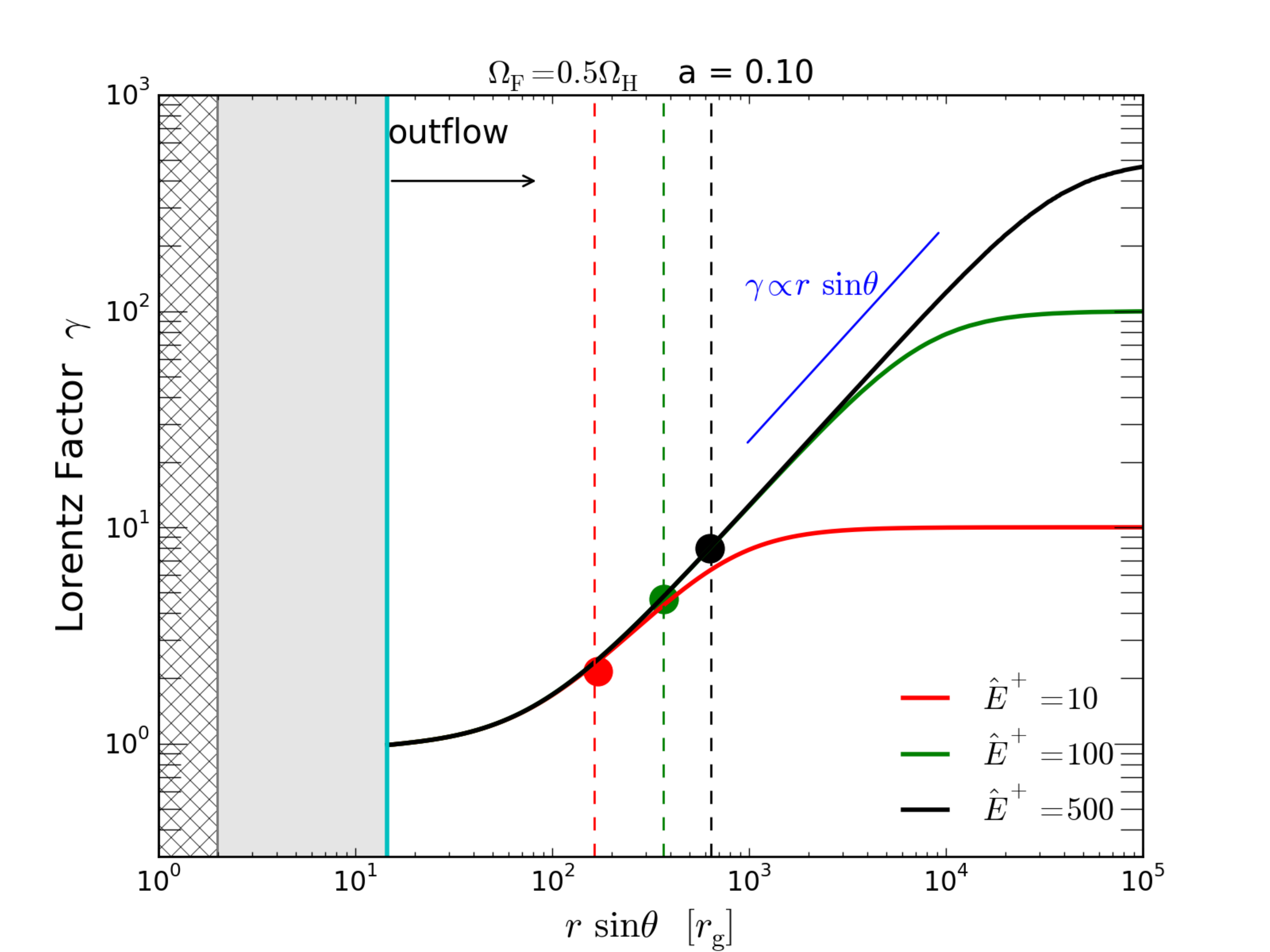}\\
\includegraphics[width=0.45\textwidth]{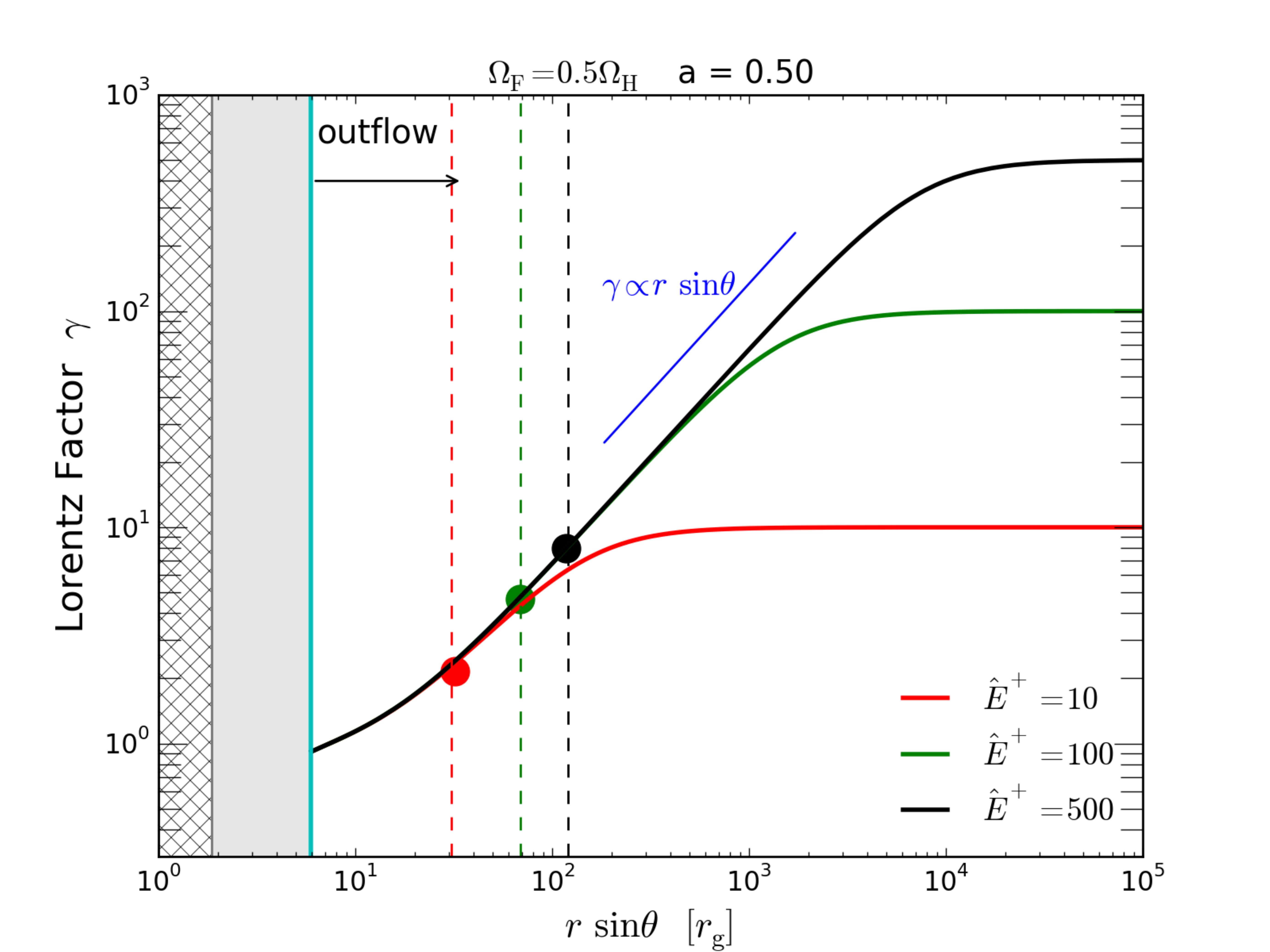}\\
\includegraphics[width=0.45\textwidth]{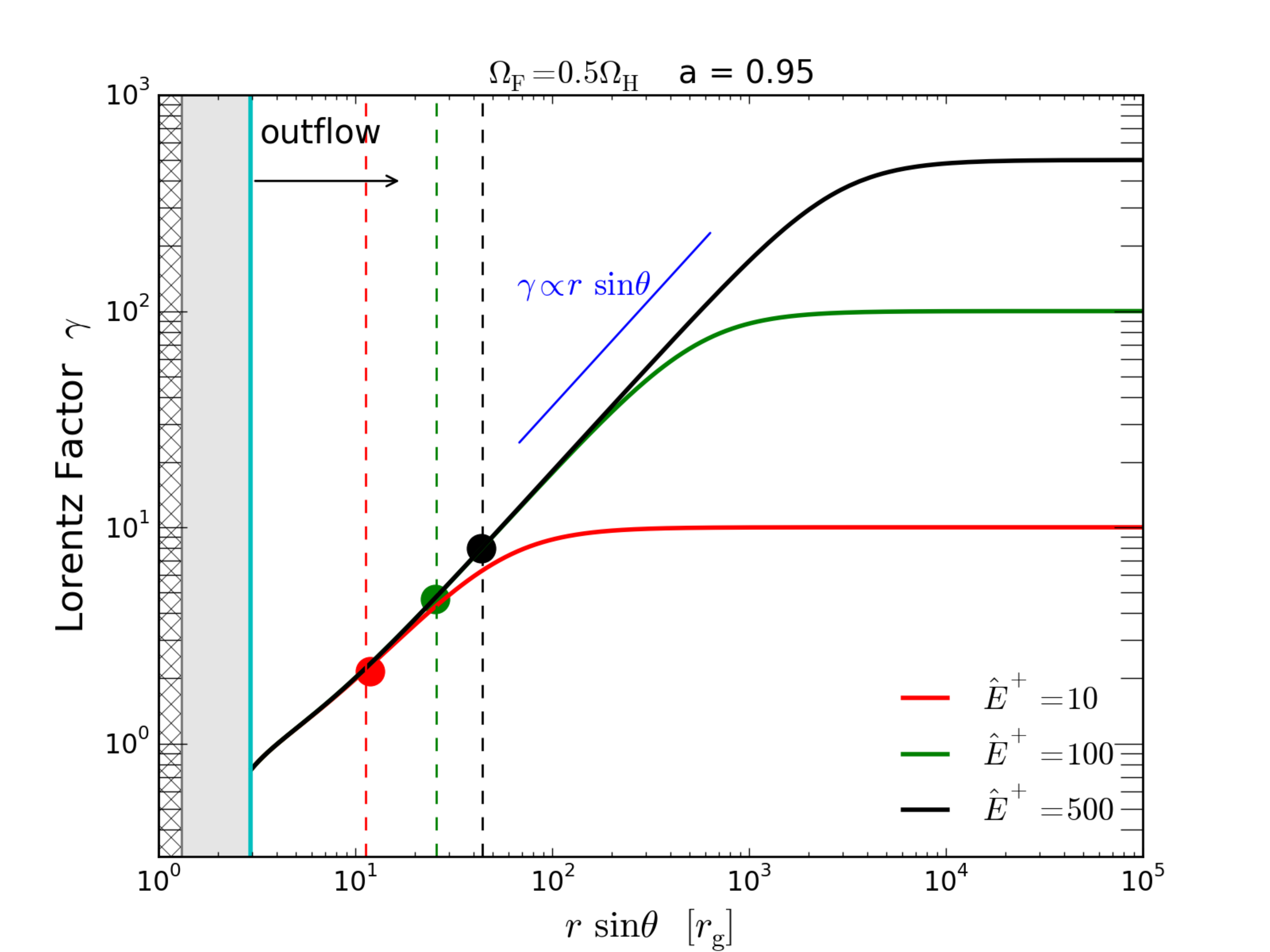}
\caption{Lorentz factor of the GRMHD outflow $\gamma$ versus the distance away from the rotational axis   ($\theta_{\rm H}=85^{\circ}$), for cases of different outflow energy $\hat{E}^{+}$ and different dimensionless black hole spin $a$.
At large distance, $\gamma_{\infty}\approx\hat{E}^{+}$ due to the efficient conversion from electromagnetic energy to kinetic energy. The vertical dashed line indicates the locations of the corresponding location of the FMS for each solution. The theoretical predicted location (
$(r\sin\theta)_{\rm FMS}^{\rm SRMHD}\approx(\hat{E}^{+})^{1/3}/\Omega_{F}$) and Lorentz factor ($\gamma^{\rm SRMHD}_{\rm FMS}\approx(\hat{E}^{+})^{1/3}$) at the FMS  for a   SRMHD flow \citep[TT03;][]{bes06} is overlapped by the color-filled circles, which  roughly fits our GRMHD solution for the outflow. The inflow region is indicated by the shaded grey area. The hatched  grey area indicates the black hole. See \S\ref{sec:result_out} for more discussions.
}  \label{fig:para_gamma}
\end{figure}

\begin{figure}
\includegraphics[width=0.45\textwidth]{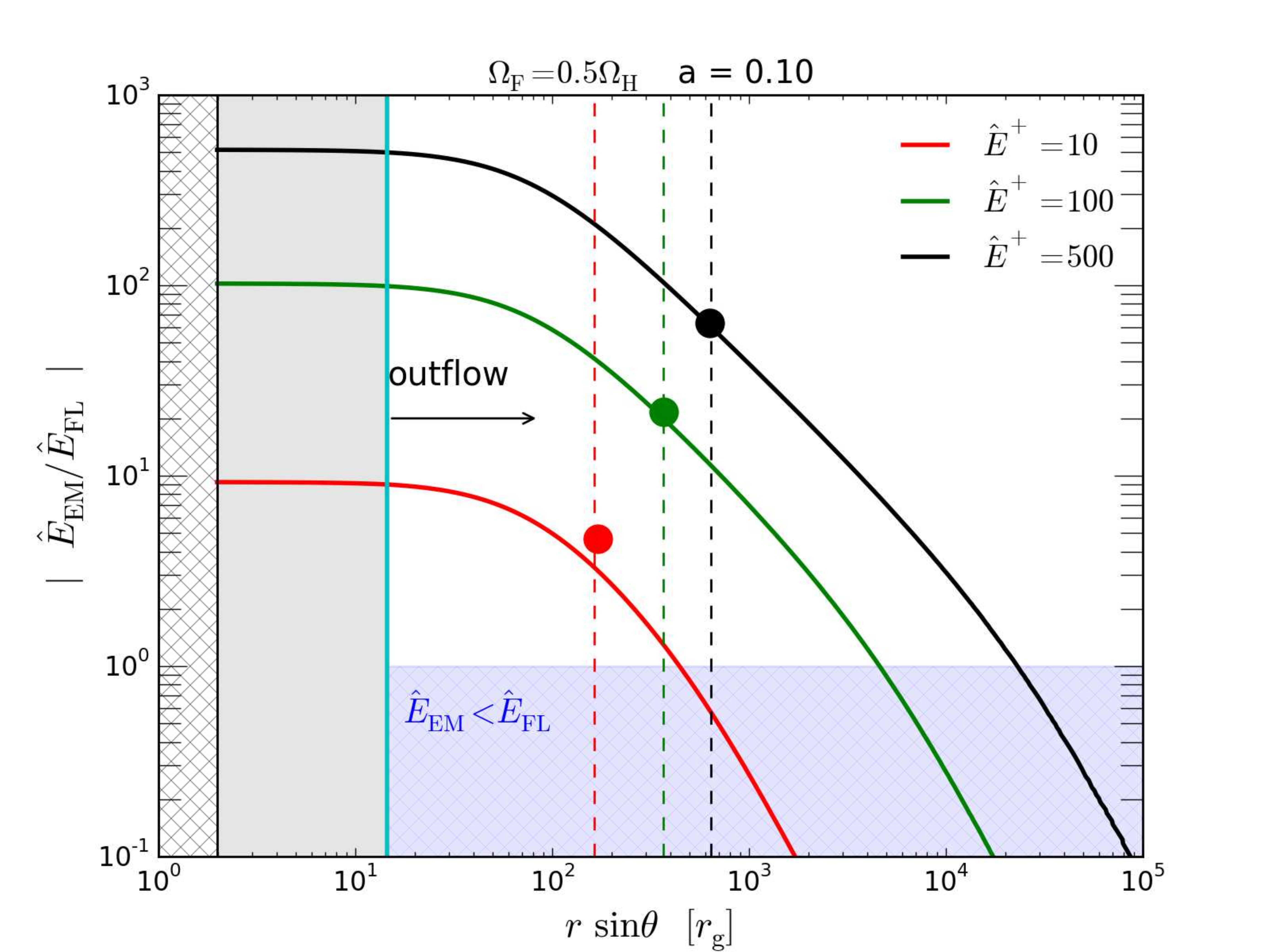}\\
\includegraphics[width=0.45\textwidth]{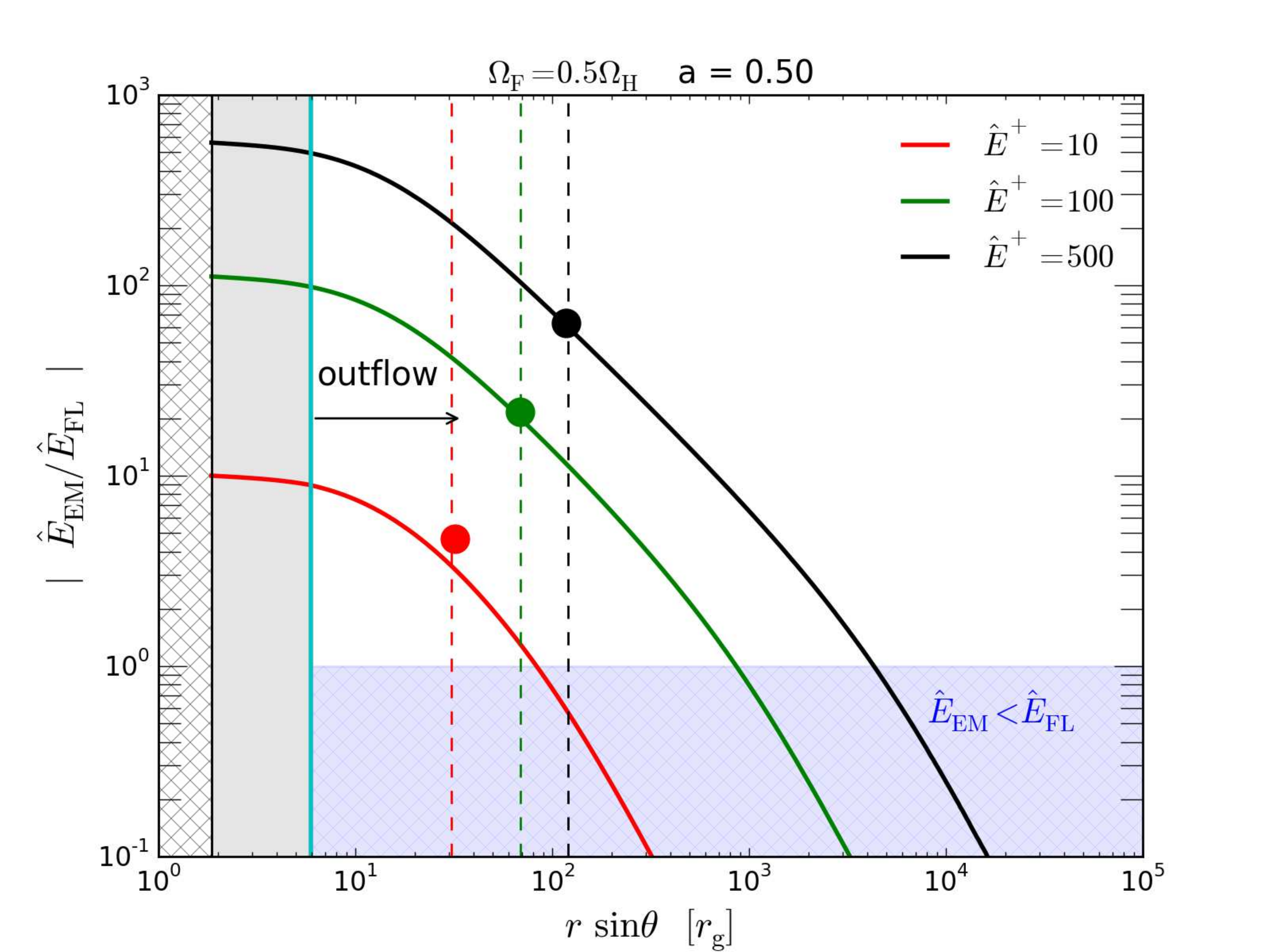}\\
\includegraphics[width=0.45\textwidth]{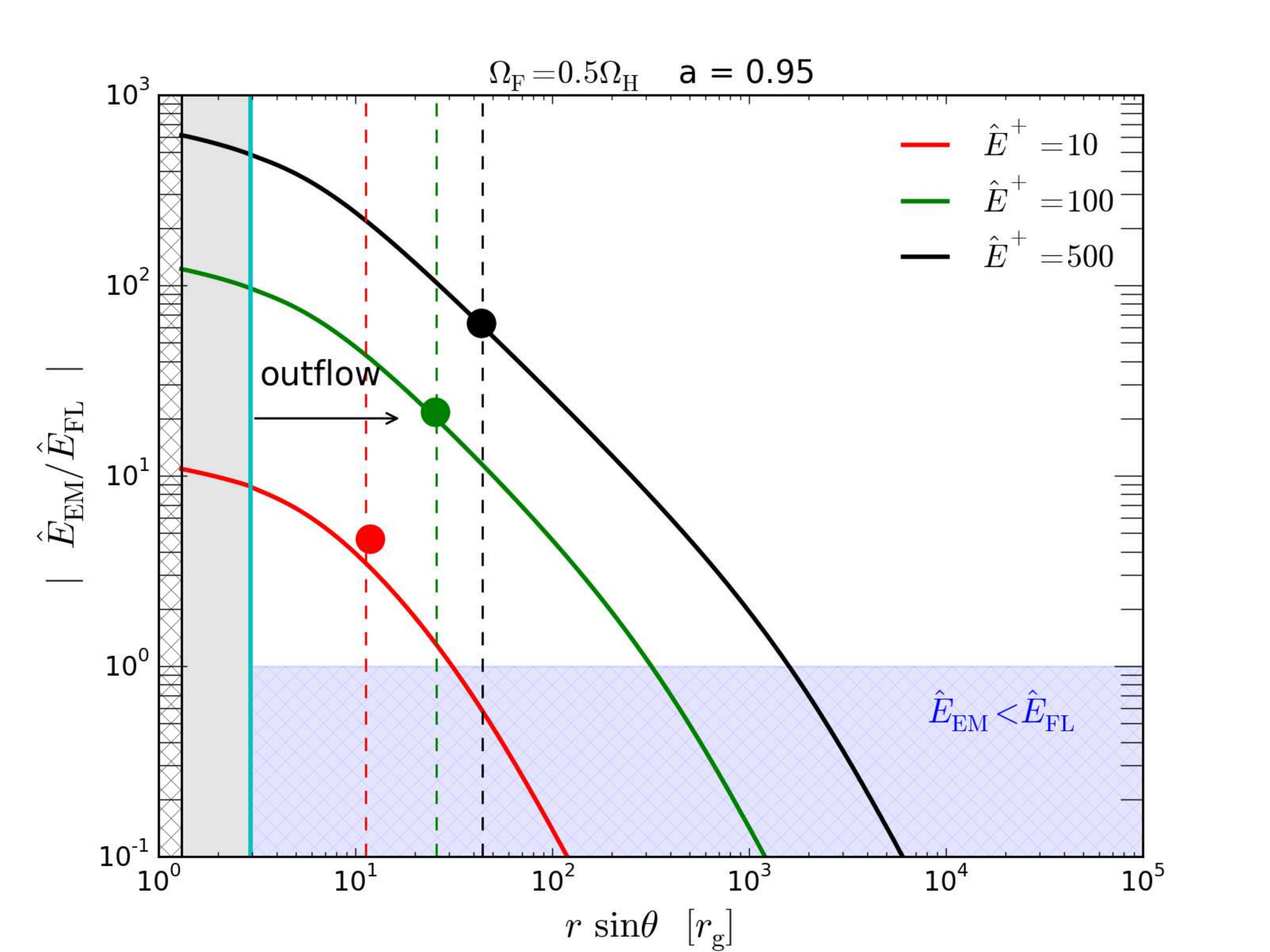}
\caption{ Conversion from the electromangetic energy $\hat{E}_{\rm EM}$ to the fluid energy  $\hat{E}_{\rm FL}$ of both the outflow solutions shown in Figure \ref{fig:para_gamma} and their corresponding inflow solutions.
The vertical dashed line indicates the locations of the corresponding location of the FMS for each solution. The theoretical predicted location and the energy conversion efficiency ($\sigma^{\rm SRMHD}_{\rm FMS}\approx(\hat{E}^{+})^{2/3}$) at the FMS for a SRMHD flow  is overlapped by the color-filled circles, which roughly fits our GRMHD solution for the outflow. The inflow region is indicated by the shaded grey area and the hatched grey area indicates the black hole. In the hatched blue region, $\hat{E}_{\rm EM}<\hat{E}_{\rm FL}$. 
}  \label{fig:para_E}
\end{figure}

\begin{figure}
\includegraphics[width=0.45\textwidth]{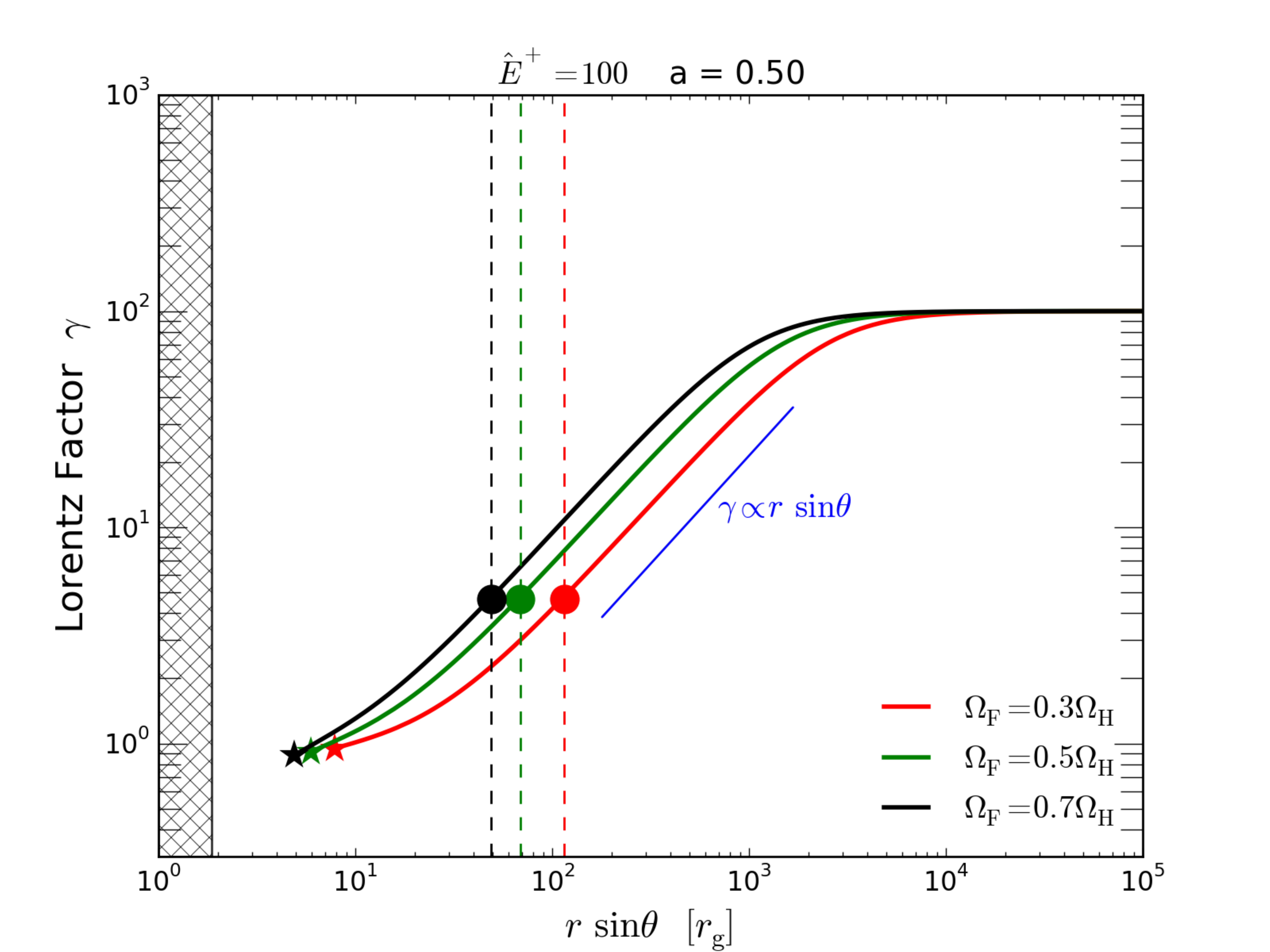}\\
\includegraphics[width=0.45\textwidth]{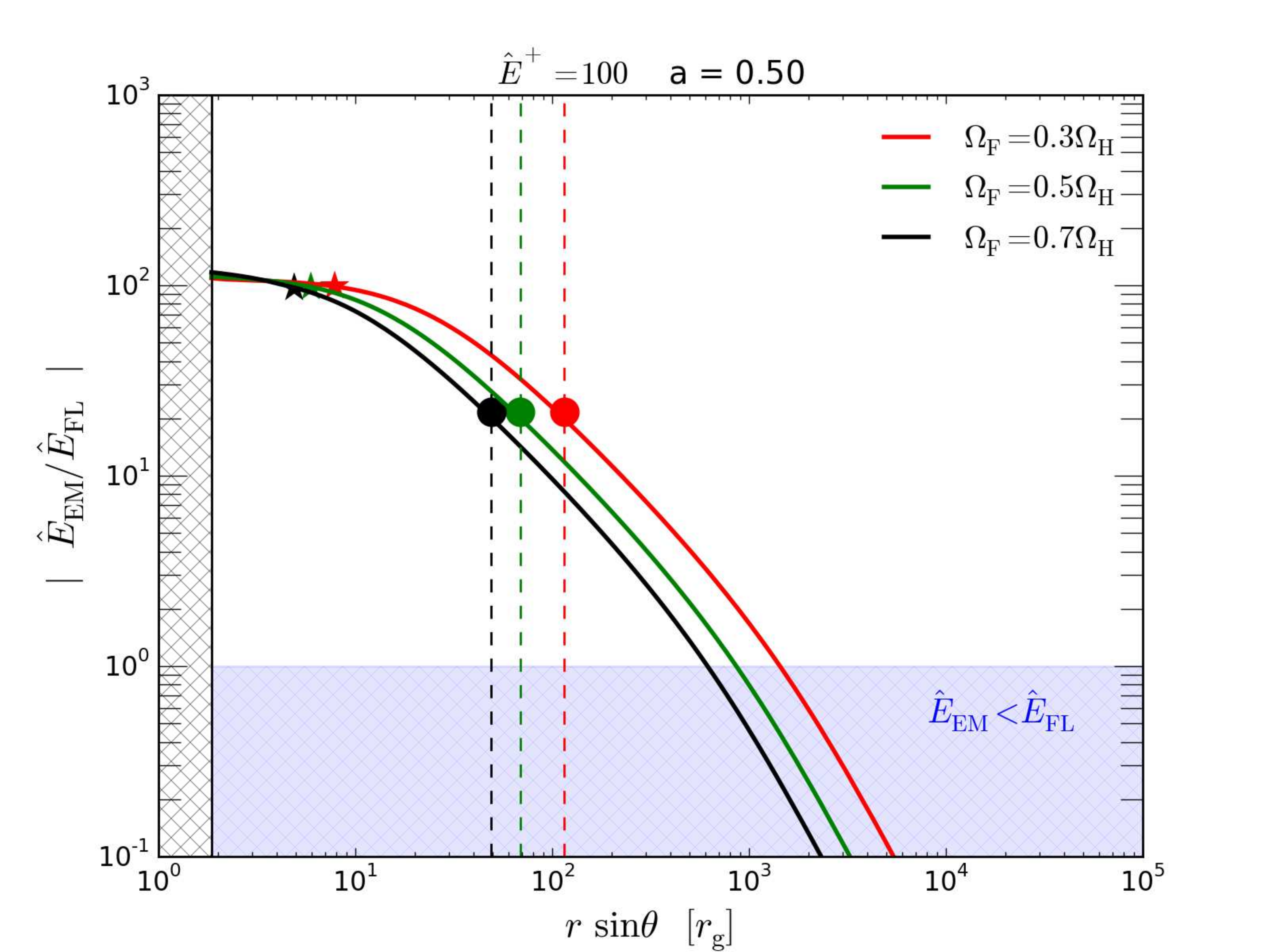}
\caption{ Lorentz factor of the outflow (top) and energy conversion (bottom) of flows along magnetic fields with different field angular velocity. The locations of the stagnation surfaces are indicated by the stars.
See also Figures \ref{fig:para_gamma} and \ref{fig:para_E} for explanations of the plot. 
}  \label{fig:para_differentF}
\end{figure}

We now focus on the outflow region and examine the jet acceleration via the evolution of the jet Lorentz factor. Results of different black hole spin ($a=0.1, 0.5, 0.95$) and the total energy of the outflow ($\hat{E}^{+}= 10, 100, 500$) are shown in Figure \ref{fig:para_gamma}. 
Again we focus on a field line which penetrates the black hole at $\theta_{\rm H}=85^{\circ}$, and assume $\Omega_{F}=0.5\Omega_{\rm H}$.
The cases for $\hat{E}^{+}=10, 100$  and $a=0.95$ is therefore corresponds to the solutions shown in Figures \ref{fig:sol_example}.
The Lorentz factor versus the distance away from the black hole  is computed only for the outflow solution, therefore the profiles starts from the stagnation surface (the vertical cyan line). 
The location of the stagnation surfaces moves further away from the black hole as the black hole spin becomes lower, due to the resulting smaller $\Omega_{F}$ and therefore weaker magneto-centrifugal force applied on to the plasma loading onto the magnetic field line. The location of the FMS for each flow solutions is indicated by the vertical dashed lines. For all cases, the $\gamma$ profiles linearly growth  ($\gamma\propto r\sin\theta$) in the super  fast magnetosonic region until the flow reaches a value $\gamma\approx \hat{E}^{+}$. 

By examining  the perturbation on the equation of motion for the plasma outflow in a parabolic force-free field line\footnote{The unperturbed field $\Psi_{0}$ considered in \citet[][]{bes06} has the form $\Psi_{0}\propto \ln[\Omega_{F}X+\sqrt{(\Omega_{F}X)^{2}+1}]$, where $X=r(1-\cos\theta)$. By using the relation $\sinh^{-1}(x)=\ln(x+\sqrt{x^{2}+1})$, the dominate form of $\Psi_{0}$ is found to be $r(1-\cos\theta)$.} $\Psi_{0}\propto r(1-\cos\theta)$ in flat spacetime, \citet[][]{bes06} found that the flow Lorentz factor grows with the distance $z=r\cos\theta$ from the equatorial plane with\footnote{For a magnetic field $\Psi\propto r^{p}(1-\cos\theta)$, the magnetic field line shape $z\propto (r \sin\theta)^{2/(2-p)}$ \citep[see, e.g.,][]{tch08}. For $p=1$, the relation $\gamma\propto z^{1/2}$ is consistent with our outflow solutions shown in Figure \ref{fig:para_gamma}, which satisfy $\gamma\propto r\sin\theta$.} $\gamma\propto z^{1/2}$, until the flow converts all its electromagnetic energy into kinetic energy. In addition, the FMS is located at the $(r\sin\theta)_{\rm F}^{\rm SRMHD}\approx\sigma_{0}^{1/3}/\Omega_{F}$, and the Lorentz factor there is $\gamma^{\rm SRMHD}_{\rm F}\approx\sigma_{0}^{1/3}$, where  $\sigma_{0}$ is the Michel's magnetization parameter \citep[][]{mic69}. The same conclusion is also obtained in \citet[][]{tom03}.

To compare with SRMHD theory in the distant flat spacetime, the above predicted location and Lorentz factor at the FMS are overlapped in figure \ref{fig:para_gamma} (the colored circles), by considering the magnetization at the stagnation surface, where the outflow starts: $\sigma_{s}\approx\hat{E}^{+}$ (see also Equation (\ref{eq:Eatrs})). A good agreement of the jet acceleration between  SRMHD and GRMHD flows for all different black hole spin are found. Such interesting feature seems resulting from that the outflow region is far away enough of the black hole.

To show the energy conversion from the electromagnetic component $\hat{E}_{\rm EM}$ to fluid component $\hat{E}_{\rm FL}$ along the flow, the ratio of the two components, $\sigma$, is shown in Figure  \ref{fig:para_E}. The inflow part is also shown in the plot, yet note that $\hat{E}_{\rm EM}<0$ (and $\hat{E}_{\rm FL}>0$) in the inflow region \citep[see also][]{pu15}, and the black hole rotational energy is extracted by an outgoing Poynting flux dominated GRMHD inflow which has a negative total energy ($\hat{E}=\hat{E}_{\rm EM}+\hat{E}_{\rm FL}<0$) \citep[][]{tak90}. The energy conversion for the outflow starts from $\sigma_{\rm s}\approx\hat{E}^{+}$ at the stagnation surface, and gradually decreases when $\hat{E}_{\rm EM}$ components converts to $\hat{E}_{\rm FL}$. It is clearly shown that the flow remain Poynting-flux dominated at the FMS. The energy conversion  efficiency at the FMS is again consistent with the predicted value from magnetically dominated  SRMHD flows \citep[][]{bes06,tom03}, $\sigma^{\rm SRMHD}_{\rm FMS}\approx(\hat{E}^{+})^{1/3}$, as indicated by the colored circles in Figure  \ref{fig:para_E}.

We further consider cases of different field angular velocity in Figure \ref{fig:para_differentF}, with fixed outflow energy, $\hat{E}=100$ and black hole spin $a=0.5$. While the decrease of the angular velocity results in closer stagnation surfaces (indicated by the stars) and location of FMS (indicated by the dashed vertical lines), the resulting flow solutions are also in good agreement with the predicted values from magnetically dominated  SRMHD flows \citep[TT03,][]{bes06}.

\subsection{Jet properties near the Black Hole}\label{sec:near}
\begin{figure*}
\begin{center}
\includegraphics[trim={0 1cm  0 0},width=0.3\textwidth]{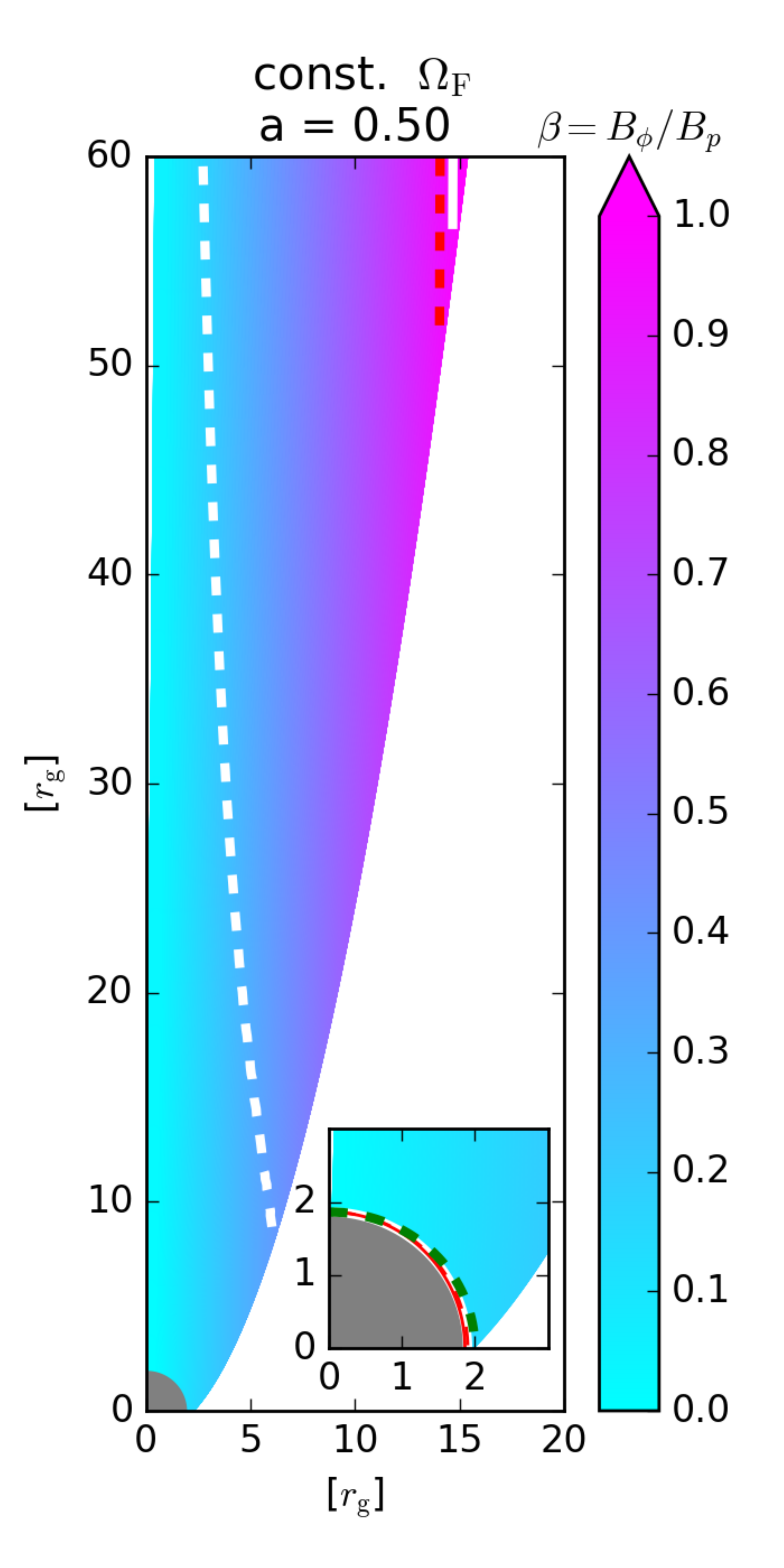}
\includegraphics[trim={0 1cm  0 0},width=0.3\textwidth]{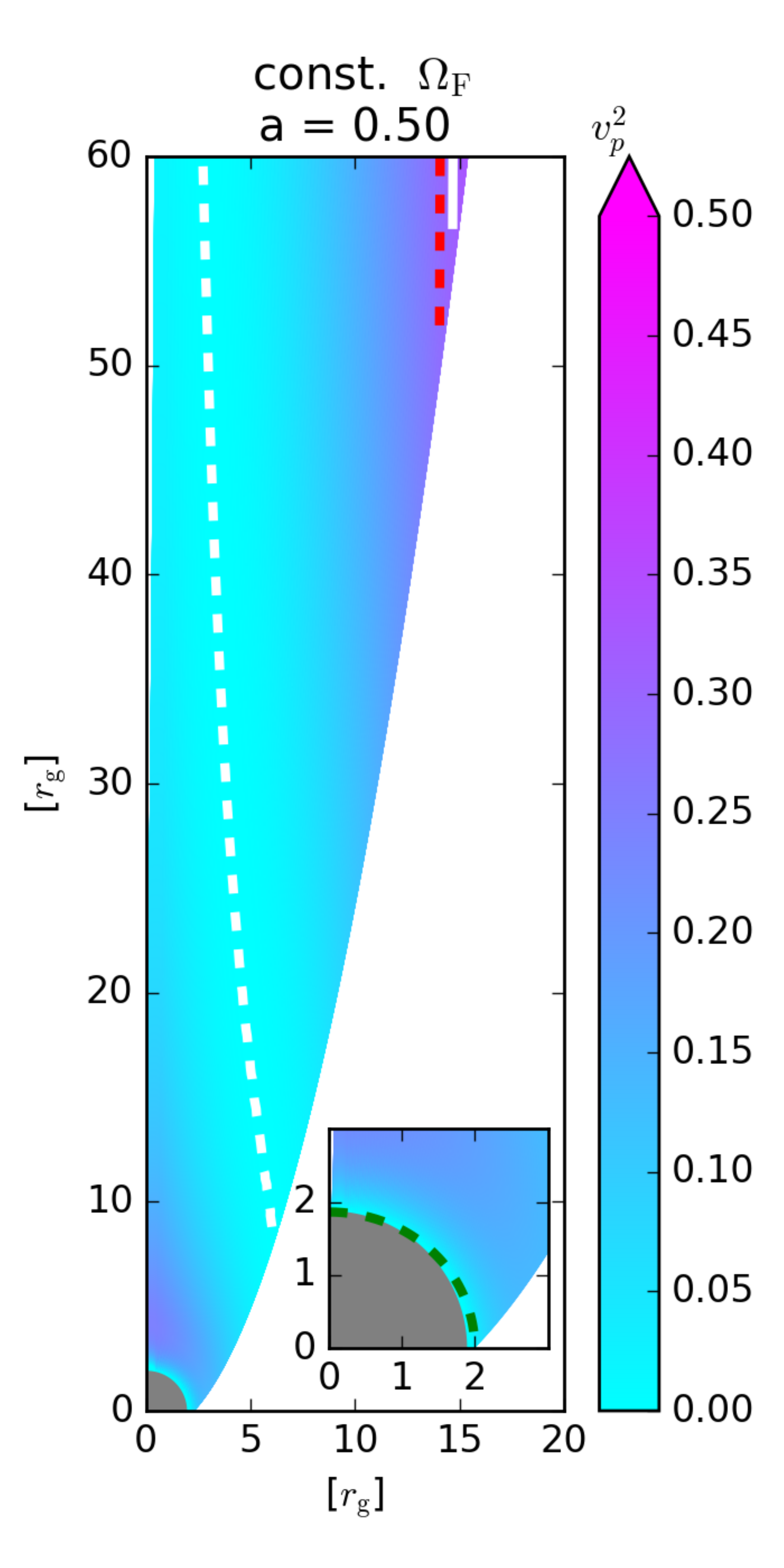}
\includegraphics[trim={0 1cm  0 0},width=0.3\textwidth]{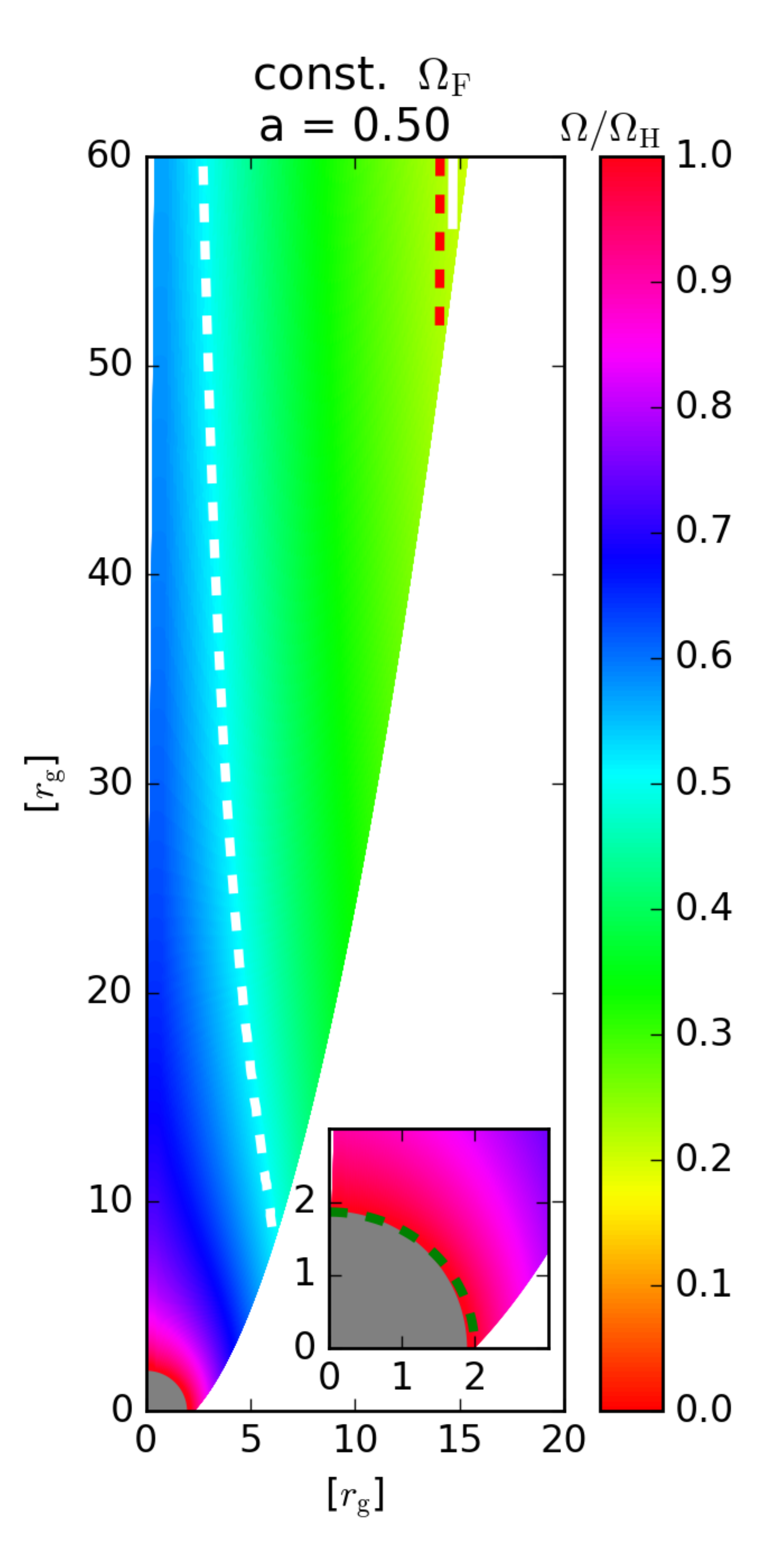}\\
\includegraphics[trim={0 1cm  0 0},width=0.3\textwidth]{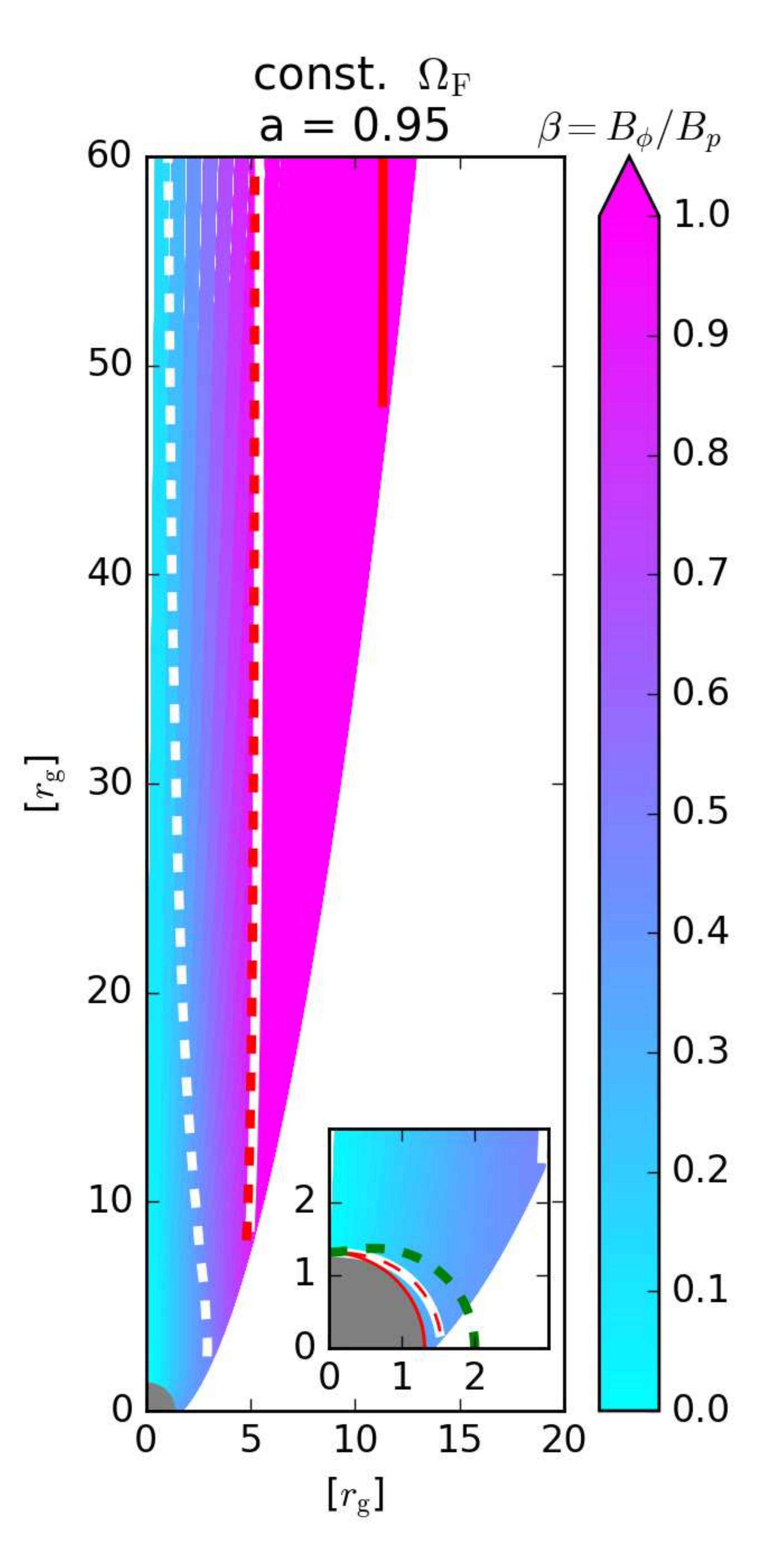}
\includegraphics[trim={0 1cm  0 0},width=0.3\textwidth]{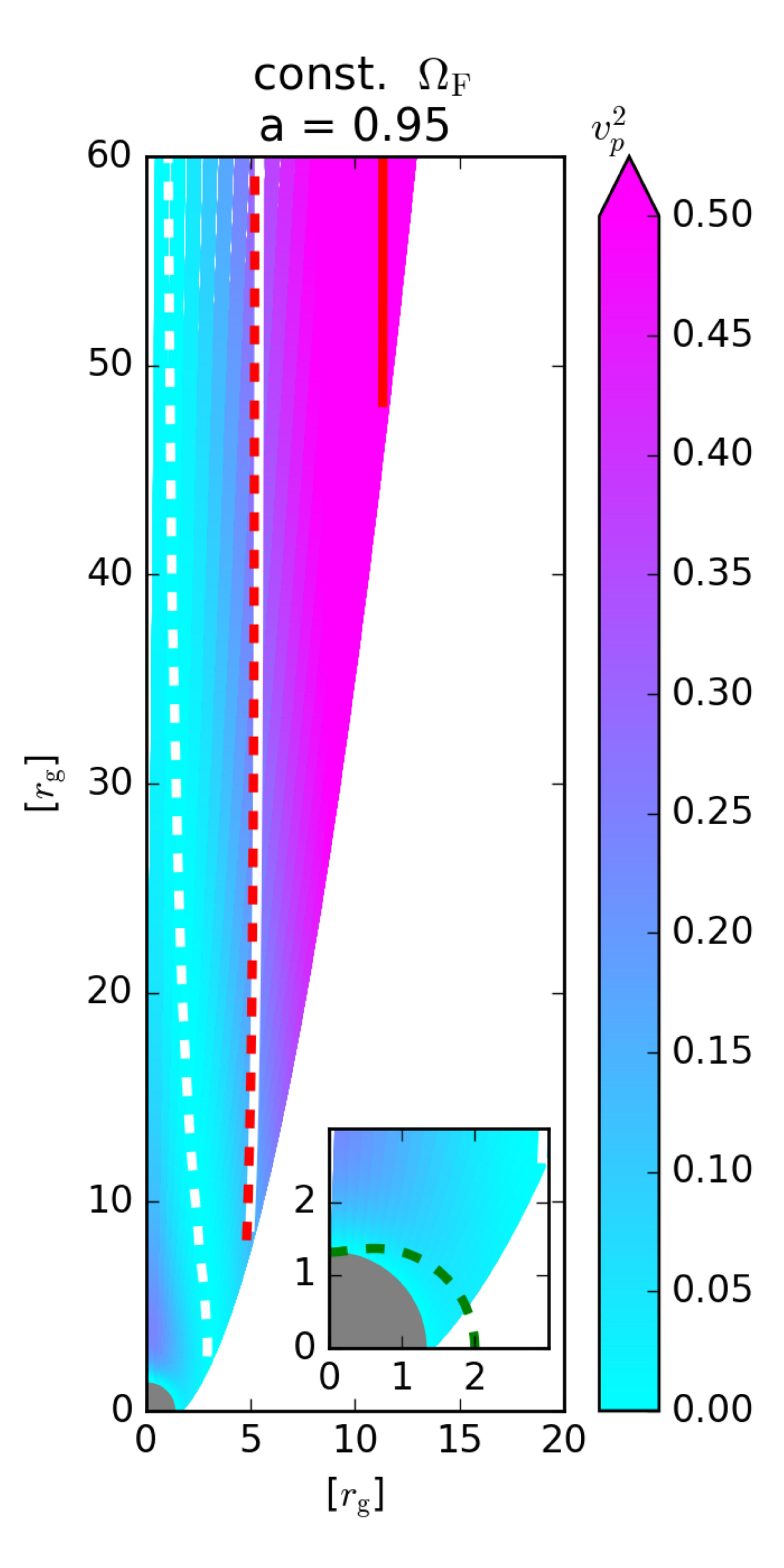}
\includegraphics[trim={0 1cm  0 0},width=0.3\textwidth]{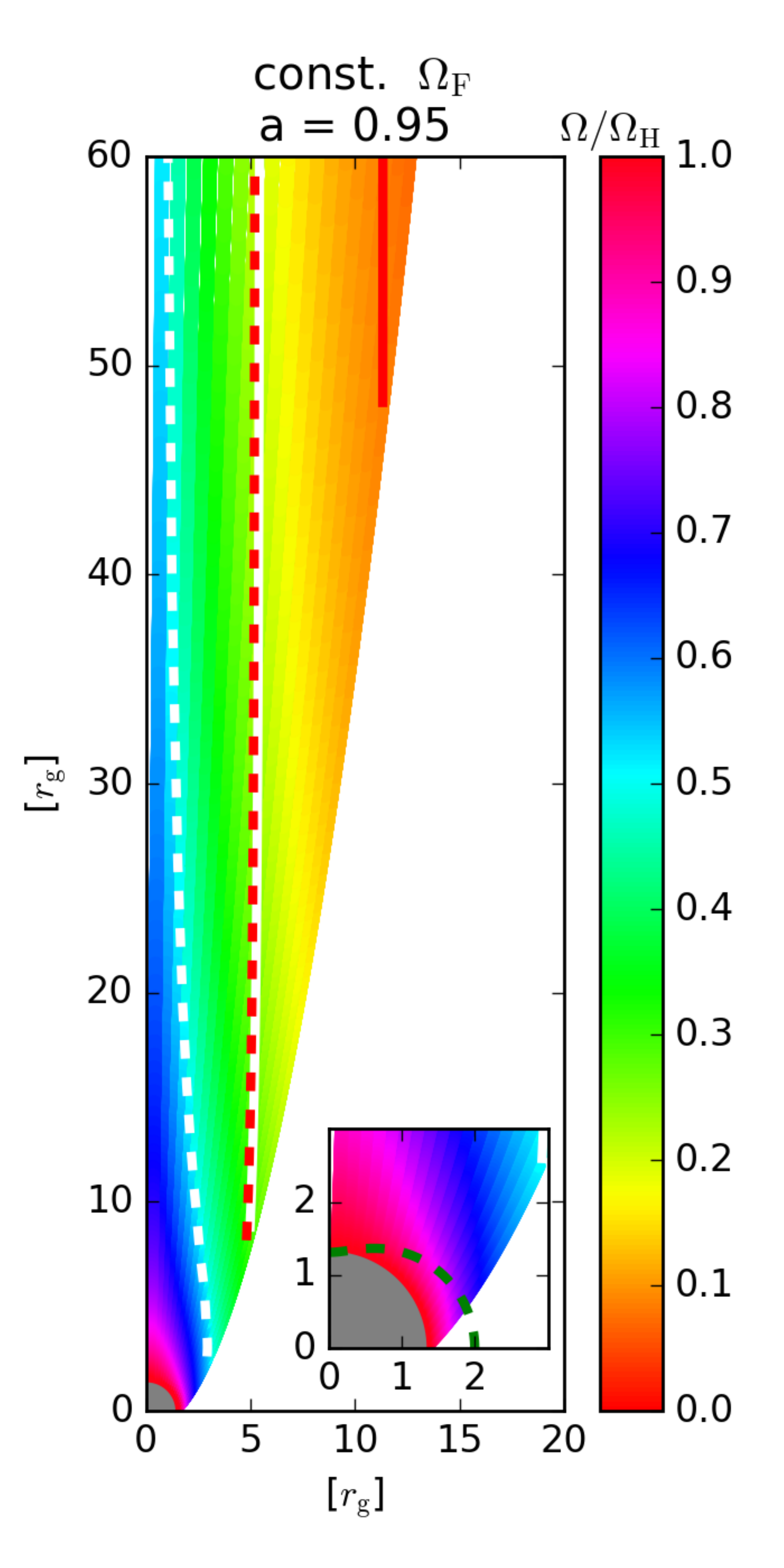}\\
\end{center}
\caption{GRMHD flow structure of a parabolic magnetosphere with a constant field angular velocity $\Omega_{F}(\Psi)=0.5\Omega_{\rm H}$, for cases of different dimensionless black hole spin $a$. The total energy of the outflow is assumed to be  $E^{+}(\Psi)=10$. Left panel: $\beta$, the ratio between the toroidal and poloidal magnetic field. Middle panel, square of poloidal flow velocity $v_{p}^{2}\equiv u_{p}^{2}/(u^{t})^{2}$. Right: angular velocity of the flow $\Omega\equiv u^{\phi}/u^{t}$, in terms of black hole angular velocity $\Omega_{\rm H}$.
For all panels, the the outer FMS (red solid line), the outer light surface (white solid line), the the outer Alfv\'en surface (red dashed line), and the stagnation surface (white dashed line) are shown. The static limit surface are indicated by the green dashed line in the insets. The inner light surface (white solid line), inner Alfv\'en surface (red dashed line;  which almost coincide with the white solid line) and inner FMS (red solid line) are shown in the inset of the left panel. The central shaded area indicates the black hole. See \S\ref{sec:near} for more discussions.
}  \label{fig:par_linear_f1}
\end{figure*}

\begin{figure*}
\begin{center}
\includegraphics[trim={0 1cm  0 0},width=0.3\textwidth]{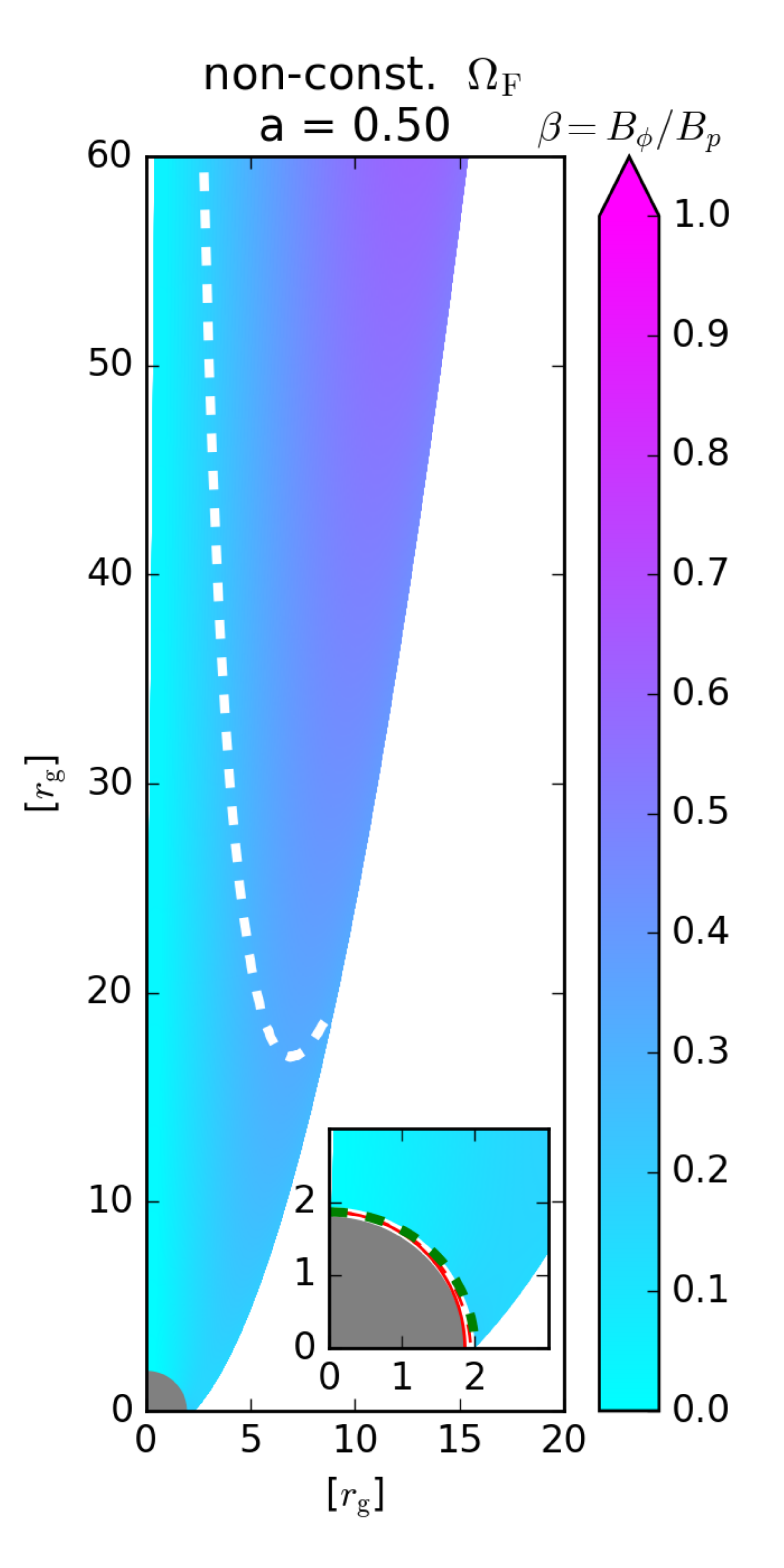}
\includegraphics[trim={0 1cm  0 0},width=0.3\textwidth]{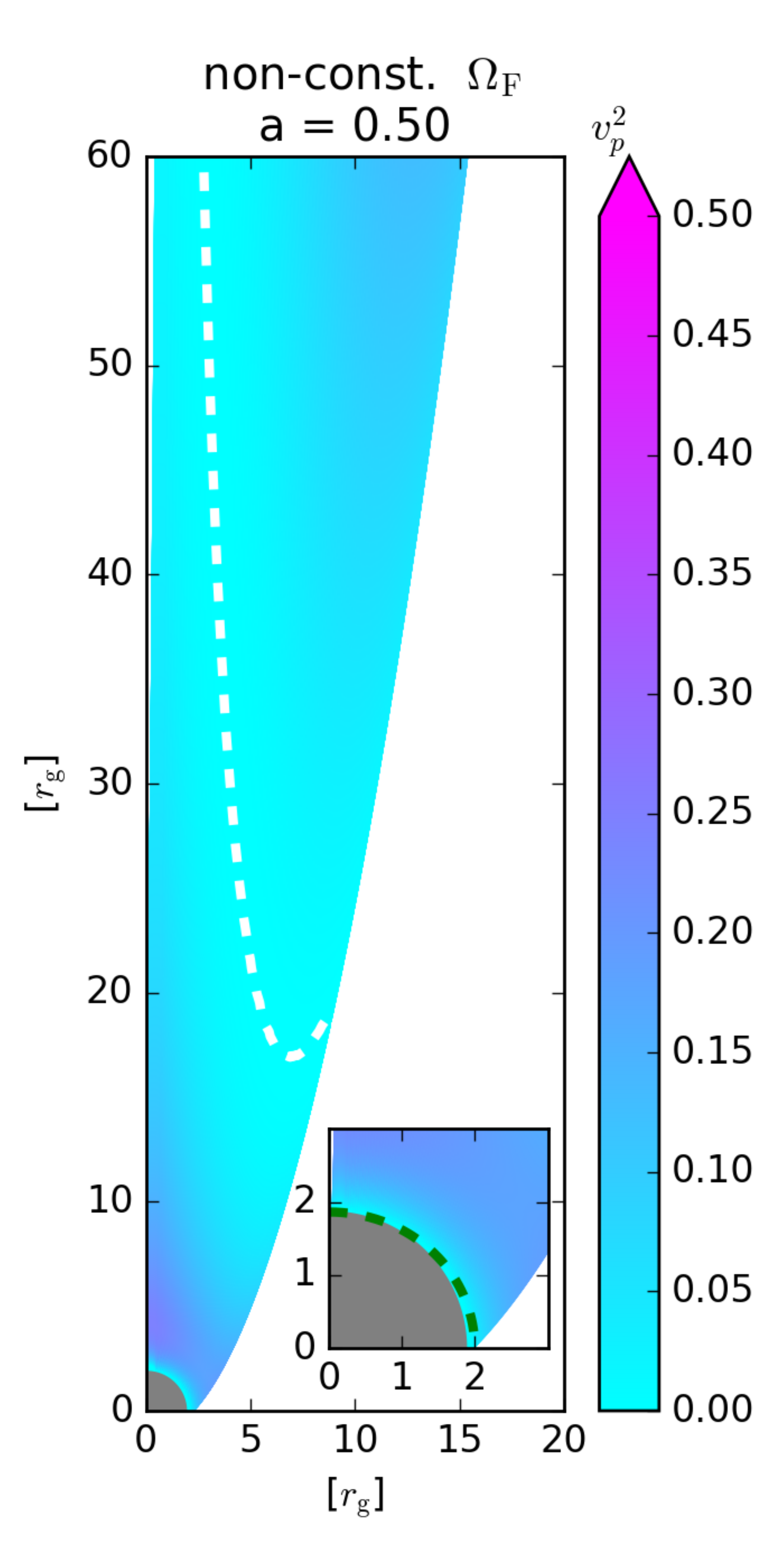}
\includegraphics[trim={0 1cm  0 0},width=0.3\textwidth]{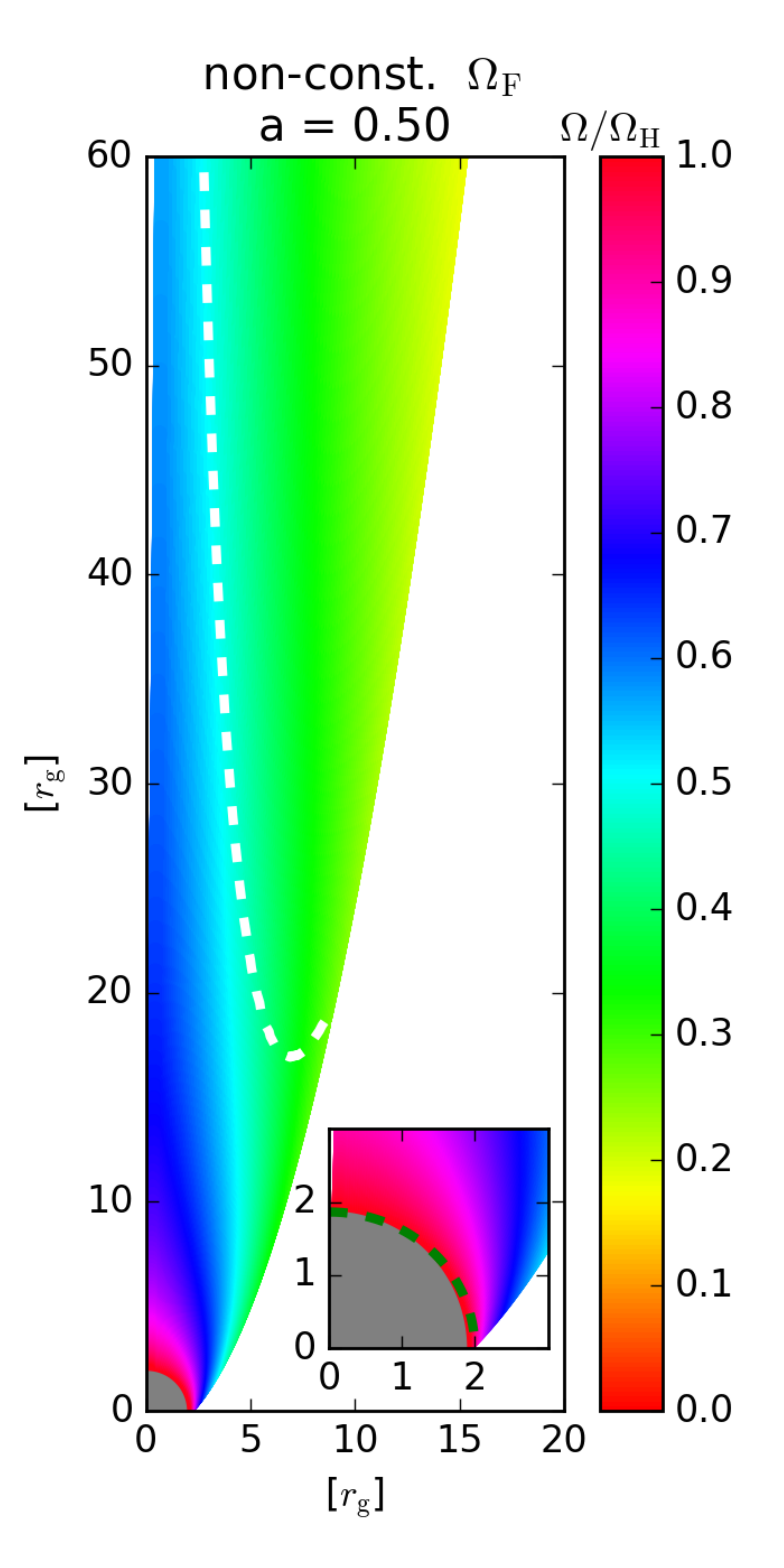}\\
\includegraphics[trim={0 1cm  0 0},width=0.3\textwidth]{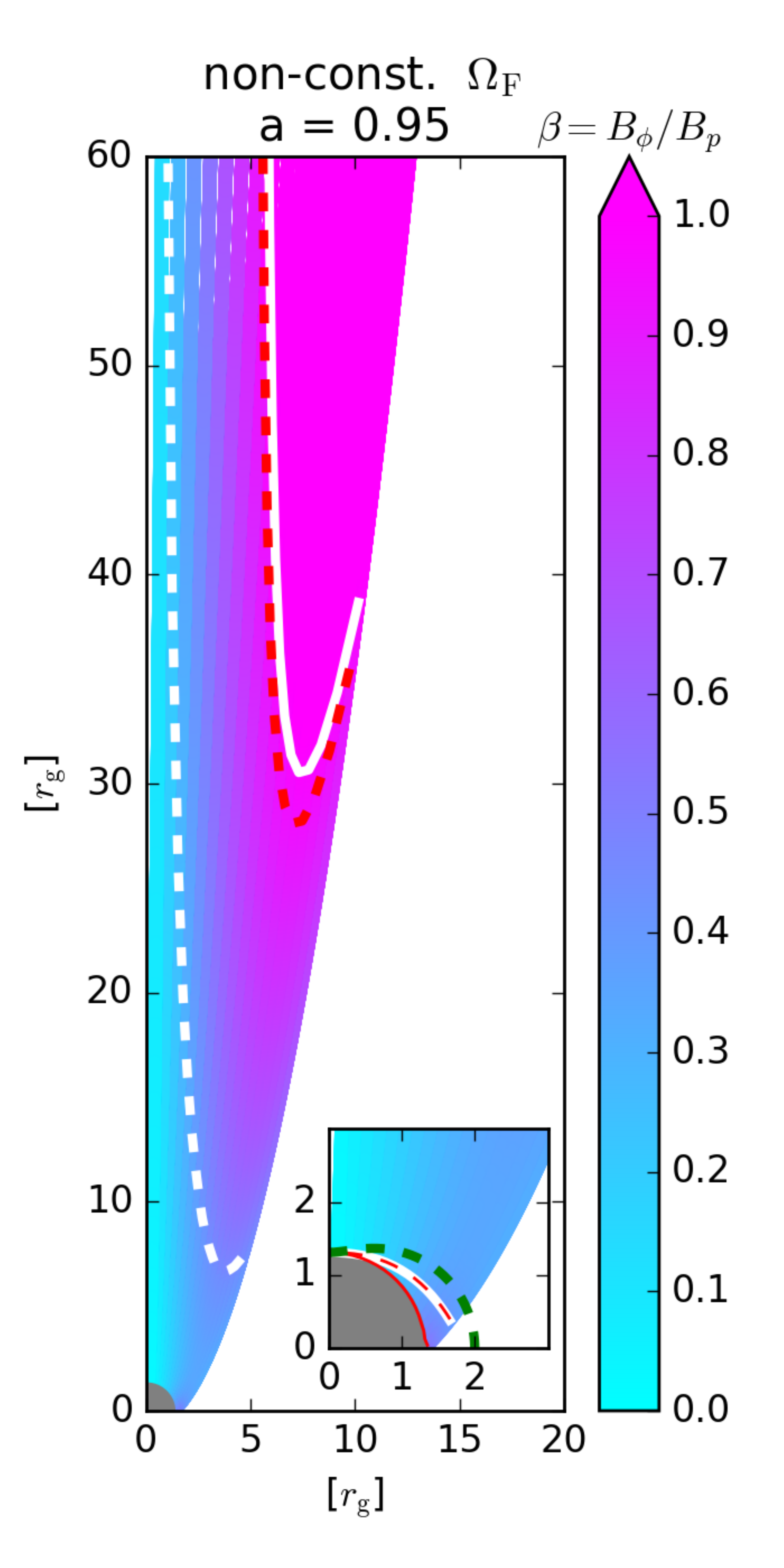}
\includegraphics[trim={0 1cm  0 0},width=0.3\textwidth]{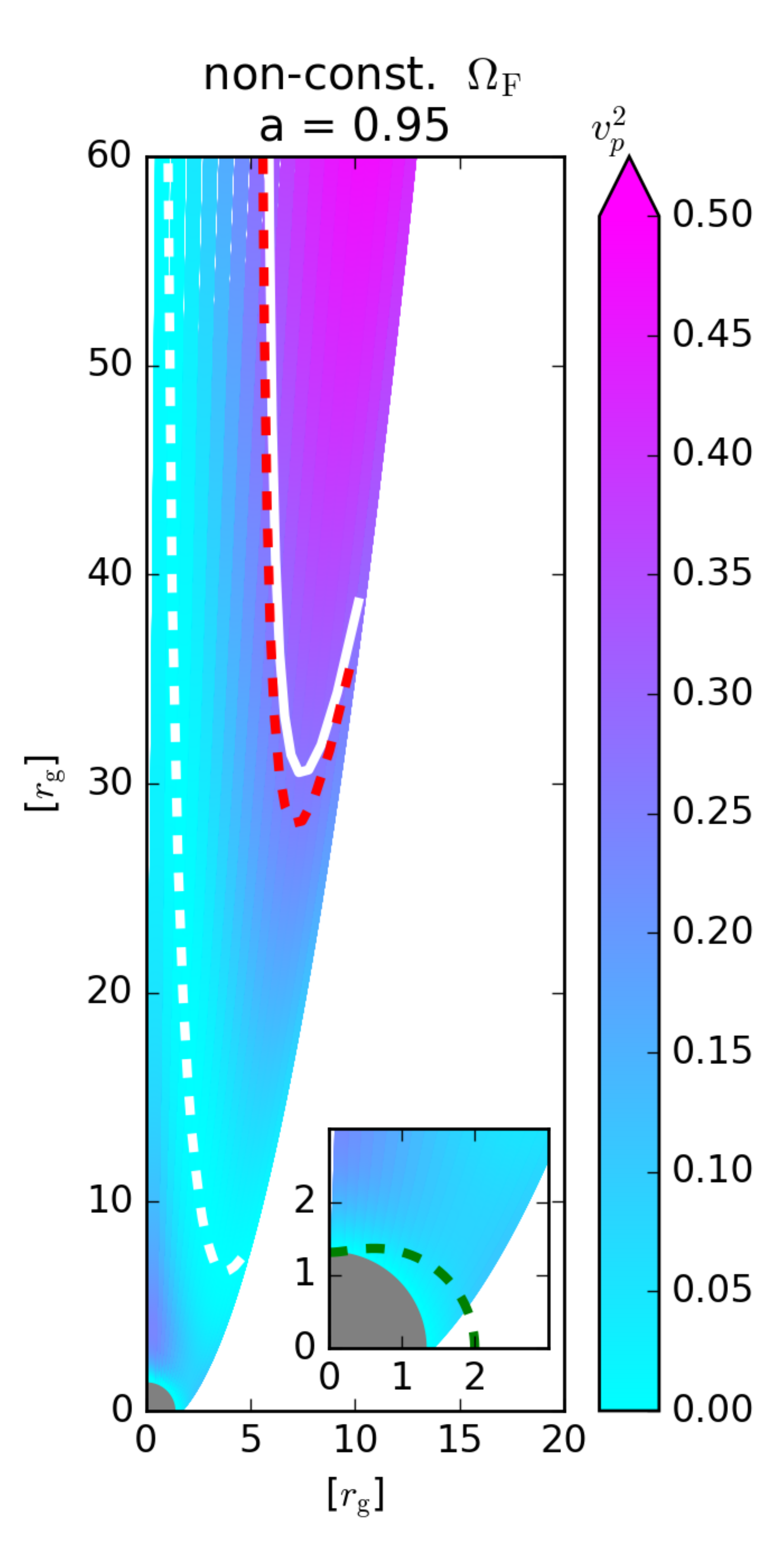}
\includegraphics[trim={0 1cm  0 0},width=0.3\textwidth]{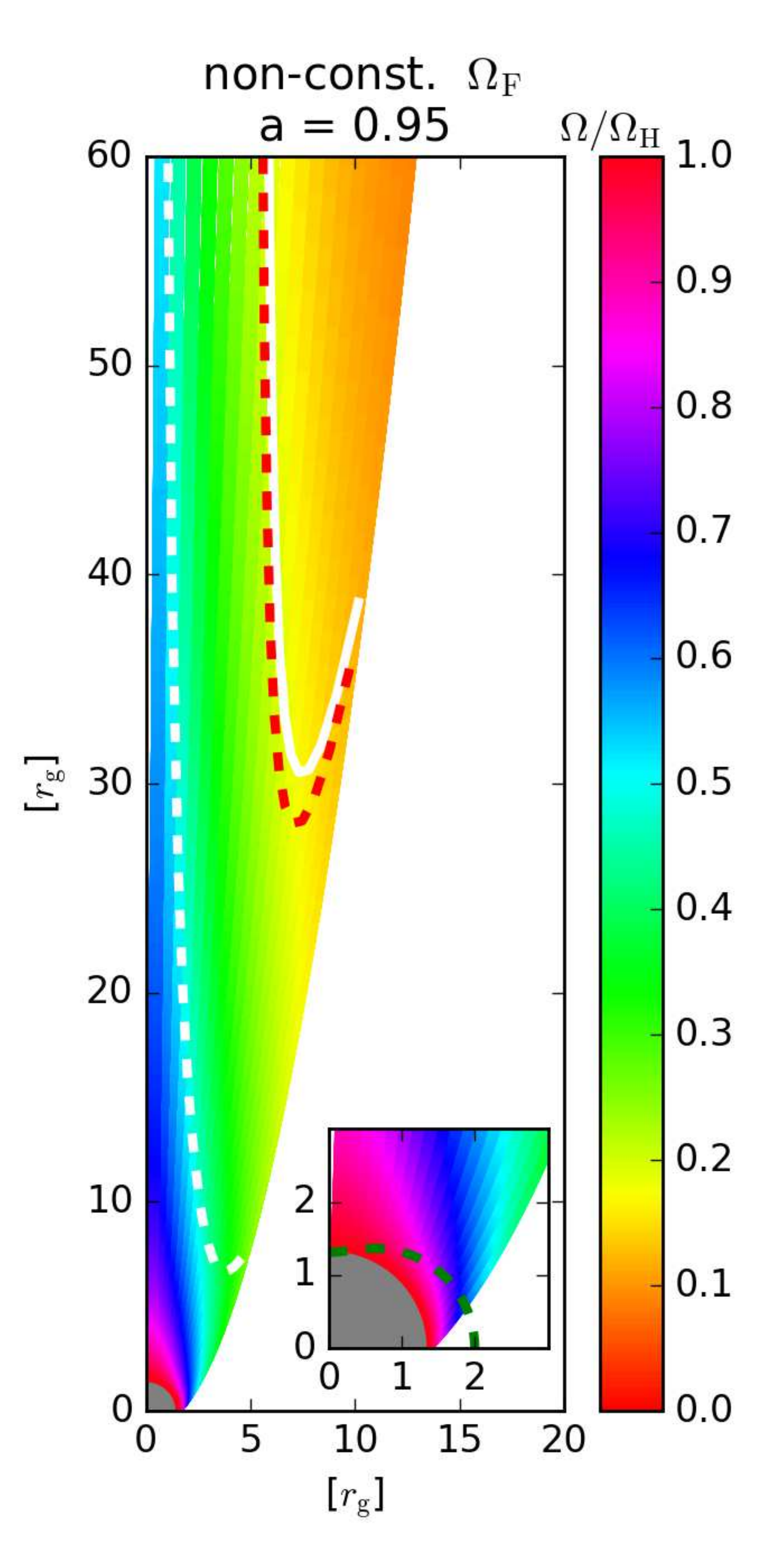}\\
\end{center}
\caption{Same as Figure \ref{fig:par_linear_f1} but for the cases of the non-constant field angular velocity profile $\Omega_{F}(\Psi)$ shown in Figure \ref{fig:para_omega_choice}.}  \label{fig:par_linear_f2}
\end{figure*}
The global properties of GRMHD jet near a black hole can be examined by solving the trans-fast magnetosonic solutions for the whole magnetic field lines within  the funnel region. The GRMHD jet solutions in the black hole magnetosphere with constant  magnetic field angular velocity is shown in figure \ref{fig:par_linear_f1}, for the case $\hat{E}(\Psi)=10$ and  black hole spins $a=0.5$ (top) and $a=0.95$ (bottom). The flow boundary shown here corresponds to the flow along the magnetic field line which penetrating the event horizon at $\theta_{\rm H}=90\degr$.
For all panels, the stagnation surface (dashed white line), the outer light surface (solid white line), and the outer Alfv\'en (dashed red line) and FMS  (solid red line) are plotted. As shown in the inset of the left panel, the inner Alfv\'en and FMSs are located inside the static limit surface, satisfying the necessary condition to extract black hole rotational energy outward \citep[][]{tak90}. For a Poynting flux dominated flow, the Alfv\'en surfaces almost coincide with the light surfaces, and the inner FMSs almost coincides with the event horizon, as seen in both top and bottom panel. As also seen is figure \ref{fig:para_gamma}, the stagnation surface of a faster spinning black hole located closer to the central black hole.

The ratio of toroidal and poloidal components of magnetic field are shown in the left panel of Figure \ref{fig:par_linear_f1} 
As shown in the plot, the poloidal field dominates in the inflow region and close to the rotational axis, and become comparable to the toroidal magnetic field near the outer Alfv\'en surface.

In the middle panel of Figure \ref{fig:par_linear_f1}, the poloidal three velocity square, $v_{p}^{2}\equiv(u_{p}/u^{t})^{2}$ is shown. At the stagnation surface, $u^{r}=0$ and $u^{\theta}=0$, which is the outer boundary of the inflow region, $u^{r}<0$ and $u^{\theta}>0$ and the inner boundary of the outflow region, $u^{r}>0$ and $u^{\theta}<0$. Along the magnetic field line, for the inflow, $v_{p}$ is not monotonic; the flow three velocity increases when they departure from the stagnation surface, but drops quickly before they enter the black hole due to the rapid increase of $u^{t}$ near the black hole (see also the bottom panel of Figure \ref{fig:sol_example}). For the outflow, $v_{p}$ continuously increases.

We show the angular velocity of the flow, $\Omega=u^{\phi}/u^{t}$ in the right panel of Figure \ref{fig:par_linear_f1}. Note that $\Omega(\Psi)\approx\Omega_{F}(\Psi)$ at the stagnation surface due to the vanishing poloidal velocity there. For the inflow, $\Omega$ gradually increase when they stream toward to the black hole, finally  $\Omega(\Psi)\approx\Omega_{\rm H}$ due to the black hole rotation.

In Figure \ref{fig:par_linear_f2} we plot the properties of GRMHD jet with magnetic field with the non-constant angular velocity. 
In addition to that the flow also share the above-mentioned general flow features shown in Figure \ref{fig:par_linear_f1}, the location of the stagnation surfaces in Figure \ref{fig:par_linear_f2} move further away from the black hole due to a slower field angular velocity profile (see also figure \ref{fig:para_omega_choice}). It is interesting to note that a more rapid decreasing of $\Omega_{F}(\Psi)$ near the jet boundary results in the ``V-shape" of the stagnation surfaces, with its valley locates close to the jet boundary. 
Such V-shape of stagnation surface in turns modify the profile of the outer light surface, and therefore the outer Alfv\'en surface. Note how the resulting constant $v_{p}^{2}$ contours (middle panel) also have a V-shape profile, indicating a non-monotonic ``slow--fast--slow'' structure of jet Lorentz factor across the jet. The outer slow layer adjacent to the funnel is resulting from the differential rotation in the magnetosphere of a spinning black hole, instead of a slower wind region emerging from the corona of the accretion flow. The resulting effect on the radiative transfer from the noticeable difference in velocity between a constant and non-constant field angular velocity (while keeping all other parameters the same), as seen in the insets of figures  \ref{fig:par_linear_f1} and  \ref{fig:par_linear_f2}, may in principle distinguishable by horizon-scale black hole images or movies \citep[e.g.][]{jet18}. The first black hole image has recently been obtained by the Event Horizon Telescope \cite[][]{eht19a,eht19b,eht19c,eht19d,eht19e,eht19f}. In addition, if energetic electrons are continuously or intermittently injected from the stagnation surface, its location could be constraint by horizon-scale observations \citep[e.g.][]{pu17}.

\subsection{Characteristic surfaces}\label{sec:result_char}
\begin{figure*}
\begin{center}
\includegraphics[trim={1.5cm 0 1.3cm 0},width=0.49\textwidth]{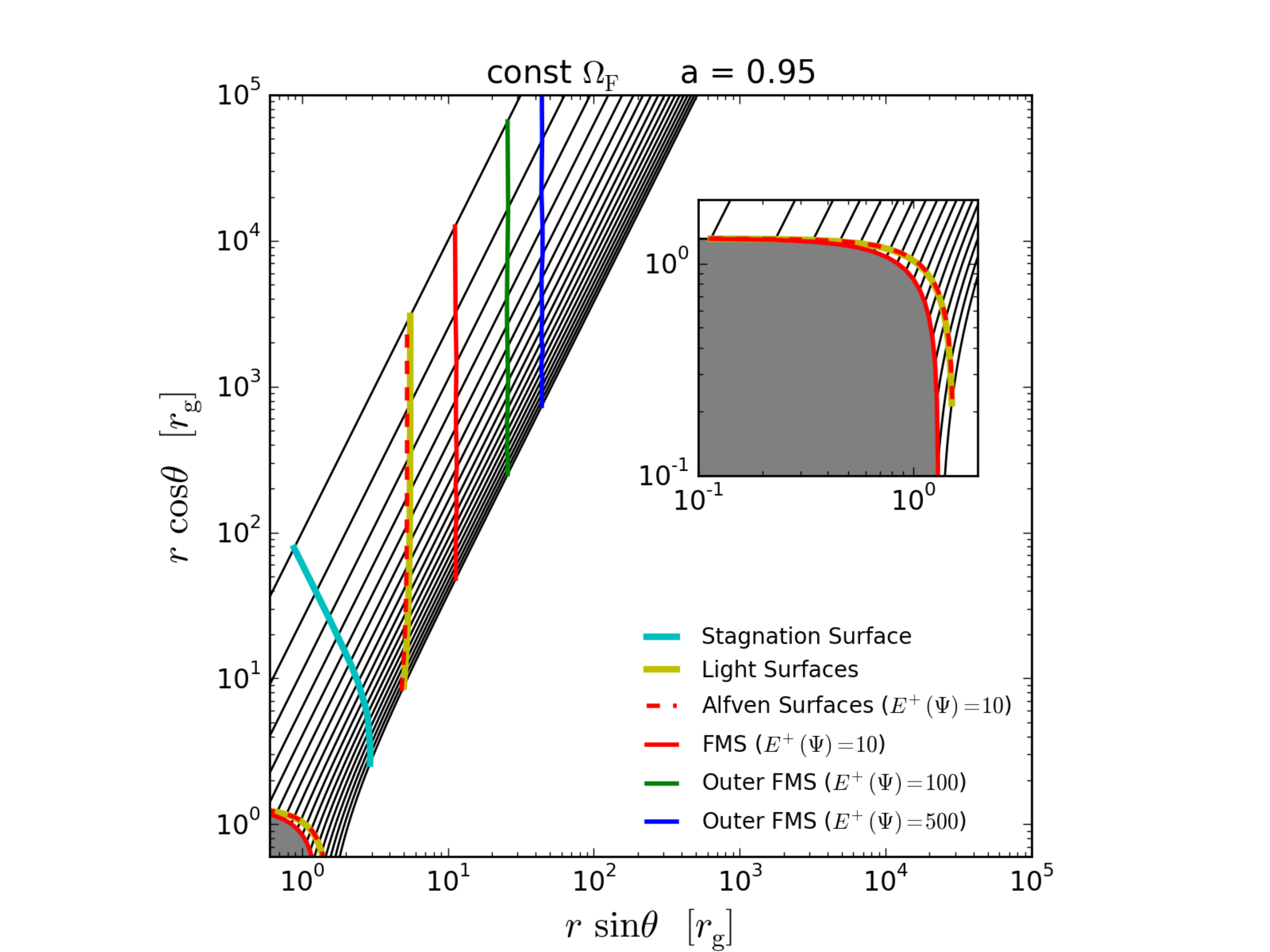}
\includegraphics[trim={1.5cm 0 1.3cm 0},width=0.49\textwidth]{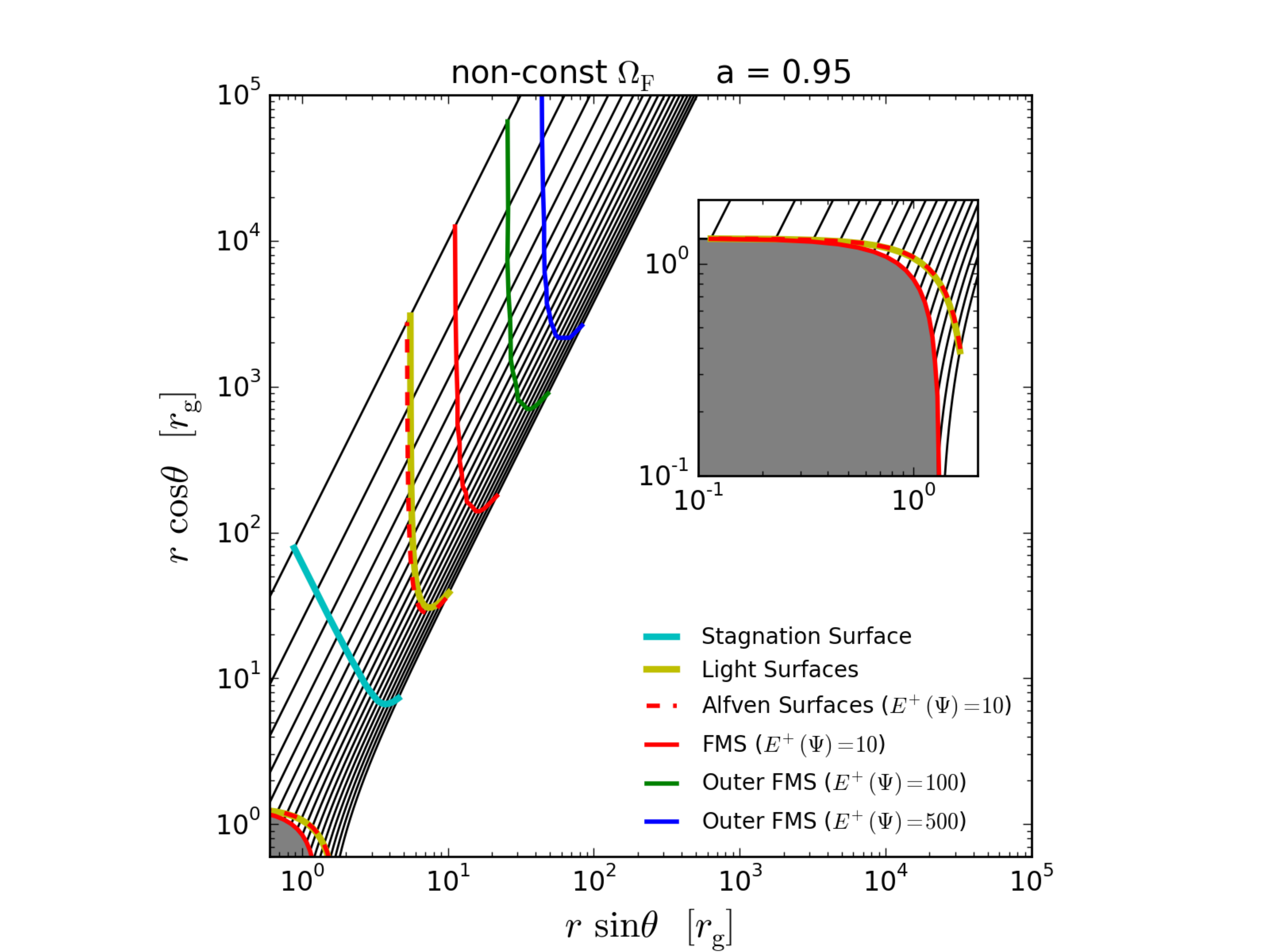}
\end{center}
\caption{Characteristic surfaces of a parabolic magnetosphere with a constant and non-constant field angular velocity (figure \ref{fig:para_omega_choice}), around a fast spinning black hole, $a=0.95$. The inner and outer light surface, and the stagnation surface are shown by the yellow and the cyan lines respectively. The outer FMSs for the total energy of outflow $E^{+}(\Psi)=(10, 100, 500)$ are shown the solid lines in different colors. The black hole is indicated by the grey shaded region. The inner fast surfaces of all different $E^{+}(\Psi)$ almost coincide with the event horizon.  In this plot, the outermost and innermost field line attaches the horizon at $\theta_{\rm H}=90^{\circ}$ and $\theta_{\rm H}=5^{\circ}$, respectively. See \S\ref{sec:result_char} for more discussions.
}\label{fig:para_loglog}
\end{figure*}

To present the distribution of characteristic surfaces: Stagnation surface, inner/outer Alfv\'en surfaces, and inner/outer FMSs, we plot in 
 Figure \ref{fig:para_loglog} for the result of three different outflow energy, $\gamma_{\infty}\approx \hat{E}^{+}(\Psi)=(10 , 100, 500)$ in a constant (left panel) and differential (right panel) black hole magnetosphere with $a=0.95$.
The cases for $\hat{E}^{+}=10$ corresponds for the solutions shown in the bottom panels of Figures \ref{fig:par_linear_f1} and \ref{fig:par_linear_f2}.
The location of the stagnation surface (in cyan) and inner/outer light surfaces (in yellow) are strongly related to $\Omega_{F}(\Psi)$ and independent of $\hat{E}^{+}$. 
Because the resulting locations of the inner FMSs (which locates almost coincides with the event horizon) and inner/outer Alfv\'en surfaces (which locate almost coincide with the inner/outer light surfaces) are all similar for all different $\hat{E}^{+}$ considered here, we only show the location of these surfaces for the case of $\hat{E}^{+}(\Psi)=10$ (in red). 
The outer FMSs for different $\hat{E}^{+}(\Psi)$, which relates closely to the jet acceleration process (as discussed in \S\ref{sec:result_out}), is shown by the solid lines in different colors. 
In general, the location of the FMS  of the outflow move further away from the black hole for a larger $\hat{E}^{+}$. The resulting V-shape of the stagnation surface and outer Alfv\'en and FMSs for a differential $\Omega_{F}(\Psi)$ (right panel) are also clearly shown.

While the surfaces shown in figure \ref{fig:para_loglog} corresponds to constant $\hat{E}^{+}(\Psi)$, the case for non-constant $\hat{E}^{+}(\Psi)$  distribution can be qualitatively inferred by the combination of the locus of different $\hat{E}^{+}(\Psi)$ at different $\Psi$. For example, for the GRMHD simulation of an accreting black hole system with $a=0.9375$ presented in \citet{mck06}, the  velocity at the jet core is more slower compared to the jet boundaries. By linking the surfaces with a higher outflow energy $\hat{E}^{+}\sim 100$ near the jet boundary to a lower outflow energy $\hat{E}^{+}<100$ toward to the core (the jet axis direction), the resulting profiles of the Alfv\'en and FMSs thus have a concave profile bending towards to the jet axis, qualitatively explains the result of \citet{mck06}, in which the locus of FMS  gradually become horizontal near the pole region in a log-log plot \citep[figure 11 of][]{mck06}. 

 Recently, notes has been added to the result of \citet{mck06}. \citet[][]{chat19} has preformed a number of large scale simulations with $a=0.9375$, and compare with the result of \citet{mck06}. In general, the overall characteristic surfaces shown in figure 14 of \citet[][]{chat19}  are similar to that in figure 11 of \citet[][]{mck06}. Intriguingly, the surfaces show "V-shape"-like profiles, a similar feature as shown in the right panel of figure \ref{fig:para_loglog}. They also found that pinch instabilities of the magnetic field  are developed around the boundary-layer region between the jet and the surrounding wind/conorna reigon play an important role to convert electromagnetic energy into heat energy.
 Alouth we assume the magnetic field configuration by the parameter $p$, the GRMHD simulaionts indicate that field geometry mildly deviate from the parabolic magnetic field line of $p=1$ (e.g. figure 11 of \citet[][]{mck06} and figure 14 of \citet[][]{chat19}).  Note that,
in our model of cold GRMHD flow, the slow-magnetosonic speed is zero  everywhere. In contrast, the pressure of the flow is properly considered in the GRMHD simulation, and therefore the slow magnetosonic surface appears. 
The location of the stagnation surface would vary from the cold fluid limit as we considered here, depending on the thermodynamical properties of the plasma (e.g., the electron temperature). Nevertheless, as the thermodynamical properties play little role in jet acceleration except when the pitch effect take place, our semi-analytical model for cold flow is capable for providing  qualitative and (a rough) quantitative insights to a magnetically dominated GRMHD flow structures from the horizon to a large distance. Observationally, for  a black hole system with known jet acceleration across the jet cross section at different radius, the black hole spin and the characteristic surfaces may therefore be constrained \citep[see, e.g.,][for the case of M87]{nak18}.

\begin{figure}
\begin{center}
\includegraphics[width=0.5\textwidth]{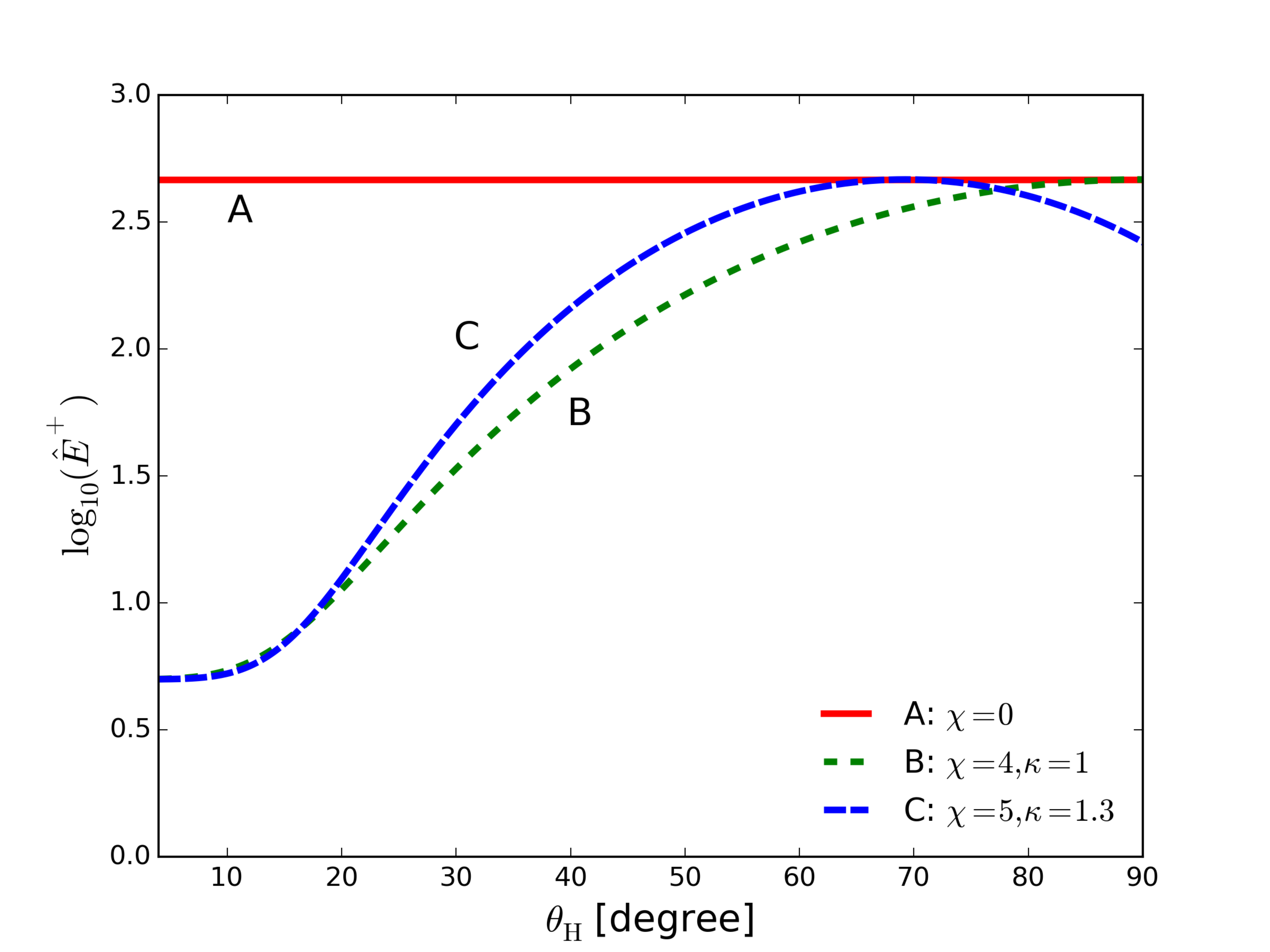}
\end{center}
\caption{Three different  cases of the outflow energy $\hat{E}^{+}(\Psi)$ along a split-monopole magnetic field line which penetrates the event horizon at different polar coordinate $\theta_{\rm H}$. See \S\ref{sec:mono} for more detatils.}  \label{fig:mono_E_choice}
\end{figure}

\begin{figure}
\begin{center}
\includegraphics[width=0.35\textwidth]{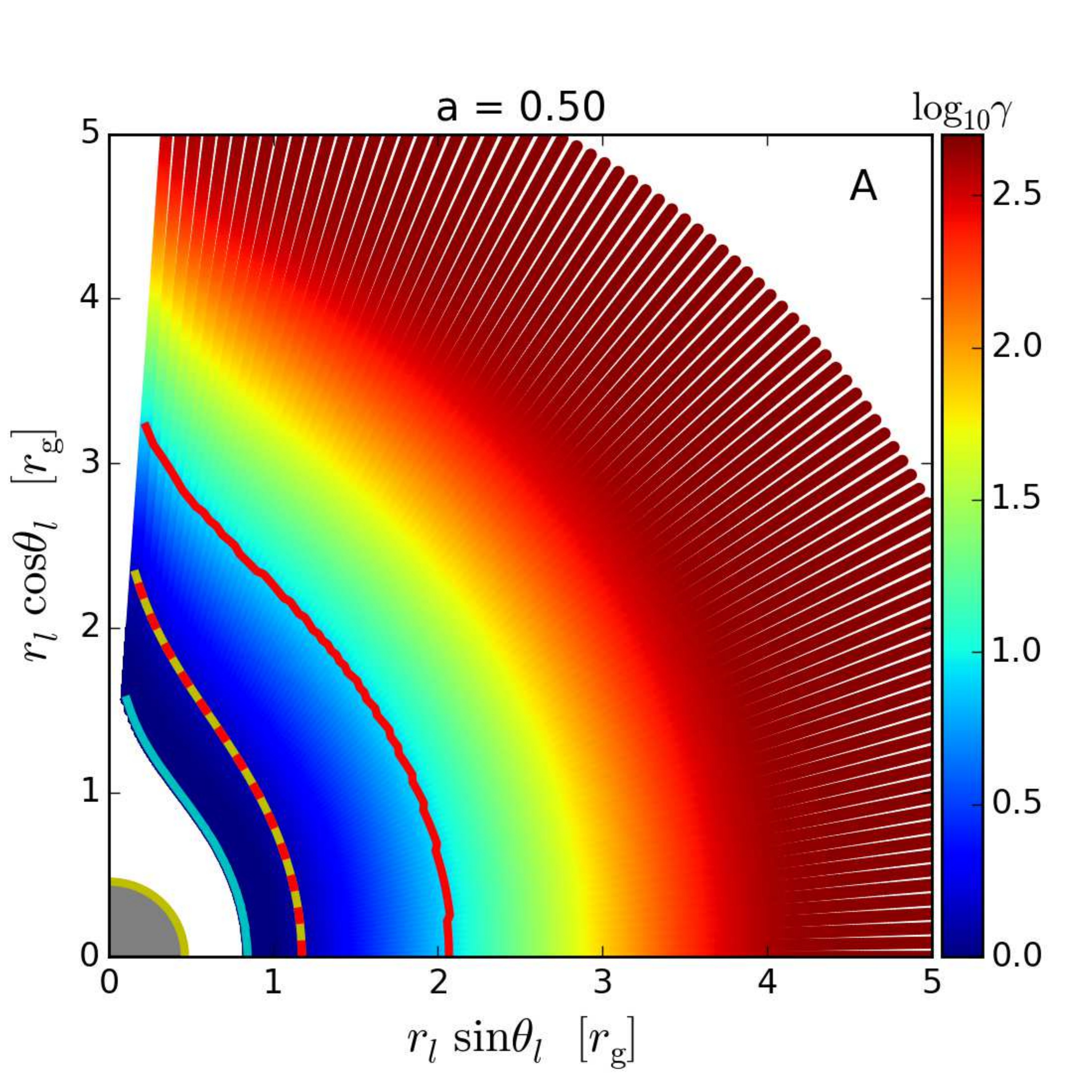}\\
\includegraphics[width=0.35\textwidth]{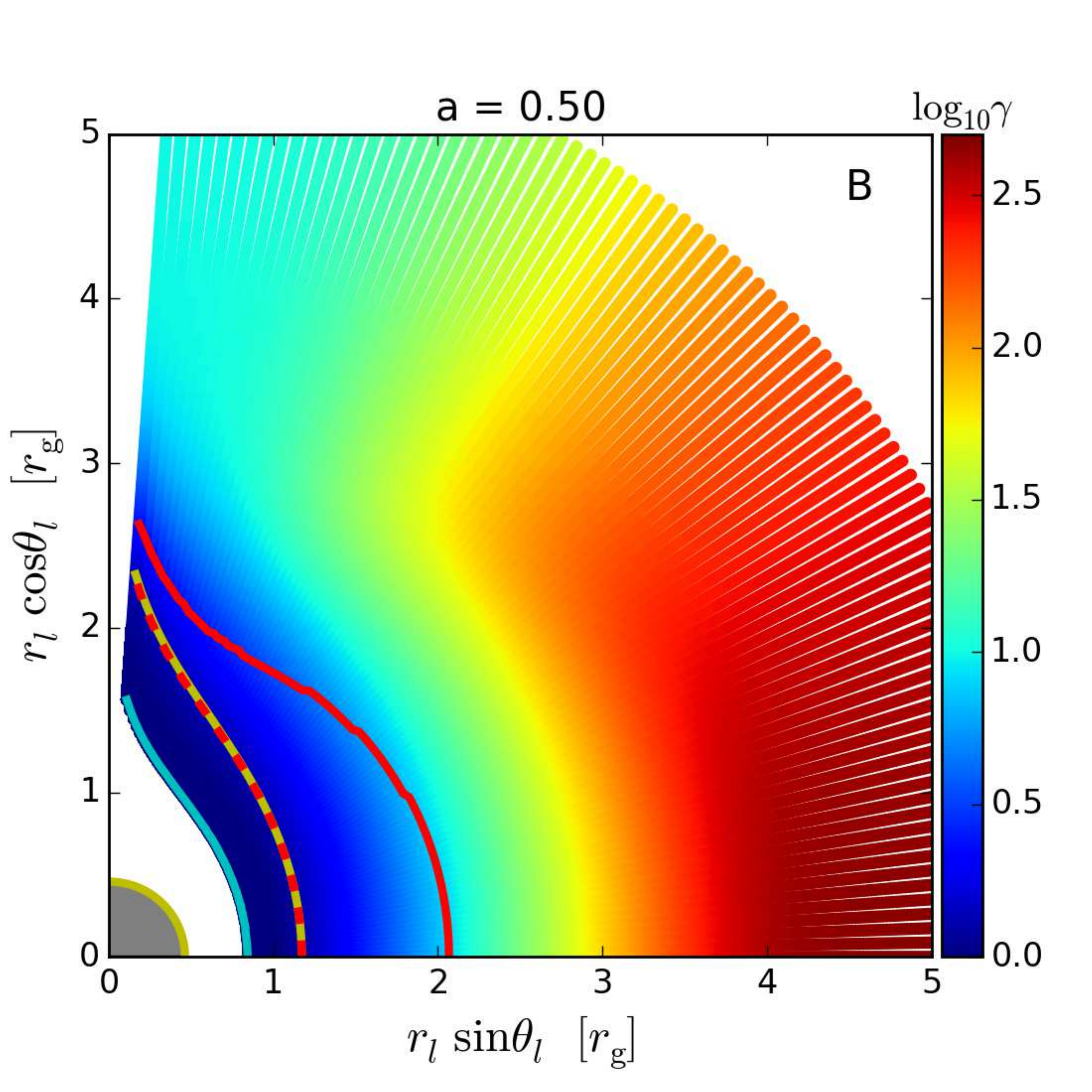}\\
\includegraphics[width=0.35\textwidth]{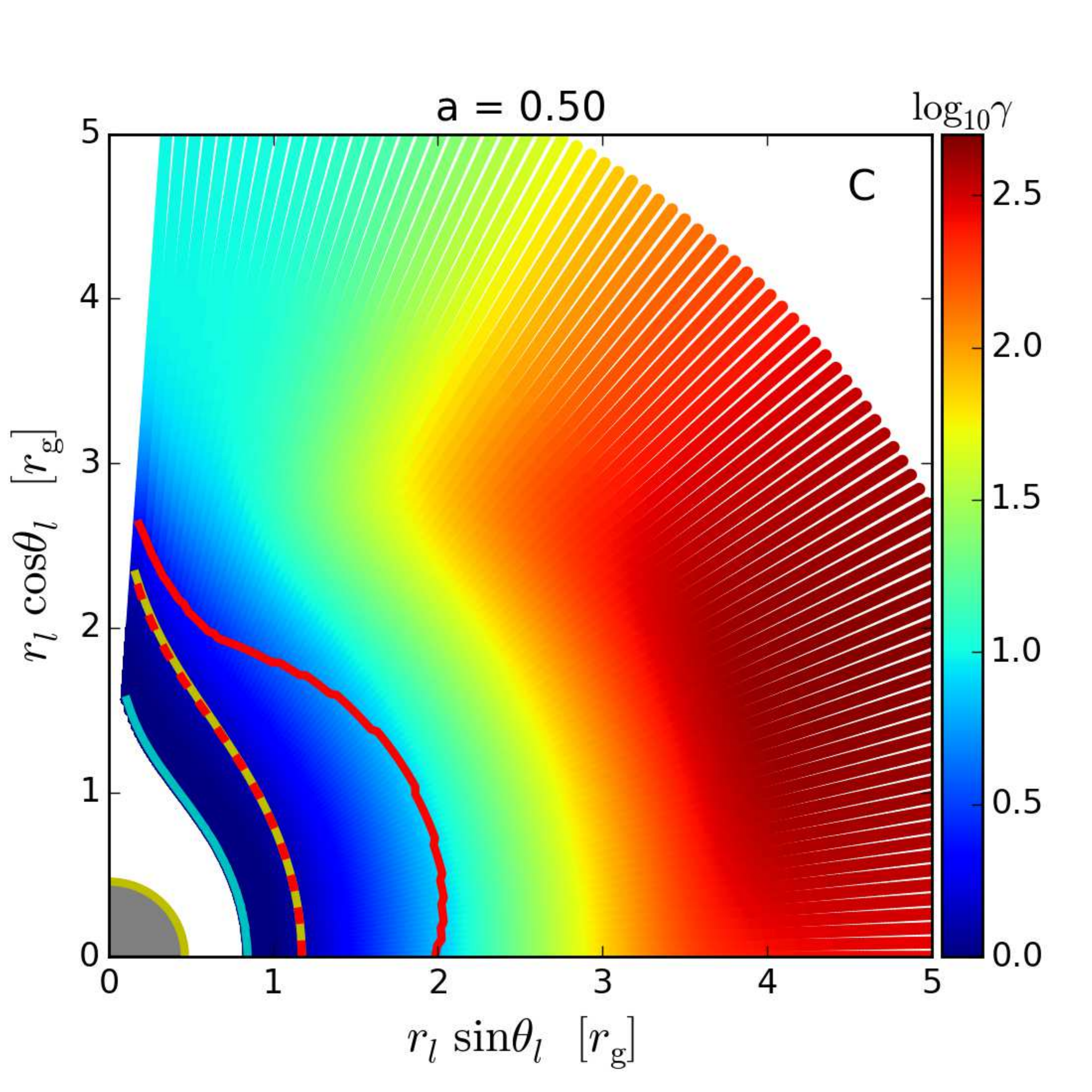}  
\end{center}
\caption{Lorentz factor of the GRMHD outflow along a split-monopole magnetosphere of a modest spinning black hole, $a=0.5$, for different flow energy choice shown in Figure \ref{fig:mono_E_choice}:  cases A (top panel), B (middle panel), and C (bottom panel). 
A ``logrithmic" spherical coordinate $(r_{l}, \theta_{l})$ defined by $(r_{l}=1+\log_{10}(r), \theta_{l}=\theta)$ is adopted.
The light surfaces (yellow lines,  which almost overlap with the red-dashed lines) and the stagnation surface (cyan line) are shown. The central shaded area indicates the black hole.  
The inner dashed and outer solid red lines are the Alfv\'en surface and FMS of the outflow. 
}  \label{fig:mono_loglog_a05}
\end{figure}

\begin{figure}
\begin{center}
\includegraphics[width=0.35\textwidth]{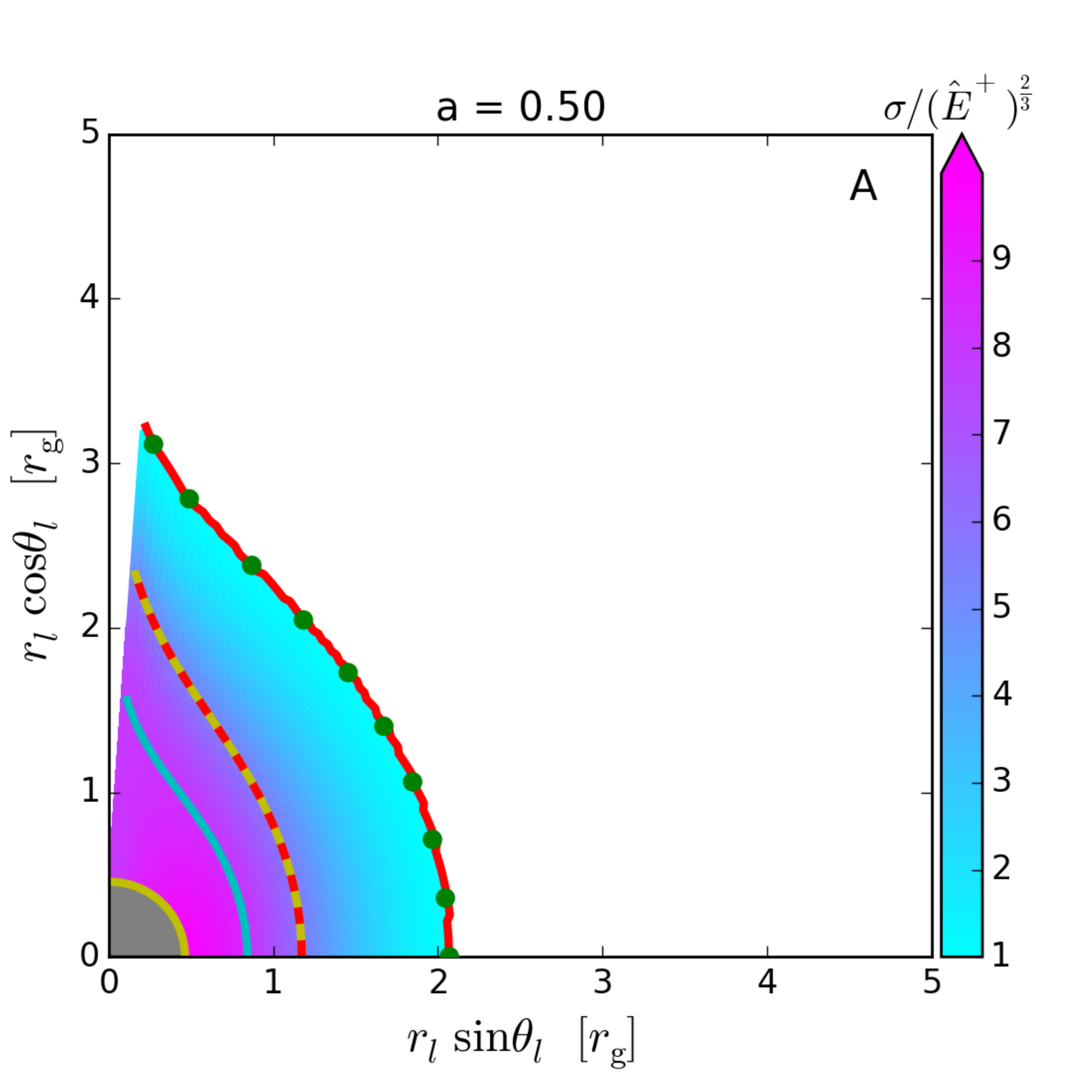}\\
\includegraphics[width=0.35\textwidth]{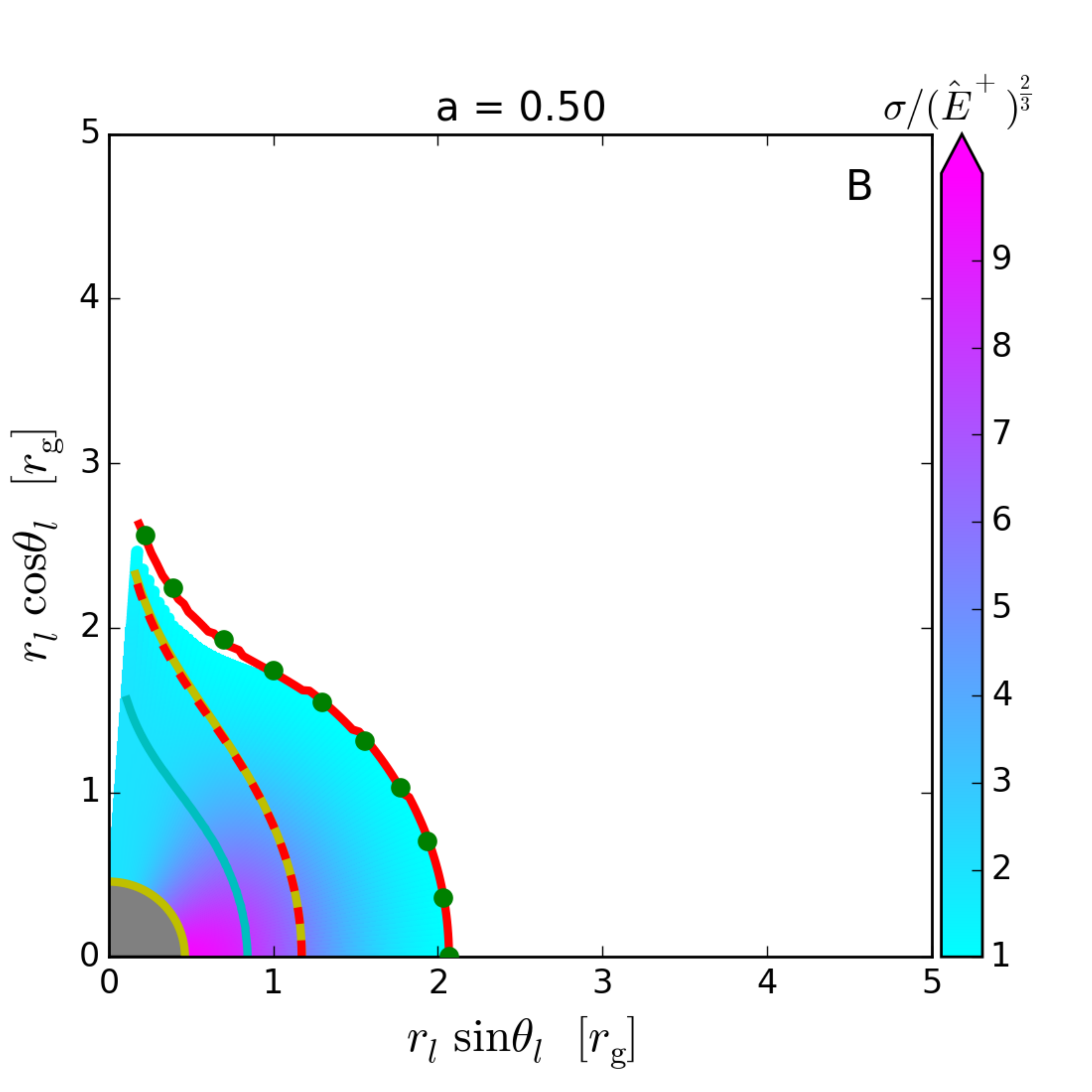}\\
\includegraphics[width=0.35\textwidth]{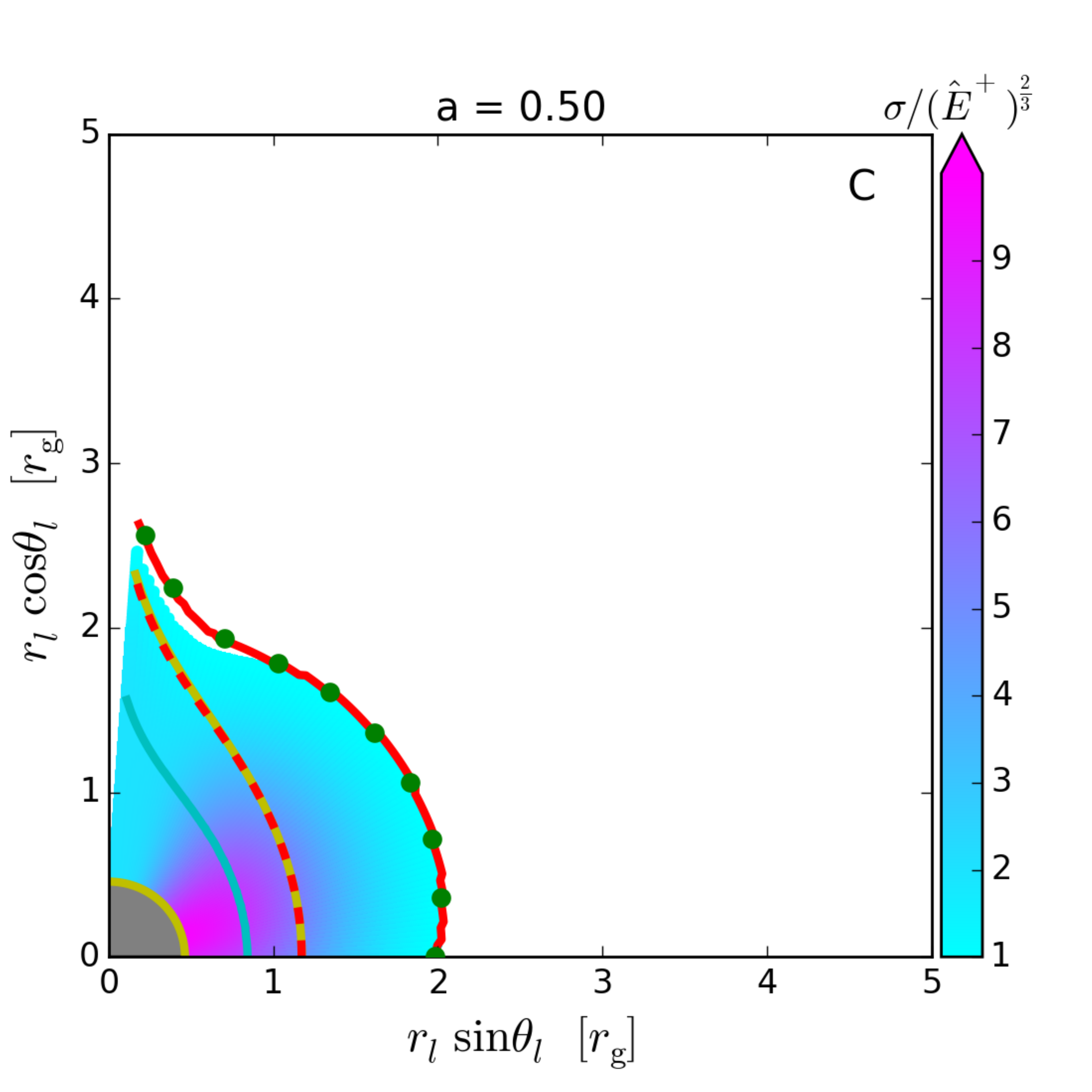}  
\end{center}
\caption{Corresponding magnetization, rescaled by $(\hat{E}^{+})^{2/3}$, for both the outflow and inflow region of Figure \ref{fig:mono_loglog_a05}. Regions for $\sigma/(\hat{E}^{+})^{2/3}<1$ are not shown, in order to highlight that contours of $\sigma/(\hat{E}^{+})^{2/3}=1$ roughly fit the outer FMS (the outer red lines), consistent with the theoretical prediction for a  SRMHD flow along a split-monopole case \citep[][]{bes98}. The green circles indicates the theoretical prediction location for selected field lines, which is also in good agreement with the location of the outer FMS.}  \label{fig:mono_loglog_sigma}
\end{figure}

\section{Trans-fast magnetosonic Flow along Split-monopole magnetic Field Lines}\label{sec:mono} \label{sec:mono}
We now explore the split-monopole magnetosphere ($p=0$). 
Motivated by the  SRMHD flow for different plasma loading  across field lines \citep[][]{tch09}, here we also release the assumption of a constant $\hat{E}^{+}(\Psi)$ and choose three representative different profiles  by
\begin{equation} 
\hat{E}^{+}(\Psi;\theta_{\rm H})=\hat{E}_{\rm max}\sin^{\chi}(\kappa\theta_{\rm H})+\hat{E}_{\rm min}\;,
\end{equation} 
with  $\hat{E}_{\rm max}$ and $\hat{E}_{\rm min}$ are the maximum and minimal energy, and $\chi$ and $\kappa$  are  parameters controlling to the corresponding angle when $\hat{E}(\Psi; \theta_{\rm H})=\hat{E}_{\rm max}$. Here we adopt $\hat{E}_{\rm max}=460$ and $\hat{
E}_{\rm min}=10$.

We consider three different energy distribution of $\hat{E}^{+}(\Psi)$ as shown in figure \ref{fig:mono_E_choice}. For case A, for reference, we adopt a constant  $\hat{E}^{+}(\Psi)$ across the magnetosphere, with $\chi=0$. 
For case B, by adopting $\chi=4$ and $\kappa=1$,  a monotonically increasing $\hat{E}^{+}=\hat{E}_{\rm min}$ from the pole region, then reaches $\hat{E}_{\rm max}$ at the funnel region of the accreting gas (the jet boundary).
For case C, the $\hat{E}^{+}$ reaches $\hat{E}_{\rm max}$ at around $\theta_{\rm H}\sim 70^{\circ}$, then deceases towards to the jet boundary, as described by using $\chi=5$ and $\kappa=1.3$. 
The constant field angular velocity, equation (\ref{eq:field_vel_1}), is applied for the split-monopole magnetic configuration considered here \citep[][]{bla77,phi83}. However, among the three representative cases, A, B, and C, the last case seem more physical \citep[see also][]{tch09}.

As in \citet[][]{tch09},  it seems convenient to show the Lorentz factor $\gamma$ in the ``logrithmic" spherical coordinate.
In Figure \ref{fig:mono_loglog_a05}, we consider a black hole with $a=0.5$, and present 
the Lorentz factor of the GRMHD outflows.  
The shapes of the stagnation surface (the cyan line), the inner and outer light surfaces (the yellow lines), and the outer Alfv\'en  surface and FMS (the dashed and solid red lines, respectively) are all elongated in the direction toward to the polar direction. 
We should note that the efficient energy conversion is not efficient for a split-monopole configuration \citep[see][]{tch09}, and the efficient jet acceleration (i.e. $\gamma\to\hat{E}^{+}(\Psi)$)  far away from the FMSs is artificial due to 
the our assumption for the  parameter $\zeta_{0}$ (see also \S\ref{sec:model_par} and  Appendix \ref{app:geo_effect}). Therefore the region far away from the outer FMS for a split-monopole configuration is beyond our interest and approach.

The perturbation on the equation of motion for the plasma loading onto a split-monopole force-free magnetic field line $\Psi\propto (1-\cos\theta)$ in flat spacetime has been well studied in \citet[][]{bes98}. The authors found that the  the FMS is located at\footnote{Although the transverse distance to the FMS, $(r\sin\theta)_{\rm FMS}^{\rm SRMHD}\approx\sigma_{0}^{1/3}/\Omega_{F}$,  are the same for the split-monopole case \cite[][]{bes98} as mentioned in this section (\S \ref{sec:mono}), and for the parabolic field line case  \cite[][]{bes06} as mentioned in \S\ref{sec:result_out}, the radial distance to the FMS, $(r)^{\rm SRMHD}_{\rm FMS}$, is much shorter for the former case compared to the latter due to different field geometry.} 
$(r\sin\theta)_{\rm FMS}^{\rm SRMHD}\approx\sigma_{0}^{1/3}/\Omega_{F}$\, and the Lorentz factor there is $\gamma^{\rm SRMHD}_{\rm FMS}\approx\sigma_{0}^{1/3}$, where  $\sigma_{0}$ is the Michel's magnetization parameter \citep[][]{mic69}. 
The comparison between our GRMHD solutions and the above predicted properties are shown in Figure \ref{fig:mono_loglog_sigma}. The green circles represents the location computed by $(r\sin\theta)_{\rm FMS}^{\rm SRMHD}\approx(\hat{E}^{+})^{1/3}/\Omega_{F}$, which shows good agreement with the FMS locations of the GRMHD outflow solutions (the solid red lines). It is also become clear that, for a given magnetic field line, the location of the outer FMS is further away for a larger $\hat{E}^{+}$. Therefore different choice of $\hat{E}^{+}(\Psi)$ correspondingly results in different profile of the FMS.
The color map in Figure \ref{fig:mono_loglog_sigma} shows the magnetization. It is verified by the contour $\sigma/(\hat{E}^{+})^{2/3}=1$ that $\gamma^{\rm SRMHD}_{\rm FMS}\approx(\hat{E}^{+})^{1/3}$ is a good approximation of the energy conversion ratio at the FMS of the outgoing GRMHD flow.

\section{Limitation of the Model}\label{sec:limit}
Although the presented semi-analytical model  provides a fast and intuitive to explore the steady trans-fast magnetosonic outflow along magnetic field lines in the black hole magnetosphere, there are some limitations for this approach. Here we enumerated several cautions for the utility of our model.

In our outflow model, we assume a magnetically dominated black hole magnetosphere at the jet formation region, so that we apply the solution of force-free magnetic fields there, by construction. The obtained trans-fast magnetosonic flow  solution  becomes fluid-kinetic energy dominated at the radius about 10 times the radius of the FMS, so that the effect of plasma inertia  on the magnetic field shape can not be ignored  (e.g. TT03). 
In this case, it is necessary to evaluate the force-balance equation of the magnetic field lines, and our presented approach can not be applied.

Another limitation is the flow solution close to the pole region, $\theta\to0$. Such region requires more cautions due to several reasons.
First, our model scheme applies for flow which passes the outer light surface. However, close to the rotational axis of the magnetosphere, the location of the outer light surface move towards to infinity ($r_{\rm L}\gg1$ for $\theta_{\rm L}\ll1$, where L denotes the outer light surface). Second, field lines near the pole tend to bunch up around the axis due to mass loading \citep[e.g.][]{tch09,tch10}, which is beyond the application of our working assumption since the deformation of the force-free field is completely ignored. In general, the flow energy, the mass loading, and the deformation of the magnetic field lines are all related \citep[e.g.][]{pu15}.

\section{Summary and Outlooks}\label{sec:summary}
A semi-analytical approach for modeling the global stationary trans-fast magnetosonic jet structure is presented in the paper, by the following working assumptions: (i) adopting a prescribed poloidal force-free magnetic field configuration and ignore the deformation of the fields due to the plasma loading. (ii)prescribing sophisticated relation between the poloidal and toroidal components of the magnetic field (TT08), to preserve the key physics introduced by the plasma inertia,  including the jet acceleration and the existence of the FMS.
The  trans-fast magnetonic outflow model   by introducing a regular function $\beta(r;\Psi)$ thus easily integrates all the key process for a black hole powered jet   acceleration at different scales: from horizon scales ($\approx r_{\rm g}$) to large scales ($> 10^{4-5}r_{\rm g}$; $\sim$ pc scale).  As demonstrated in this paper, we have discussed the jet acceleration along a magnetic field line and the distributions of the jet velocity, magnetization parameters in the funnel region, for a regular function of toroidal and poloidal magnetic field configuration. Then we show their dependence of the black hole spin $a$, the magnetic field geometry $\Psi$, the angular velocity of the magnetic field line $\Omega_{F}(\Psi)$, and the total energy of the outflow $\hat{E}^{+}(\Psi)$.

The processes includes the extraction of black hole rotational energy via the inflow, a continuous propagation of Poynting energy flux at where the inflow and outflow matches at the plasma source  (i.e. the stagnation region),  where a simple relation of the inflow energy $\hat{E}^{-}(\Psi)$ and outflow energy $\hat{E}^{+}(\Psi)$ is assumed by the matching condition of the ingoing and outgoing trans-fast magnetosonic flow solution. Then, we find that a minimal outflow energy is required for the extraction of the rotational energy of a rotating black hole. Beyond the light surface and/or FMS,
the conversion from electronmagnetic energy to plasma kinetic energy becomes effective, and hence the jet velocity reaches almost terminal velocity, in the  distant outflow region.

With flexible parameter choices, together with assumed electron heating, cooling, and distribution,
 our model is also applicable for confronting theoretical  GRMHD jet properties with observations at different scales, such as polarized jet emission \citep[e.g.][]{bro10, por11}, jet morphologies at large-scale \citep[e.g.][]{tak18,ogi19} and at horizon-scale \citep[e.g.,][]{bro09,dex12,cha15,mos16,pu17,rya18,cha19,eht19e}.  From observations, preferred parameters in an accreting black hole system can also be constraint by exploring the parameter space of a physical motivated semi-analytical or phenomenological models \citep[e.g.][]{bro09b,bro11,bro16,tak18}.

\acknowledgments
\label{ack}
We thank the anonymous referee for valuable comments. H.Y.P. is supported by  Perimeter Institute
for Theoretical Physics. 
Research at Perimeter Institute is
supported by the Government of Canada through Industry Canada and by the Province of Ontario through the Ministry of Research and Innovation. 
M.T. is supported by KAKENHI Grand Number 17K05439 and DAIKO FOUNDATION. 
This research has made use of NASA Astrophysics Data System.

\appendix
\section{Flow chart for the semi-analytical Model}\label{app:flow_chart}
\begin{figure*}
\begin{center}
 \includegraphics[width=0.6\textwidth]{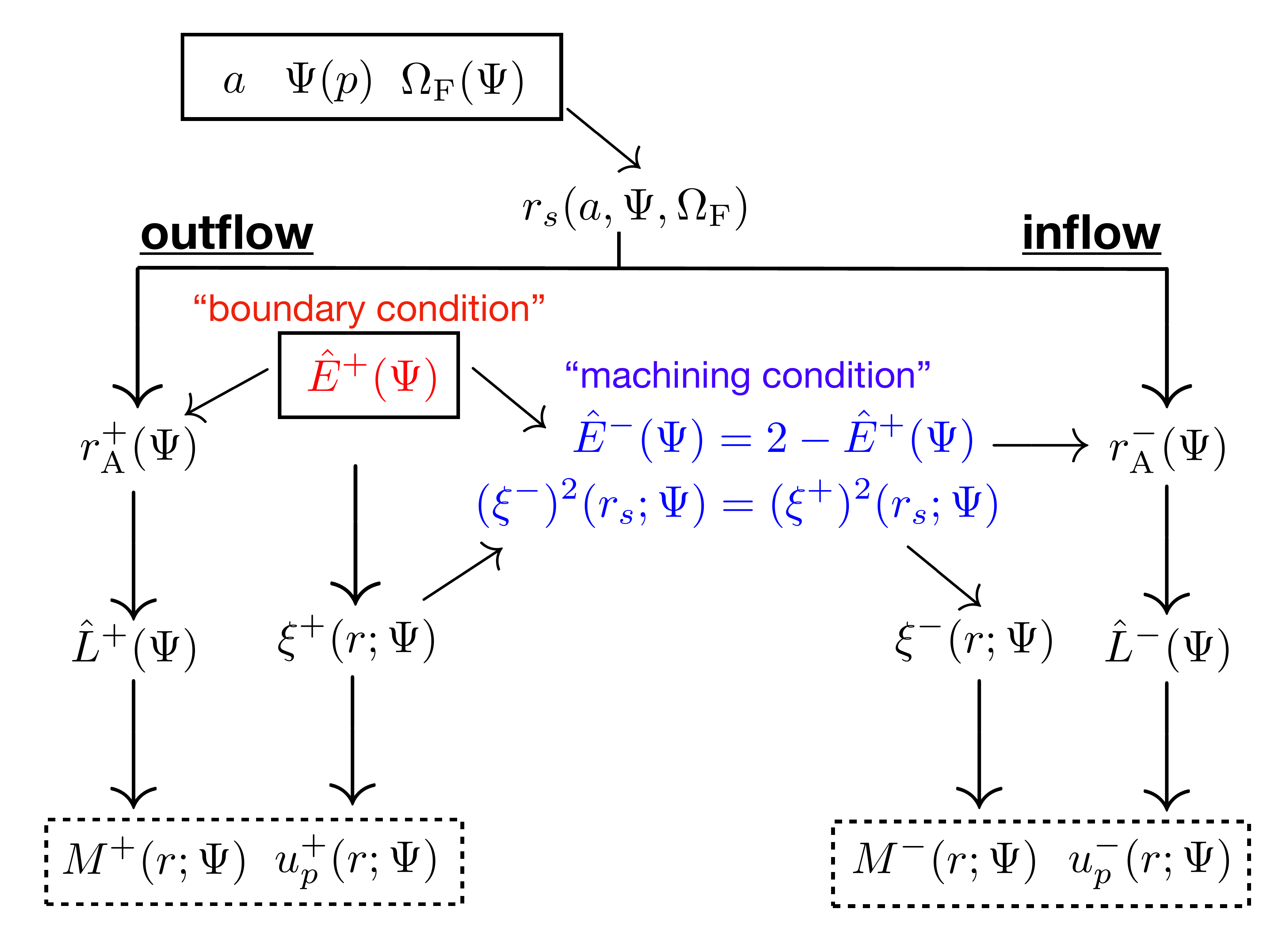}
 \end{center}
\caption{Overview of modelling Poynting flux dominated trans-fast  magnetosonic flow powered by a rotating black hole, as described in \S\ref{sec:model}.   The flow structure is semi-analytically computed after assigning four parameters in the model: black hole spin $a$, the magnetic field configuration $\Psi$ , as well as the magnetic field angular velocity $\Omega_{F}(\Psi)$ and total energy of the outflow $\hat{E}^{+}(\Psi)$ across the magnetic field lines $\Psi$. The solid boxes indicates the input parameters, and the dashed boxes indicates the output GRMHD flow solution. The symbol $A\to B$ indicates that $B$ can be computed from $A$. See Appendix \ref{app:flow_chart} for details.}  \label{fig:overview}
\end{figure*}

A flow chart for the semi-analytical approach described in \S\ref{sec:model} is summarized in Figure \ref{fig:overview}.
By describing the black hole magnetosphere with the black hole spin $a$, the field configuration $\Psi(p)$, and the magnetic field angular velocity $\Omega(\Psi)$, we can determined the stagnation surface $r_{\rm s}(a,\Psi,\Omega_{F})$, which separates the regions of the inflow (indicated by the superscript  ``--") and the outflow (indicated by the superscript  ``+").
We then use the specific flow energy of the outflow $\hat{E}^{+}(\Psi)$, a streamline conserved quantity,  as the boundary condition of the GRMHD solution, with which the location of the Alfv\'en surface $r_{\rm A}^{+}(\Psi)$ and the angular momentum of the flow $\hat{L}^{+}(\Psi)$, another streamline conserved quantity, are simultaneously determined, and the  Alfv\'en Mach number $M^{+}(r; \Psi)$ and the polodial velocity $u_{p}^{+}(r; \Psi)$ of the outflow can be algebraically solved by  using equation (\ref{eq:mach_eqn}).
Along each magnetic field line, the inflow solution is consistently solved by applying the matching condition for the flow energy $\hat{E}^{-}(\Psi)$ and the pitch angle $\xi^{-}(r; \Psi)$.  The relativistic jet powered by a rotating black hole should satisfy the following  conditions:  $0<\Omega_{F}<\Omega_{\rm H}$ and $\hat{E}^{+}>2$.  Note that this procedure is the case of a stationary solution, so it is also easy to reverse ($B\to A$). By using detailed observational data in the distance in near feature, it will be possible to estimate our theoretical model parameters around the black hole.

\section{Flow Acceleration and their Dependence on the Field Configuration}\label{app:geo_effect}
The relation of the efficiency of  SRMHD flow acceleration and the  magnetic field geometry has been extensively discussed in previous works. It is shown that for an  infinite magnetically-dominated plasma, the flow velocity is similar to the drift velocity \citep[][]{nar07, tch08}, and  the Lorentz factor $\gamma$ can be decomposes into \citep[][]{tch08,tch09,kom09}
\begin{equation}\label{eq:gamma_decomp}
\frac{1}{\gamma^{2}}=\frac{1}{\gamma_{1}^{2}}+\frac{1}{\gamma_{2}^{2}}\;,
\end{equation}
and
\begin{equation}
\gamma_{1}^{2}=\frac{B^{2}}{B_{p}^{2}}\;,
\end{equation}
\begin{equation}
\gamma_{2}^{2}=\frac{B^{2}}{B_{\phi}^{2}-E_{p}^{2}}\;,
\end{equation}
where  $E_{p}$ is the strength of the poloidal electric field and $B_{p}$ and $B_{\phi}$ are the poloidal and toroidal component of the magnetic field $B$.
Two types of acceleration  regime exist: the first term in Equation (\ref{eq:gamma_decomp}) corresponds to a {\em linear} (faster) acceleration, and the second term in Equation (\ref{eq:gamma_decomp}), which is related to the field configuration and the poloidal radius of the curvature of the field lines,  corresponds to a {\em logarithmic} (slower) acceleration.
In the detailed analysis in \citet[][]{tch08}, it is shown that for a field configuration $\Psi\propto r^{p}(1-\cos\theta)$, the second term is negligible $p\ge1$, and all the electromagnetic energy will eventually convert to the kinetic energy of the flow. 
It is further pointed out in \citet[][]{tch09} that, while the first term always dominant close to the compact object, the second term becomes dominant beyond a ``causality surface''  introduced in \citet[][]{tch09}.
While the FMS marks the boundary beyond which the flow can no longer communicate with its upstream {\it along} a stream line, the acceleration is also related with the communication {\it across} the stream lines. The causality surface is therefore defined by beyond which  the jet can no longer communicate with the jet rotation axis at where the magnetic field are usually bunched up.
However, the causality surface is always located beyond the FMS. Therefore, including the consideration of the causality surface effect and the transition of the two different acceleration regime would {\it only modify the acceleration properties beyond the FMS}.  

\begin{figure*}
\begin{center}
\includegraphics[width=0.4\textwidth]{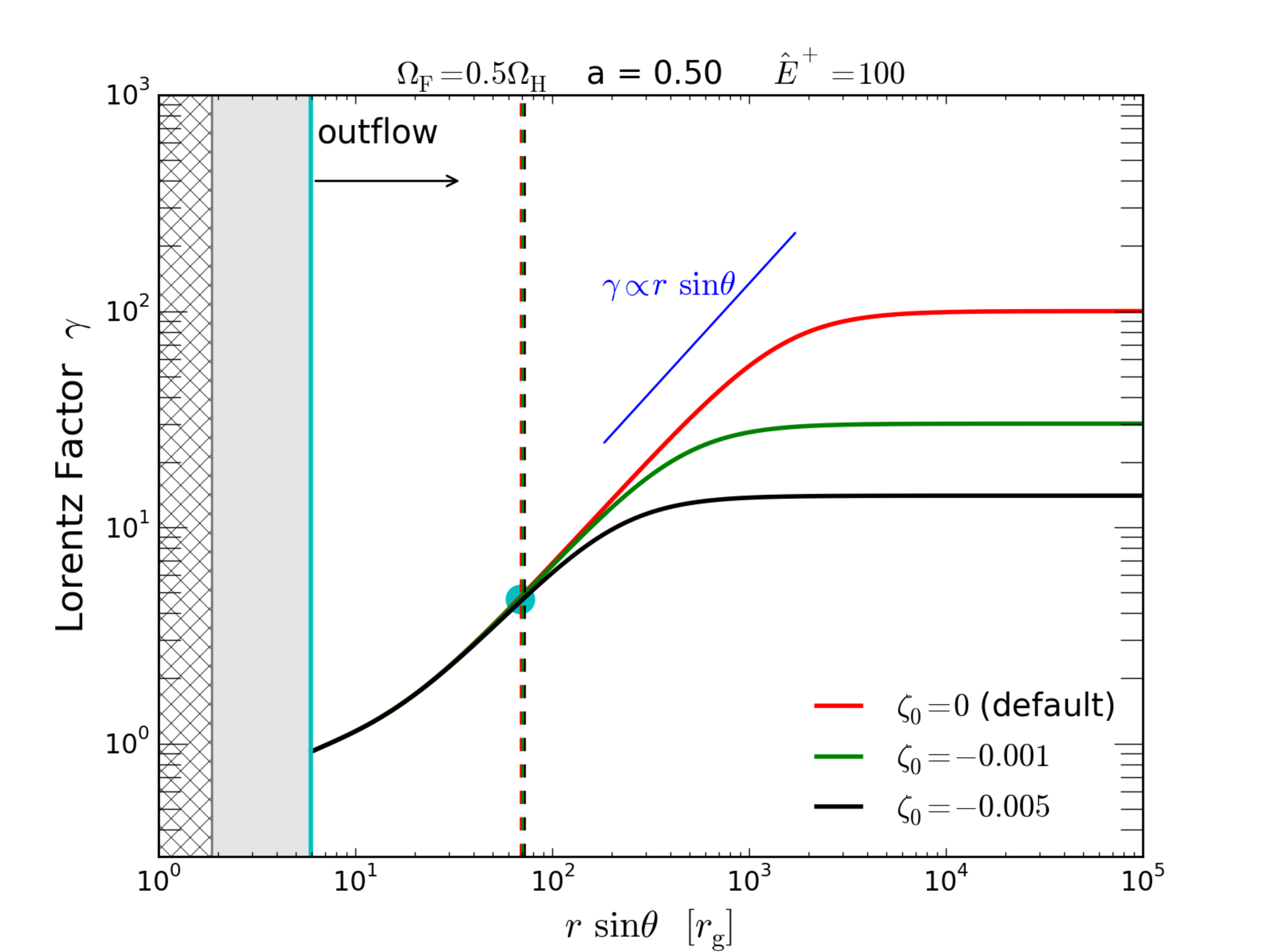}
\includegraphics[width=0.4\textwidth]{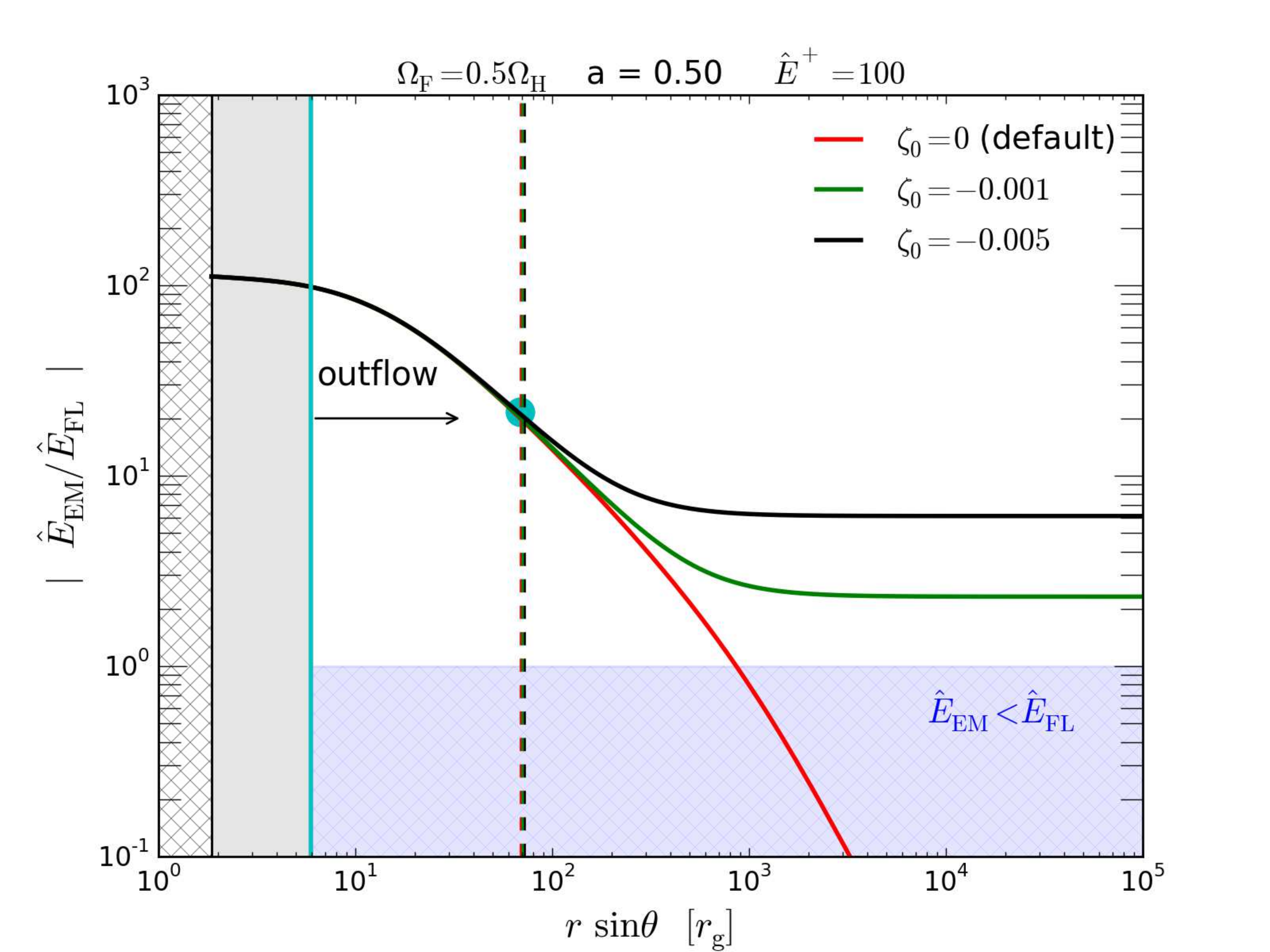}\\
\end{center}
\caption{Outflow solutions of different choice of $\zeta_{0}$, with the same setup in figures \ref{fig:para_gamma} and \ref{fig:para_E}.
The red profile shows the case $\zeta_{0}=0$,  the default value in our model.
Left panel: Lorentz factor $\gamma$ of the flow.
Right panel: The ratio between the electronmagnetic energy to particle energy as function of distance.
The location of the FMS for each cases are shown by the corresponding vertical dashed lines, which are almost overlapped with each other. 
The theoretical predicted value at the FMS from the perturbation method \citep[][]{bes06} for a SRMHD flow is overlapped by the filled cyan circle. The choice of $\zeta_{0}=0$ ensure a linear acceleration and an efficient conversion from electron magnetic energy to kinetic energy. A slightly deviation from $\zeta_{0}=0$ can further modify the conversion efficiency and the growth of Lorentz factor, but only beyond the FMS.}  \label{fig:par_zetas}
\end{figure*}

 In our model, the acceleration properties is associated with the term $\zeta_{0}$ in equation (\ref{eq:xi_def}). To demonstrate the effect of non-zero $\zeta_{0}$, we compare the parabolic outflow solution for the case $\hat{E}^{+}=100$ shown in figure \ref{fig:para_gamma} with other different choices of $\zeta_{0}$ in figure \ref{fig:par_zetas}. 
The right panel shows the energy conversion efficiency from electromagnetic part to particle part. 
 The vertical dashed lines (which are almost overlaped with each other) indicate the location of the FMSs for different choice of $\zeta_{0}$.
 A choice of $\zeta_{0}=0$ actually guarantees the linear acceleration (the first term second term in Equation (\ref{eq:gamma_decomp})), and therefore efficient energy conversion is obtained ($|\hat{E}^{+}_{\rm EM}/\hat{E}^{+}_{\rm FL}|<1$) far beyond the outer FMS, while for $\zeta_{0}<0$ the flows remain magnetically dominated ($|\hat{E}^{+}_{\rm EM}/\hat{E}^{+}_{\rm FL}|>1$). Note that for  $\zeta_{0}>0$ the flow velocity (or the Alfv'en Much number) becomes infinitely large at the critical radius (see, TT03).
 We therefore ignore the effect of the second term second term in Equation (\ref{eq:gamma_decomp}), and adopt $\zeta_{0}=0$ as the default value. Such choice guarantees the effect of the first term in Equation (\ref{eq:gamma_decomp}), and therefore
 $\gamma\approx\gamma_{1}(\propto r\sin\theta)$ \citep[][]{tch08,tch09,kom09} can be properly realized  (see Figures \ref{fig:para_gamma} and \ref{fig:para_differentF}).

It is possible to include the transition from linear grow to logarithmic grow by a further fine-tune of $\zeta_{0}$ with a generalized function $\zeta_{0}(\Psi)=\zeta_{0}(p, \hat{E}^{+}(\Psi), \eta (\Psi))$, depending on the location of the causality surface for a specific black hole magnetosphere. 
Nevertheless, as explained before, including the consideration of the causality surface effect and the transition of two different acceleration would only modify the acceleration properties beyond the FMS. This is shown by that the flow solutions of different choice of $\zeta_{0}$ are similar before passing the cyan circle. Thus, consideration of where the FMS will occur will be key in estimating the jet acceleration region.

\end{document}